\documentclass[useAMS,usenatbib]{mn2e}
\usepackage{morefloats}
\usepackage{subcaption}
\usepackage{graphicx}
\usepackage{hyperref}
\usepackage{natbib}
\usepackage{pdflscape}
\usepackage[above]{placeins}
\usepackage{lastpage}
\usepackage{longtable}
\title[A Near-Infrared Survey of the Inner Galactic Plane for Wolf--Rayet Stars III.  New Methods: Faintest WR Stars]{A Near-Infrared Survey of the Inner Galactic Plane for Wolf--Rayet Stars III.  New Methods: Faintest WR Stars}
\author[G. Kanarek et al.]{G.~Kanarek,$^1$ M.~Shara,$^2$ J. Faherty,$^3$ D.~Zurek,$^2$ A.~Moffat$^4$\\
$^1$Columbia University, 116th St \& Broadway, New York, NY 10027\\
$^2$American Museum of Natural History, 79th Street and Central Park West, NewYork, NY, 10024\\
$^3$Department of Terrestrial Magnetism, Carnegie Institution of Washington, 5241 Broad Branch Road NW, Washington, DC 20015\\
$^4$D\'{e}partment de Physique, Universit\'{e} de Montr\'{e}al, CP 6128 Succ. C-V, Montr\'{e}al, QC, H3C 3J7, Canada}

\makeatletter
\newcommand{\ion}[2]{#1$\;$\textsc{\rmfamily\@roman{#2}}\relax}
\makeatother
\newcommand{\fnm}[1]{\ensuremath{^{#1}}}
\newcommand{\fnt}[2]{\footnotesize{\fnm{#1} #2}}
\newcommand{\pres}[1]{\, { }#1\! }

\pagerange{\pageref{firstpage}--\pageref{LastPage}}

\begin{document}
\label{firstpage}

\maketitle

\begin{abstract}
A new method of image subtraction is applied to images from a $J$, $K$, and narrow-band imaging survey of 300~deg$^2$ of the plane of the Galaxy, searching for new Wolf--Rayet stars.  Our survey spans 150\degr in Galactic longitude and reaches $b=\pm1\degr$ with respect to the Galactic plane.  The survey has a useful limiting magnitude of $K=15$ over most of the observed Galactic plane, and $K=14$ (due to severe crowding) within a few degrees of the Galactic centre.  The new image subtraction method described here (better than aperture or even point-spread-function photometry in very crowded fields) detected several thousand emission-line candidates.  In 2011 and 2012 June and July, we spectroscopically followed up on 333 candidates with MDM--TIFKAM and IRTF--SPEX, discovering 89 emission-line sources.  These include 49 Wolf--Rayet stars, 43 of them previously unidentified, including the most distant known Galactic WR stars, more than doubling the number on the far side of the Milky Way.  We also demonstrate our survey's ability to detect very faint PNe and other NIR emission objects.
\end{abstract}


\section{Introduction}
In the more than 140 years since their first identification \citep{Wolf:jk}, Wolf--Rayet (WR) stars have remained one of the most interesting (and, at times, baffling) classes of stars.  Their huge masses ($\ge25$~M\sun\ for initial stellar mass) and short lifetimes (typically $\sim3\times10^5$~years in the WR phase) make them excellent tracers of recent star formation, and their position in the stellar evolutionary chain is important to both stellar astrophysics and supernova theory.  The intense stellar winds of WR stars add a significant amount of C and O, and some N, to the interstellar medium (ISM), and contribute a significant fraction of the ISM's energy and momentum budget; they also produce characteristic emission lines which give astronomers a probe into the atmospheres of these very hot, evolved stars.

The most reliable method to date of WR detection has been to look for narrowband excess due to emission lines in the optical, particularly the strong \ion{He}{2} 4686~\AA\ line.  However, probing the Milky Way using this technique is difficult beyond $\sim$3~kpc due to extreme dust extinction ($\sim$30~visual magnitudes across and through the centre of the Galaxy); in the near-infrared (NIR) only $\sim$3~magnitudes of extinction occurs across the Galactic plane \citep[see][section 7]{1999AJ....118..390S}.  Clearly the NIR is the wavelength range of choice for searching out the vast majority of WR stars in the Milky Way.

Over the last 5 years, advances in NIR selection techniques have led to the identification of more than 250 new WR stars, using a variety of methods. Two earlier papers in this series have focused on photometric techniques, and identified 112 new WR stars in the Galactic Plane \citep[][hereafter Paper I and Paper II, respectively]{2009AJ....138..402S,2012AJ....143..149S}. In addition, various colour-selection criteria in the near- and mid-infrared have been developed; see \citet{2007MNRAS.376..248H}, \citet{2009AAS...21460509M}, \citet{2011AJ....142...40M}, and \citet{2012A&A...537A..10M} (as well as \citet{2014AJ....147..115F} for a discussion of the specific techniques used in this paper).

Simple models of the WR distribution in the Galaxy predict high concentrations near the Galactic centre (Paper I), where extremely crowded images cause usual photometric techniques to fail. However, using a new method of image subtraction, along with NIR and MIR colour-cuts we are able to significantly improve the selection criteria in the most crowded parts of the Galaxy; follow-up of candidates with this new technique produced 89 new emission-line sources in the Galactic Plane, including 49 new Wolf--Rayet stars.  In section~\ref{sec:obs} we briefly describe the imaging survey and data reduction pipeline. Spectrographic follow-up data reductions are described in section~\ref{sec:spec}, and the new candidates are presented in section~\ref{sec:res}. The resulting overall distribution of all known WR stars is presented and discussed in section~\ref{sec:dist}. We briefly summarize our conclusions in section~\ref{sec:end}.

\section{Survey \& Data Reduction Pipeline}\label{sec:obs}
This survey was previously described in Paper I.  More than 88,000 exposures were taken of the Galactic plane on the CTIO 1.5-m telescope over approximately 200 nights during 2005-2006.   The survey covers $1^\circ$ above and below the Galactic plane, from longitudes $-90^\circ$ to $60^\circ$.  The images are $35\times35$~arcmin$^2$, with $1.03$~arcmin pixel$^{-1}$ plate scale, in four narrowband filters as well as the $J$ and $K$ bands.  The narrowband filter set is described in Paper I, with the central wavelengths and FWHMs of the emission-line filters repeated in table~\ref{tab:cal}.  The justification for the narrowband filters is: WCs can be discriminated using the ratio of the \ion{C}{4} 2.081~\micron\ filter to a 2.112~\micron\ \ion{He}{1} line, and WNs with the ratio between the \ion{He}{2} 2.192~\micron\ filter and each of the 2.112~\micron\ line and the 2.169~\micron\ Br $\gamma$ filters.

\begin{table*}
\caption{Predicted candidate strength in survey filters, with central wavelength and FWHM\label{tab:cal}}
\begin{center}
\begin{tabular}{@{}lllll@{}}
\hline \noalign{\smallskip}
\textbf{Filter Name} & \ion{He}{1} & \ion{C}{4} & Br $\gamma$/\ion{He}{1} & \ion{He}{2} \\ 
\noalign{\smallskip}\hline
$\lambda$ (\micron) & 2.062 & 2.081 & 2.169 & 2.192\\
$\Delta\lambda$ (\micron) & 0.010 & 0.020 & 0.020 & 0.020\\
\hline
Early WN & none - very weak & none\fnm{a} & weak & strong - very strong\\
Late WN & very strong & none\fnm{a} & very strong & weak - strong\\
Early WC & weak - strong & very strong & none & none - very weak\\
Late WC & weak - strong & very weak - weak & very weak & none - weak\\
WO & none & weak & strong & weak\\
\hline
\end{tabular} 
\end{center}
\begin{flushleft}\fnt{a}{WNs may be detected in the \ion{C}{4} filter due to the wings of a strong \ion{He}{1} line.  This is a much more likely occurrence in WNLs than WNEs because of the relative strength of the \ion{He}{1} line.}\end{flushleft}
\end{table*}

The images from this survey were completely re-reduced for the new image-differencing pipeline described in this paper.  Super dome flats for each month were created by combining all dome flats in that month.  The images were then divided by the master dome flats.  A skyflat was created by median-combining the first dither positions from all the separate pointings.  The skyflat was then scaled to match each image in the dither sequence and subtracted.  Next, \textsc{daophot}  \citep{1987PASP...99..191S} was used to find sources in each of seven dither positions for every field. The sources were matched using \textsc{daomatch} and  \textsc{daomaster}, and finally the 7 dither positions were combined using \textsc{montage2} \citep{1994PASP..106..250S}.

World Coordinate System (WCS) astrometry for the images was computed by the \textsc{astrometry.net} package \citep{2010AJ....139.1782L}, using $J$ and $K$ index files with skymark diameter ranges from 8 to 11~arcmin.  The program was called with xy-lists obtained by the NASA \textsc{idl} Astronomy Library's version of \textsc{daophot}'s \textsc{find} procedure \citep{1993ASPC...52..246L}, using the 500 brightest stars in each image.  After including SIP distortion coefficients, mean residuals less than 0.5~arcsec were found when compared to \textit{2MASS} sources \citep{2006AJ....131.1163S}.  Then, working on a field-by-field basis, we computed photometry using the \textsc{idl-daophot} procedures \textsc{sky}, \textsc{find}, and \textsc{aper}, and the positions output by these routines were matched among the images, to a precision of $\sim1''$.  Sources were then matched to \textit{2MASS} source lists in $J$ and $K$, and a mean magnitude offset calculated to calibrate the photometry library to \textit{2MASS}.

Once WCS solutions and photometry had been obtained, we performed image differencing in wavelength-space, which provided a method to find rare emission-line sources from very crowded images.  We generated a continuum image for each emission-line-centred narrowband image (ENB) by linearly interpolating between the two continuum narrowband (CNB) images in wavelength space; this interpolated image (INB) was then subtracted from the emission line image to produce the residual.  Images were subdivided to allow modelling of the spatially-varying sky background by an array of intersecting planes.  After removal of the background, stellar positions in each subdivision were matched between images and a translation/rotation warp was applied to spatially align subdivisions as accurately as possible.  Higher-order warp solutions were not found to improve alignment significantly. 

After warping the subdivisions to match, we scaled the global brightness between the ENB and INB images.  Through trial and error we concluded that no single method of determining a scaling factor worked for all images (or even all subdivisions); thus, to remove outliers, five different scaling factors were determined for each subdivision, using different methods.  The final scaling factor was chosen as the median value of all methods on all subdivisions.  No PSF matching was performed, as (a) each image had a PSF which displayed large spatial variations, and (b) the blurring due to convolution was generally catastrophic to the resolution of the images.

\begin{figure*}
\centering
\begin{subfigure}{.45\textwidth}
  \centering
  \includegraphics[width=\linewidth]{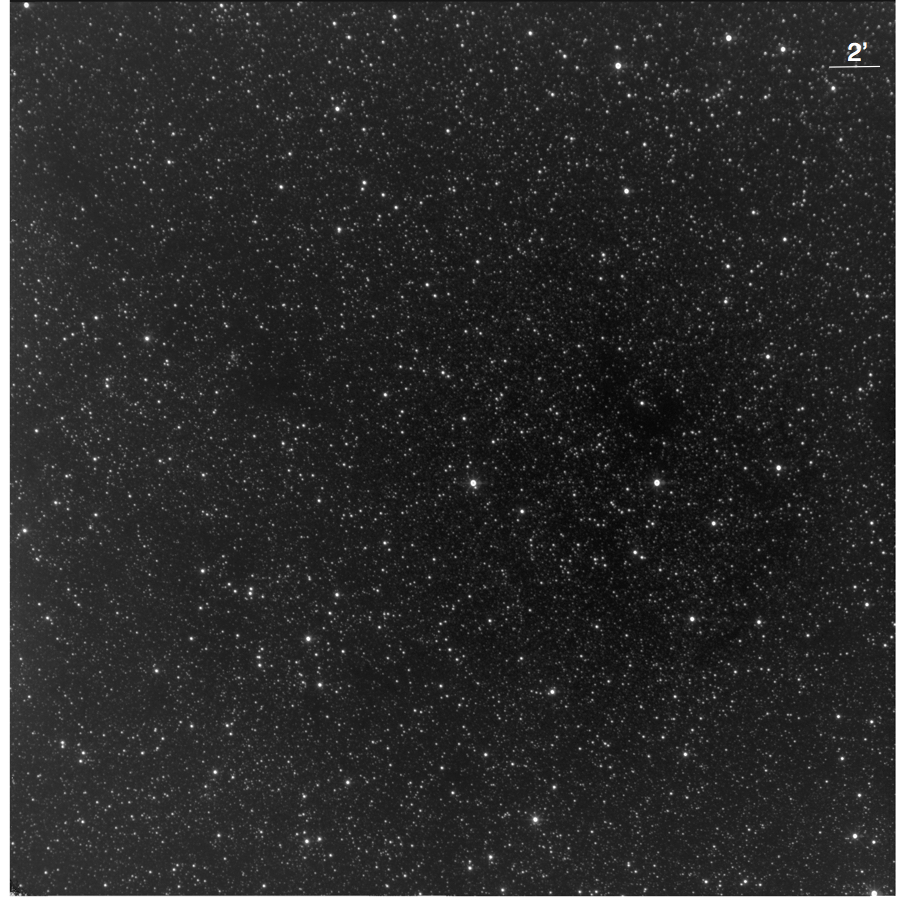}
  \subcaption{A quadrant of the \ion{C}{4} image from field 1093 (a representative survey field), before image differencing.}
  \label{fig:civ-left}
\end{subfigure}
\begin{subfigure}{.45\textwidth}
  \centering
  \includegraphics[width=\linewidth]{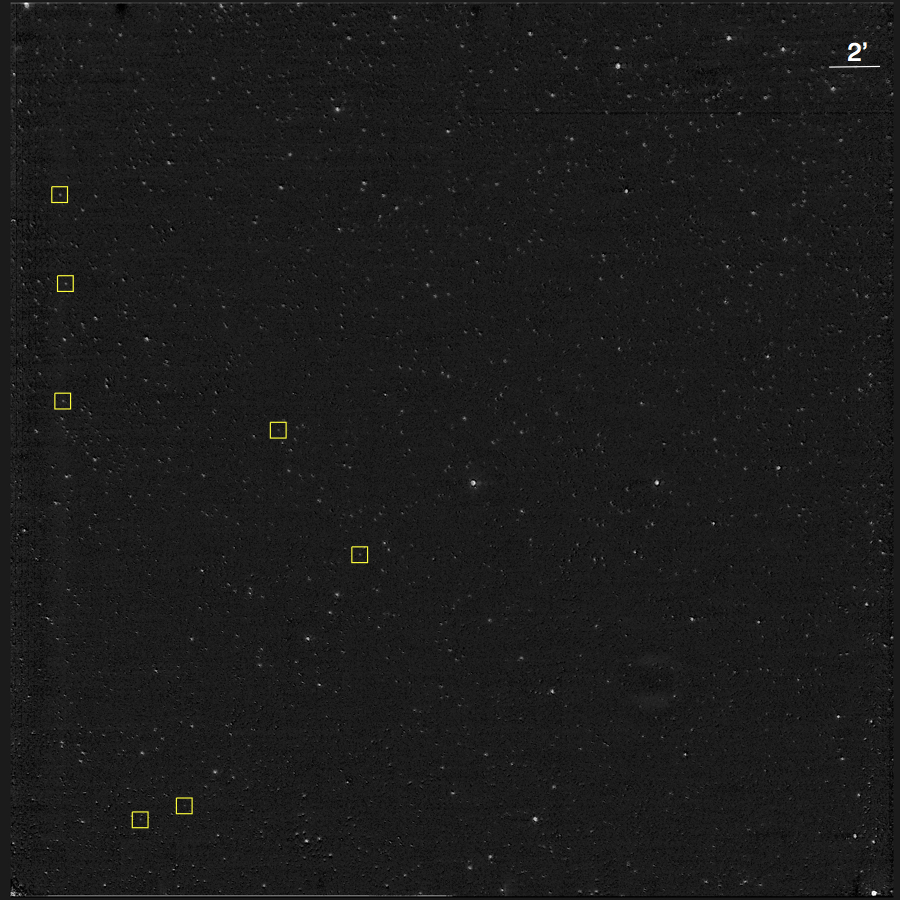}
  \subcaption{The same image, after image differencing, with known Wolf--Rayet stars indicated.}
  \label{fig:civ-right}
\end{subfigure}
\caption{\label{fig:civ}A demonstration of the results of image differencing. We subtract a narrow-band continuum image from the narrow-band emission-line-centred image; both are scaled to the same wavelength and intensity, so that most stars should subtract to zero. This image demonstrates the high degree of crowding present in the survey images (the displayed field is $\sim24\degr$ from the Galactic centre). Residuals ideally are candidate emission objects. The great majority of point sources have been completely removed; residual flux from most of the remaining non-emission stars is due to incomplete subtractions of saturated stars ($K\le9$), or from inadequate PSF matching due to high spatial variability in the PSF between and within the images. Previously-known WR stars from the literature are identified with yellow squares in figure~\ref{fig:civ-right}.}
\end{figure*}

\begin{figure*}
\includegraphics[width=\textwidth]{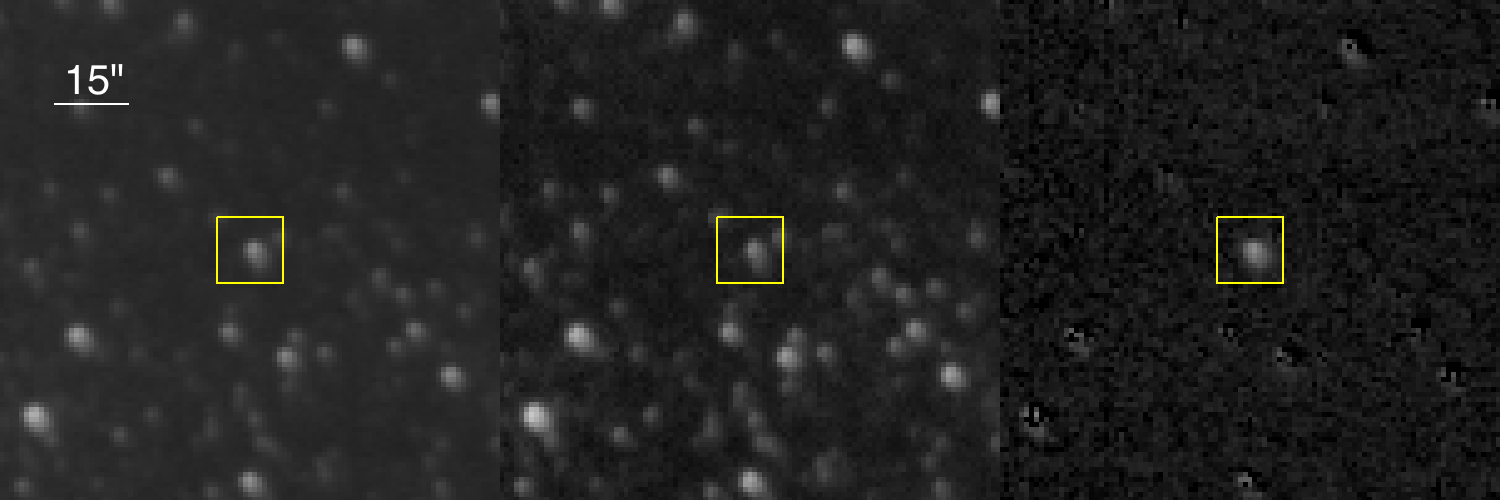}
\caption{A zoomed portion of the image in figure~\ref{fig:civ}, highlighting a known WR star in the \ion{C}{4} narrow band image (left), \ion{C}{4} interpolated continuum image (centre), and difference image (right). The WR star remains as a perfect residual PSF in the difference image, while the great majority of source pixels elsewhere are removed by the subtraction process, leaving only the brightest sources in the original \ion{C}{4} image as incomplete subtractions. \label{fig:civzoom}}
\end{figure*}

Once the ENB and INB images were matched as closely as possible, the difference image was obtained.  Figure~\ref{fig:civ} shows the \ion{C}{4} ENB image (left) and difference image (i.e. on-line intensity including continuum, minus interpolated and scaled off-line continuum-only intensity, right) for field 1093, a typical field which lies $24\degr$ from the Galactic Centre, and is relatively crowded, containing 6 previously known WR stars.  The difference image in figure~\ref{fig:civ} demonstrates how both the spatially-varying sky and more than 99 per cent of the stellar sources were removed by the subtraction process.  Figure~\ref{fig:civzoom} shows a previously known WR star in that field.  Despite the lack of PSF fitting, residual artefacts for the non-emission sources are quite minimal compared to the strong residual-emission PSF of the WR star.

The final determination of candidates from the difference images involved a 3-step process.  Each difference image contained many residuals that were the product of bad subtractions instead of true stellar-line emission; these must be removed.  First, we determined the magnitude of every star which produced a significant positive residual in the difference image ($m_{diff}$), and normalised this magnitude by the original magnitude of the star in the CNB images ($m_c =  (m_{C1}+m_{C2})/2$) to produce $\Delta m = m_{diff} - m_c$, a metric conceptually similar to equivalent width.  

We then plotted $m_c$ vs $\Delta m$ for each filter, including a large number of fields on each plot, and determined isodensity contours. Then, $m_c$ vs $\Delta m$ was plotted for each filter in each individual field, and the 99 per cent density contour was overlaid.  Only those sources with significant bright deviations outside the 99 per cent contour were considered as candidates, to eliminate field stars with no emission that survived the subtraction due to random fluctuations and poor PSF matches.

\begin{figure}
\includegraphics[width=\linewidth]{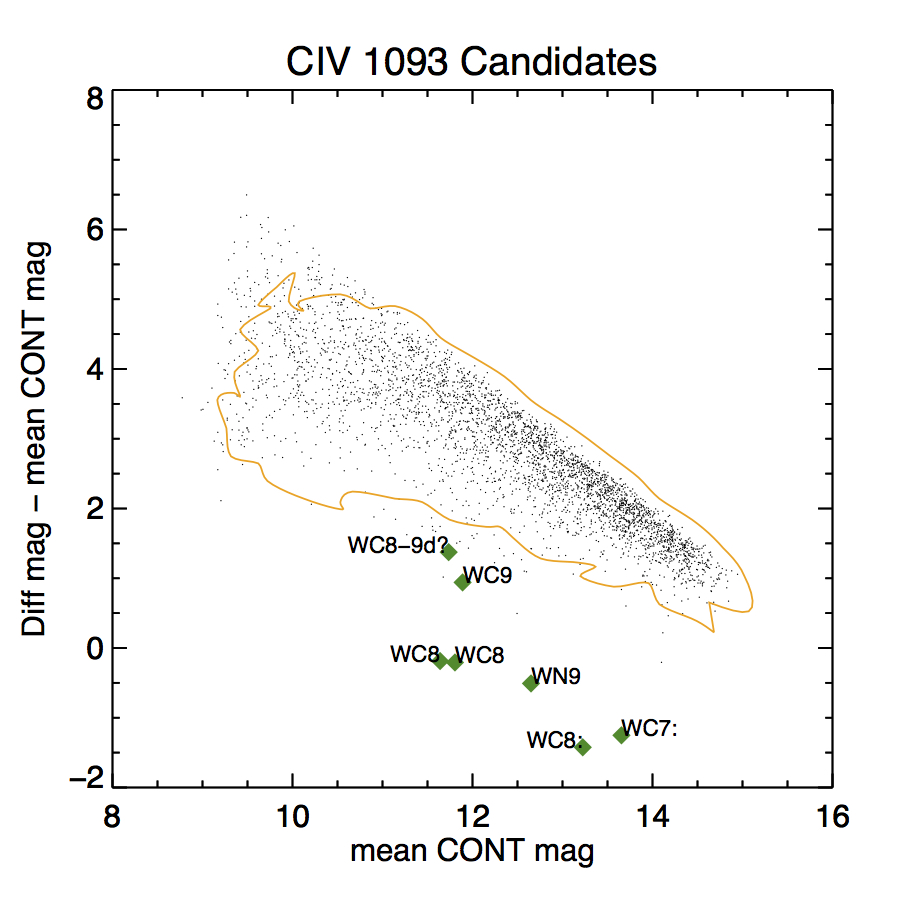}
\caption{\label{fig:dmag}The diagnostic (`$\Delta m$') plot for field 1093, in the \ion{C}{4} filter (the the image shown in figures~\ref{fig:civ} and~\ref{fig:civzoom}).  The difference image magnitude ($m_{diff}$) on the y-axis has been normalised for the original source brightness by subtracting continuum magnitude ($m_c$), leaving $\Delta m$ the normalised emission-line excess. By enclosing 99 per cent of the points in the orange isodensity contour, we isolate those sources with excess that are least likely to be incomplete subtractions; known WR stars in this field, the green labelled points, lie well outside the contour. To survive the culling process, prospective candidates must lie below the contour, in a similar region to the WRs from the literature.}
\end{figure}

Figure~\ref{fig:dmag} shows the $\Delta m$ plot for the \ion{C}{4} filter in field 1093.  Determining the optimal region of the $\Delta m$ plot for strong WR characteristics was accomplished by plotting previously known Wolf--Rayet stars, and identifying areas where particular types seemed clustered, particularly with separation from the main density of residual field stars.  By overplotting smoothed 99 per cent density contours on to these diagnostic plots, we were able to isolate strong candidates for further refinement.

\begin{figure*}
\centering
\hspace*{\fill}%
\begin{subfigure}{.48\textwidth}
  \centering
  \includegraphics[width=\linewidth]{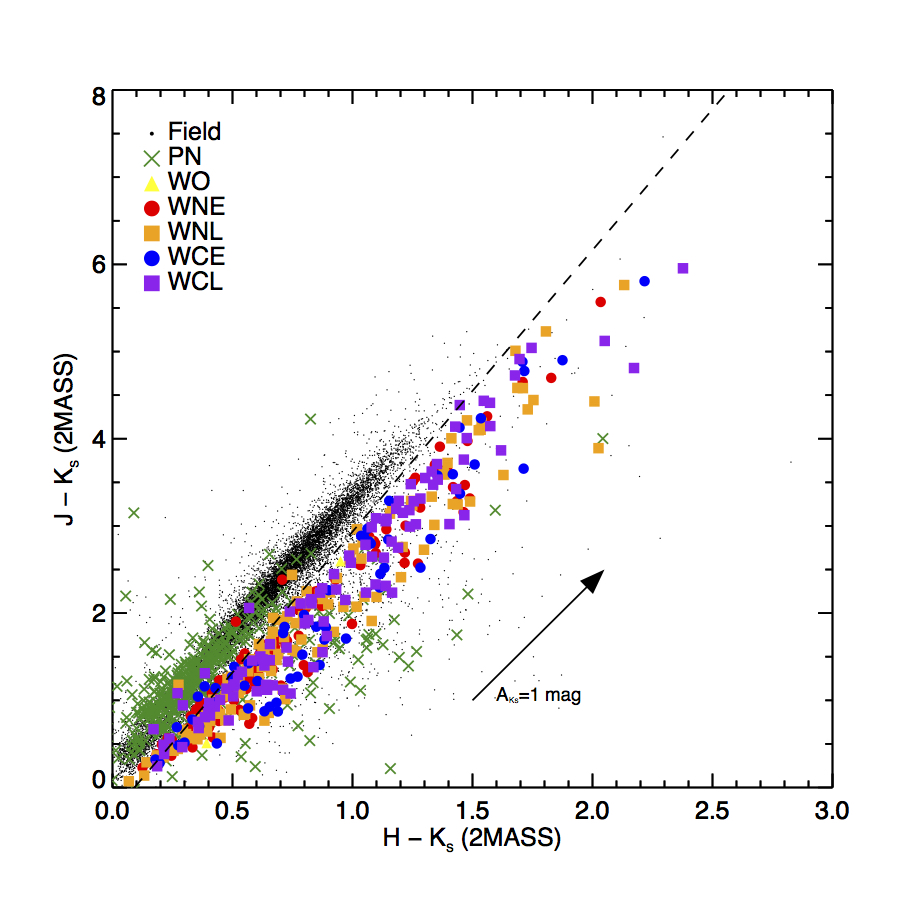}
  \label{fig:cuts-left}
\end{subfigure}\hfill%
\begin{subfigure}{.48\textwidth}
  \centering
  \includegraphics[width=\linewidth]{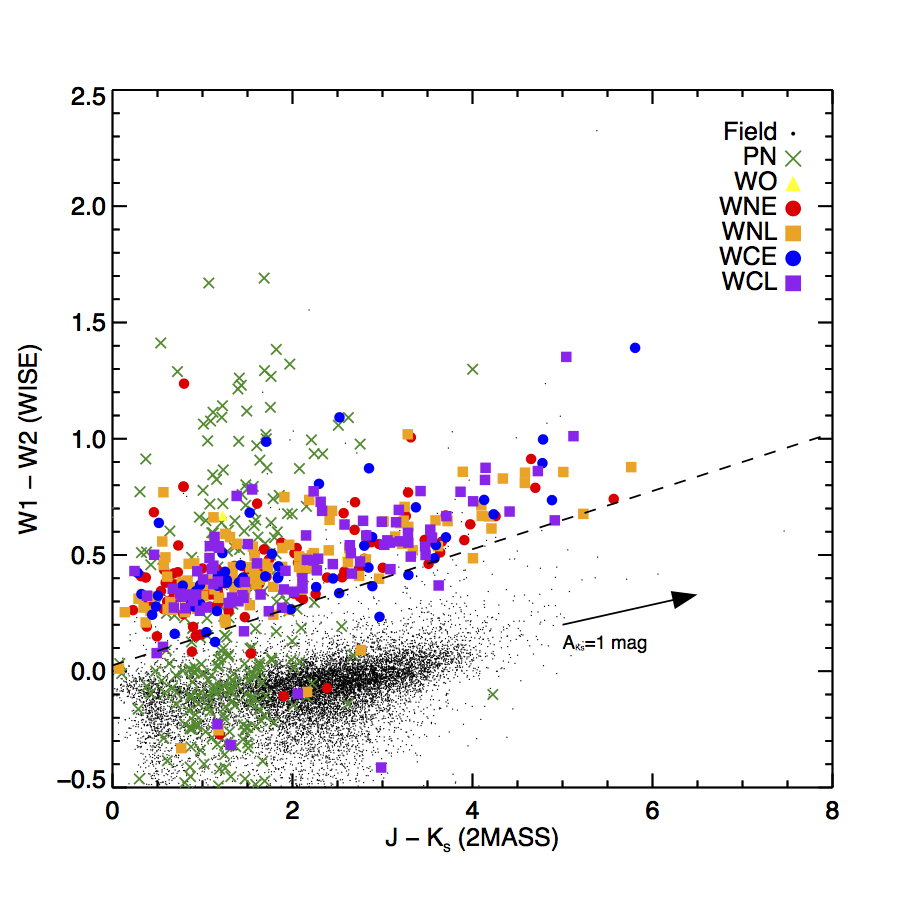}
  \label{fig:cuts-right}
\end{subfigure}%
\hspace*{\fill}%
\caption{The second diagnostic tool for candidate selection: colour cuts from NIR (\textit{2MASS}) and MIR (\textit{WISE}) magnitudes. Left shows a colour-colour diagram in the NIR only, while the right includes both NIR and MIR colours. These are analogous to the figures in \citet{2011AJ....142...40M}, described in more detail in \citet{2014AJ....147..115F}. These plots provide another layer of  differentiation during the selection process. PNe are included as common emission objects other than WR stars which can be found using these tools; 18 newly-identified PNe are presented in this paper. A$_{Ks}=1$ reddening vectors have been added, using the values for A$_{[\lambda]}/$A$_{Ks}$ from \citet{2005ApJ...619..931I}; the values for [3.6] and [4.5] were used for \textit{WISE} filters W1 and W2, as the central wavelengths are sufficiently similar. The right figure in particular shows the wide separation between emission objects (WR stars and PNe) and field stars in NIR/MIR colour space.\label{fig:cuts}}
\hspace*{\fill}%
\begin{subfigure}{.48\textwidth}
  \centering
  \includegraphics[width=\linewidth]{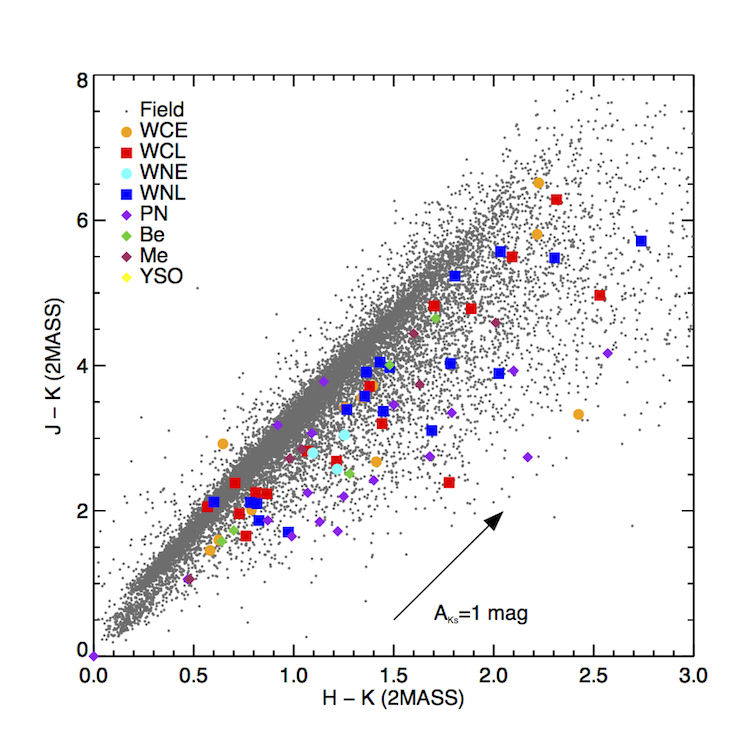}
  \label{fig:newcuts-left}
\end{subfigure}\hfill%
\begin{subfigure}{.48\textwidth}
  \centering
  \includegraphics[width=\linewidth]{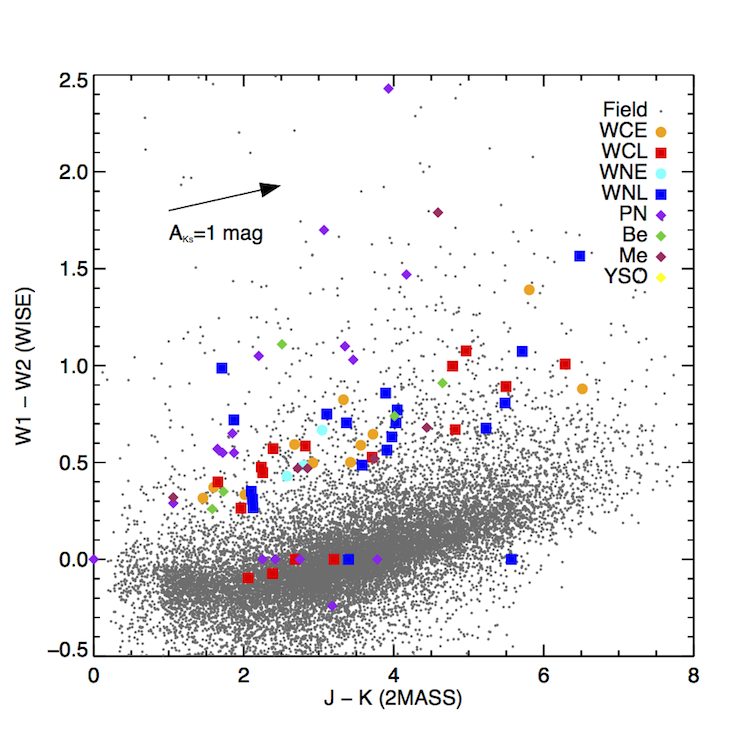}
  \label{fig:newcuts-right}
\end{subfigure}%
\caption{\label{fig:newcuts}  Colour-colour plots like those in figure~\ref{fig:cuts}, displaying the NIR (left) and MIR (right) colours for the new emission-line objects classified in this work. Grey dots are field stars; orange and cyan circles are early-type WCs and WNs respectively, while red and blue squares are late-type WCs and WNs. Purple diamonds are PNe, green diamonds are Be stars, brown diamonds are emitting M-giants and supergiants, and yellow diamonds are probable YSOs. Three new WR stars are below the cut line in the right panel; these have extremely close neighbours which lie on the same \textit{WISE} pixel and confuse the magnitudes. The reddening vectors are as in figure~\ref{fig:cuts}.}
\hspace*{\fill}%
\end{figure*}

The second step of the selection process was to apply MIR and NIR colour cuts.  We matched every source in each image to entries in the \textit{2MASS} and \textit{WISE} \citep{2010AJ....140.1868W} point source catalogues, and applied the following colour cuts, as per \citet{2014AJ....147..115F}:
\begin{eqnarray*}
J-K_s &<& 3.23(H-K_s) - 0.296 \\
W1-W2 &>& 0.125(J-K_s) + 0.025
\end{eqnarray*}
The laft panel of figure~\ref{fig:cuts} is a colour-colour diagram showing the NIR colour cut, using \textit{2MASS} colours only.  The great majority of WR stars lie below the cut line, while the great majority of field stars lie above it.  In the right panel, a colour-colour plot including colours from the \textit{WISE} photometry, the separation is even more pronounced.  WR stars are intrinsically very hot, and therefore blue, and so it should come as no surprise that no WR stars lie below the cut line, with at least a 0.2 magnitude separation from the main bulk of field stars (chosen at random from the \textit{2MASS} and \textit{WISE} catalogues).  The physical reason for the positions of WR stars on the colour-colour diagrams in figure~\ref{fig:cuts} is discussed in detail in \citet{2014AJ....147..115F}.

Figure~\ref{fig:cuts} also includes a selection of Galactic planetary nebulae (PNe), selected from the Strasbourg--ESO catalogue \citep{1992secg.book.....A}, which are one of the primary contaminants in the emission-line candidate sample.  The third and final step was to visually blink all candidates that both satisfied the colour cuts and had significant deviations on the $\Delta m$ plot; this method serves to remove PN contaminants which are resolved, and (along with other colour cuts) removes the great majority of likely PNe from the candidate list.\FloatBarrier

\begin{figure*}
\centering
\hspace*{\fill}%
\begin{subfigure}{.48\textwidth}
  \centering
  \includegraphics[width=\linewidth,clip=true,trim=0.25in 0.25in 0in 1in]{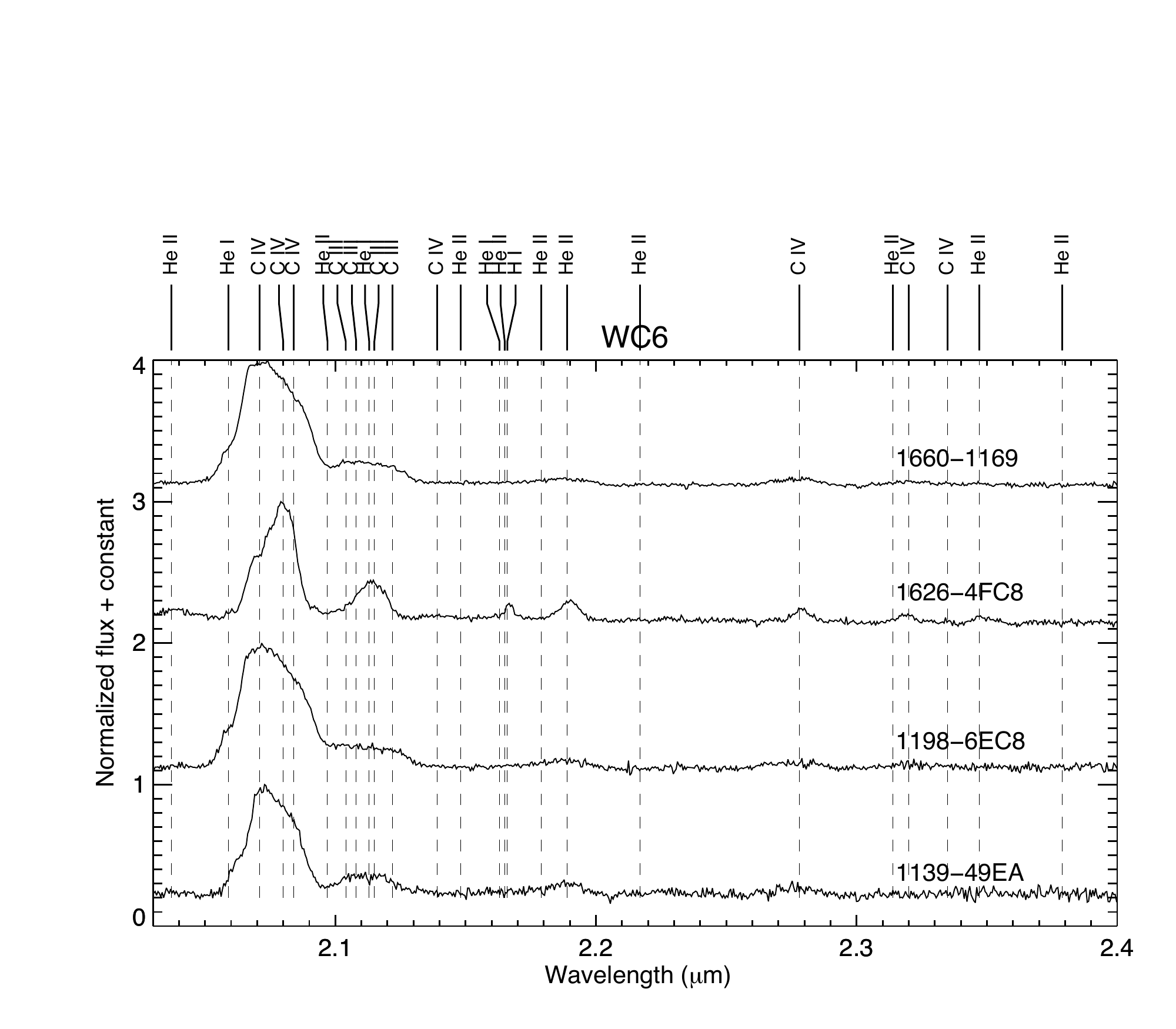}
  \includegraphics[width=\linewidth,clip=true,trim=0.25in 0.25in 0in 1.5in]{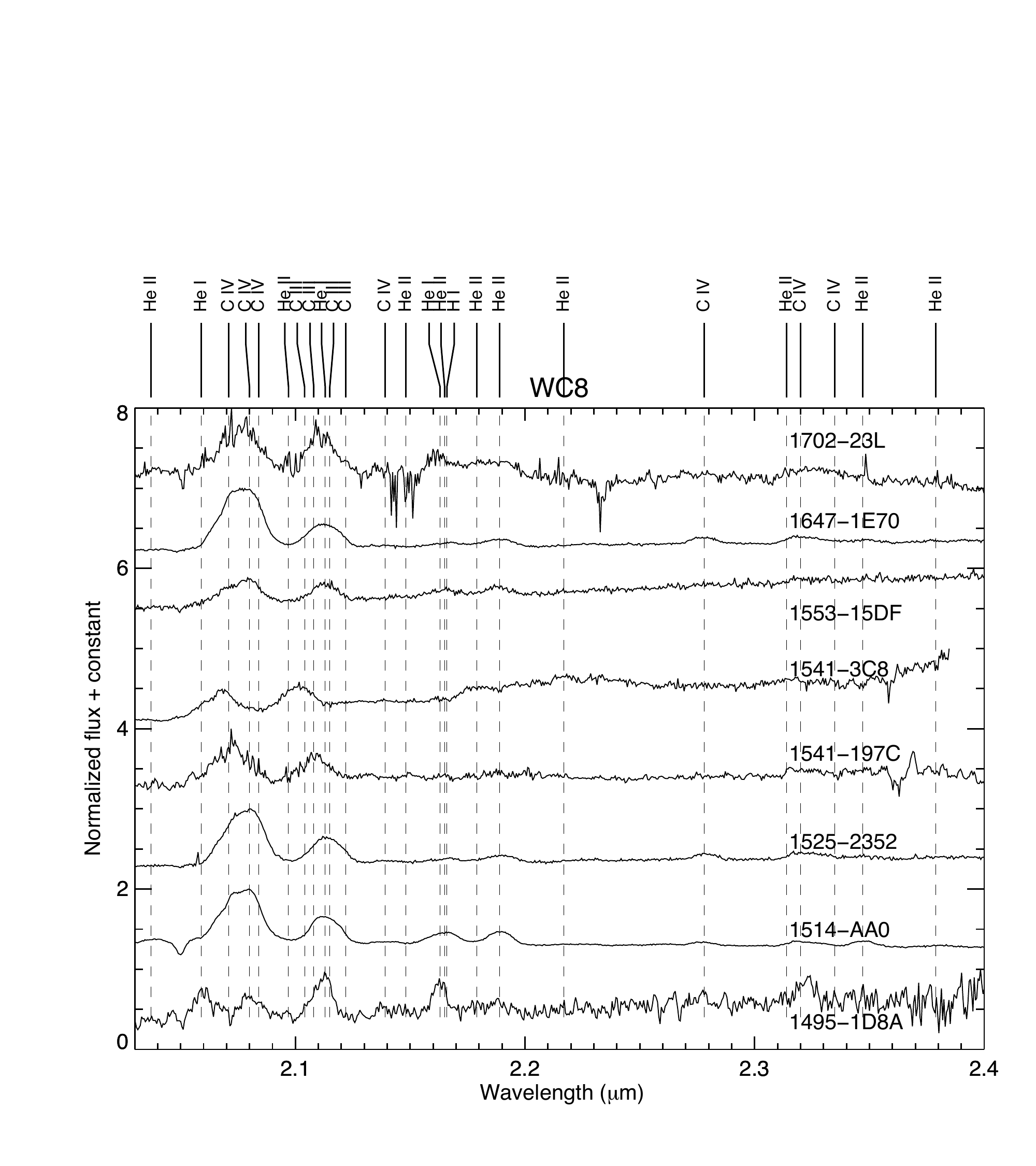}
\end{subfigure}\hfill%
\begin{subfigure}{.48\textwidth}
  \centering
  \includegraphics[width=\linewidth,clip=true,trim=0.25in 0.25in 0in 1in]{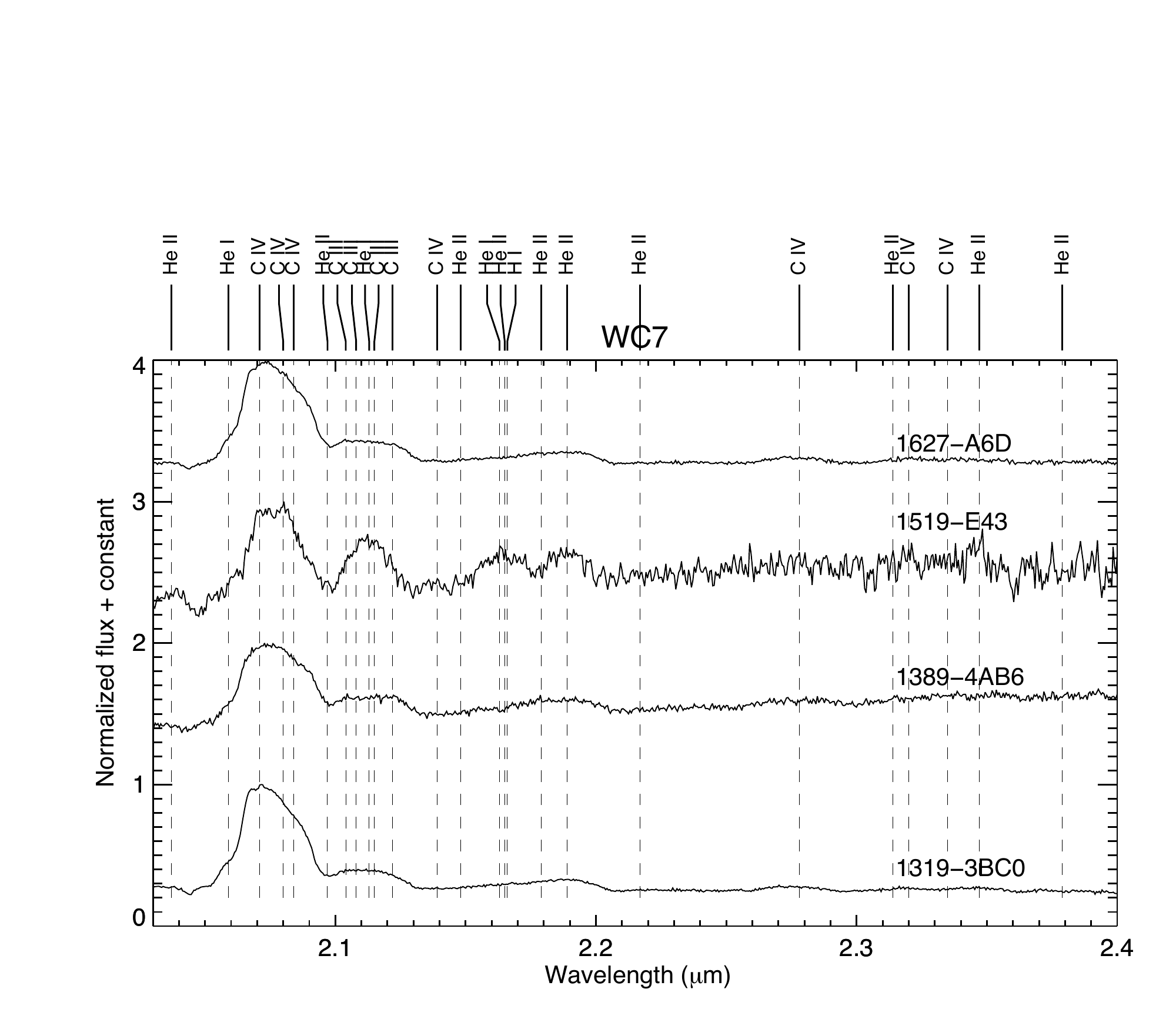}
  \includegraphics[width=\linewidth,clip=true,trim=0.25in 0.25in 0in 1.5in]{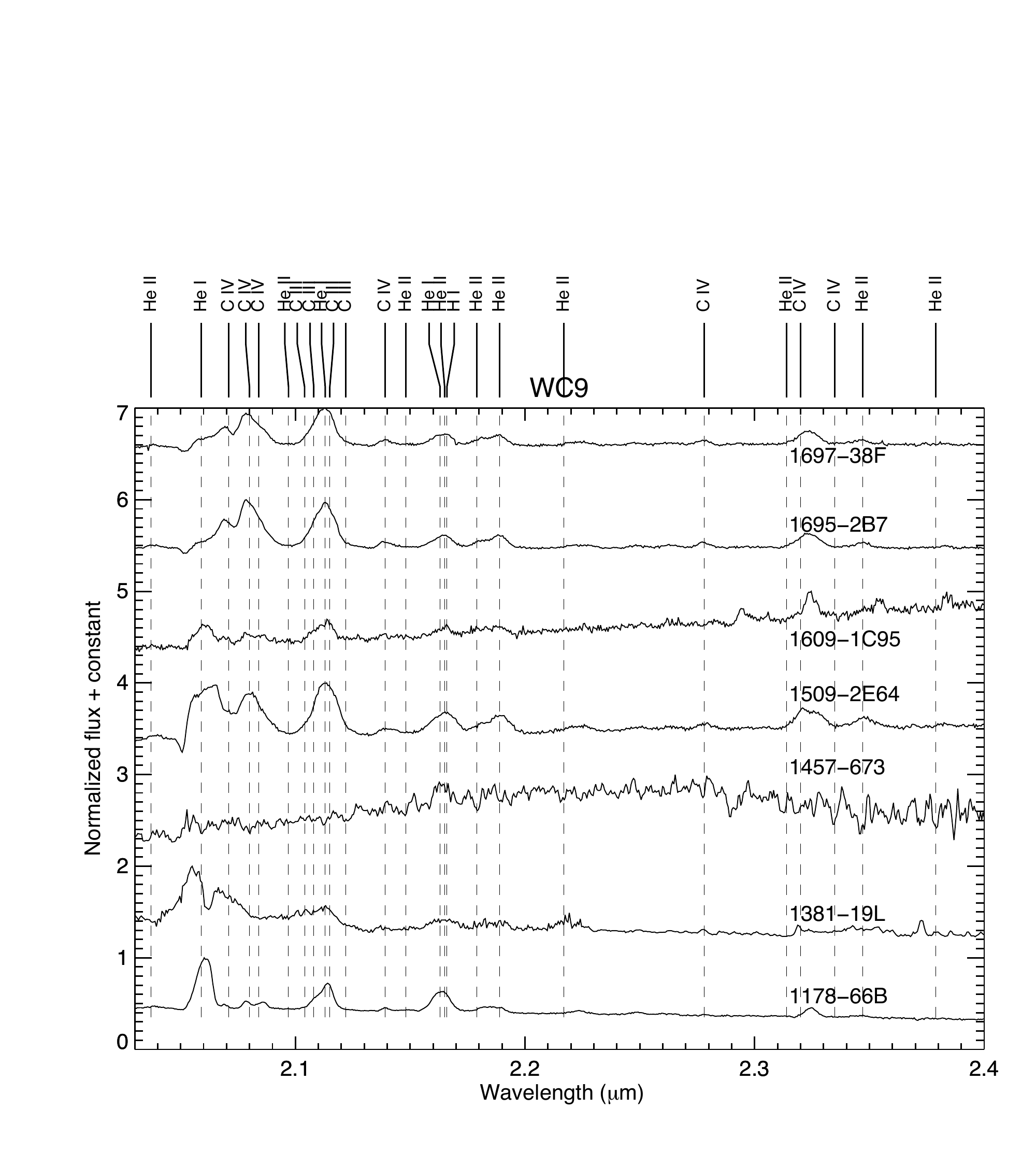}
\end{subfigure}%
\hspace*{\fill}%
\caption{\label{fig:wc} All new WC6 (top left), WC7 (top right), WC8 (bottom left), and WC9 (bottom right) objects classified in this work. Note the characteristic extremely strong \ion{C}{4} lines in the early WC6s.}
\end{figure*}

\begin{figure*}
\centering
\hspace*{\fill}%
\begin{subfigure}{.48\textwidth}
  \centering
  \includegraphics[width=\linewidth,clip=true,trim=0.25in 0.25in 0in 1.5in]{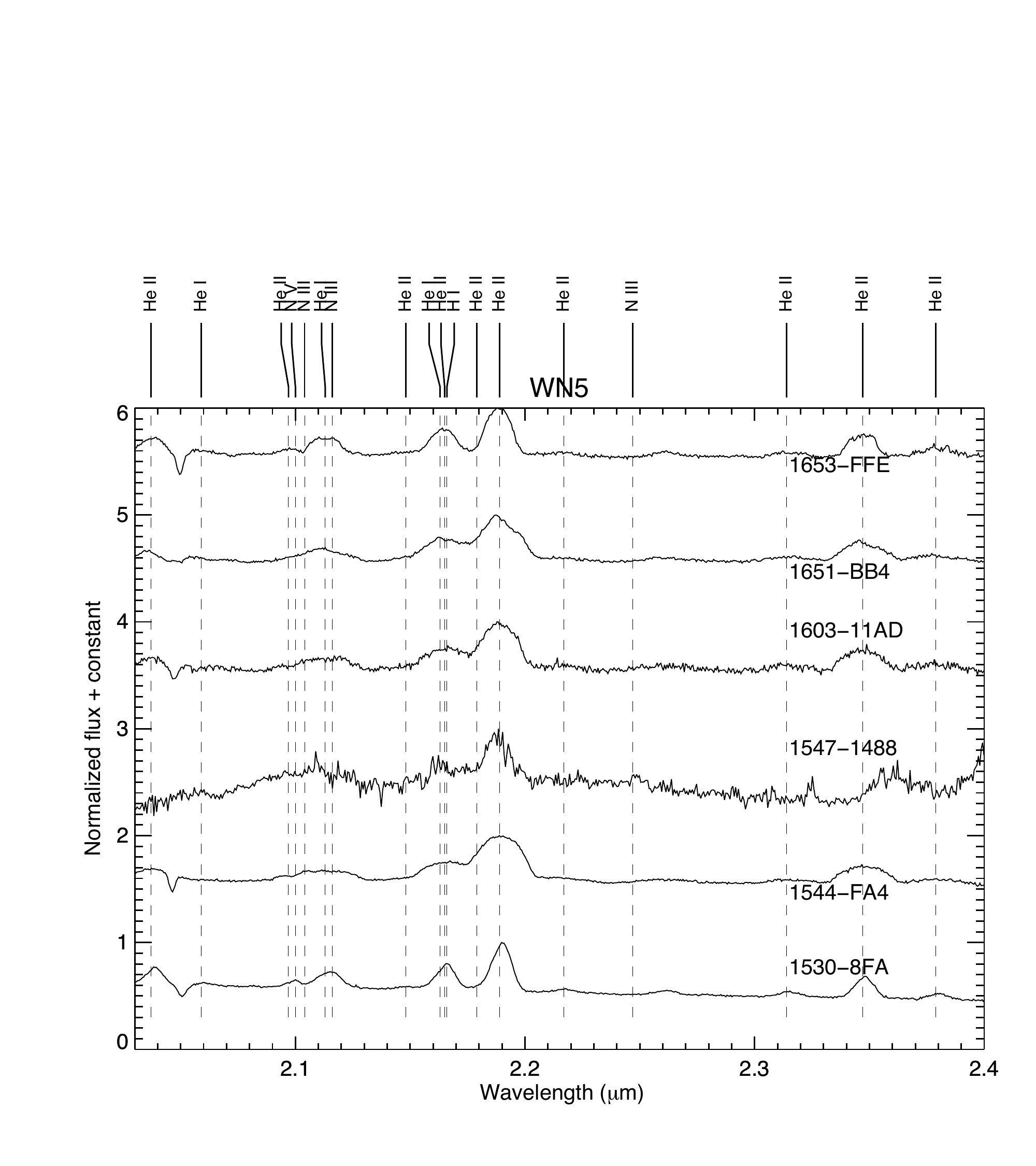}
  \includegraphics[width=\linewidth,clip=true,trim=0.25in 0in 0.15in 0.75in]{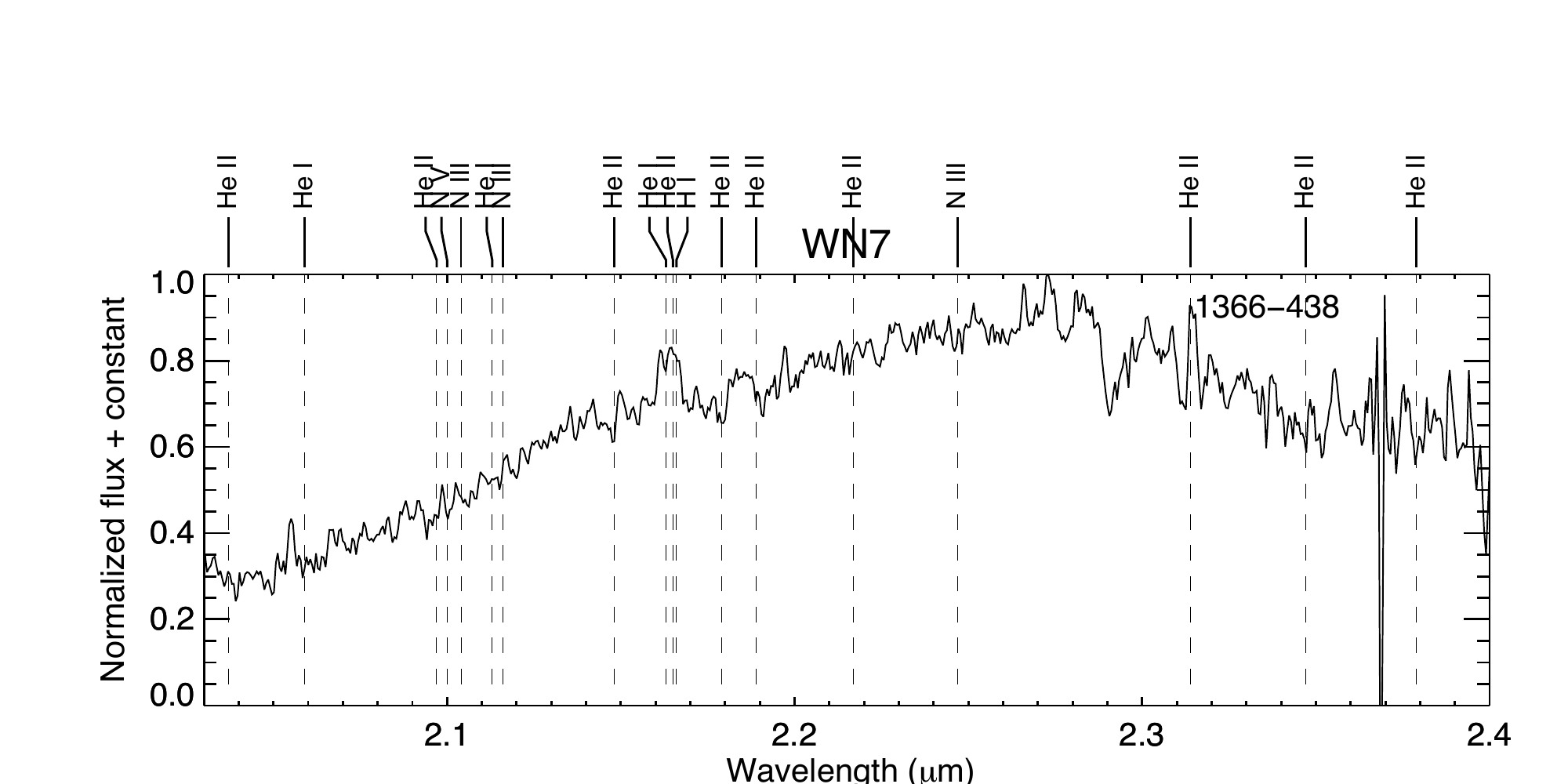}
  \includegraphics[width=\linewidth,clip=true,trim=0.25in 0.25in 0in 1.2in]{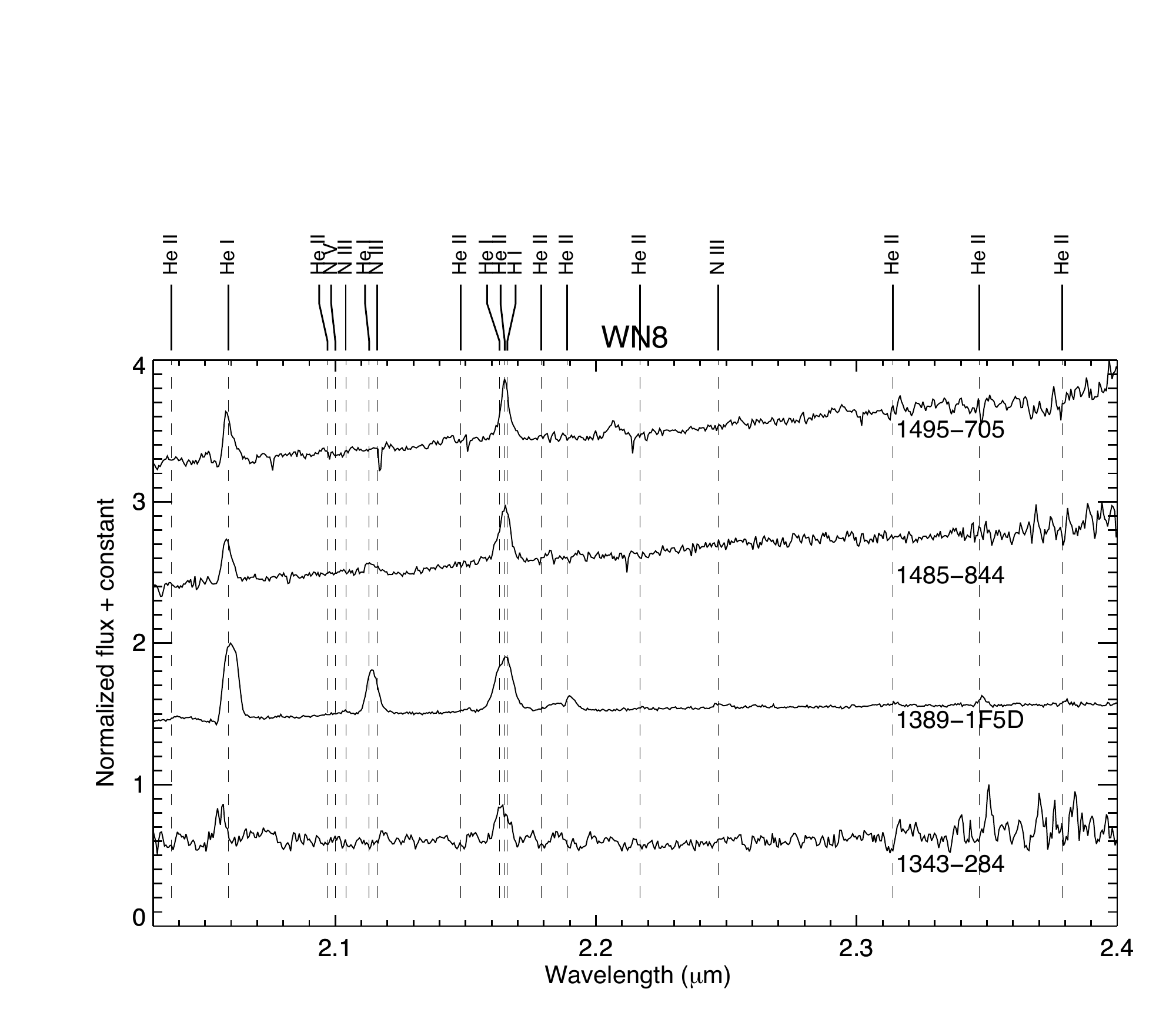}
\end{subfigure}\hfill%
\begin{subfigure}{.48\textwidth}
  \centering
  \includegraphics[width=\linewidth,clip=true,trim=0.25in 0.25in 0in 1.5in]{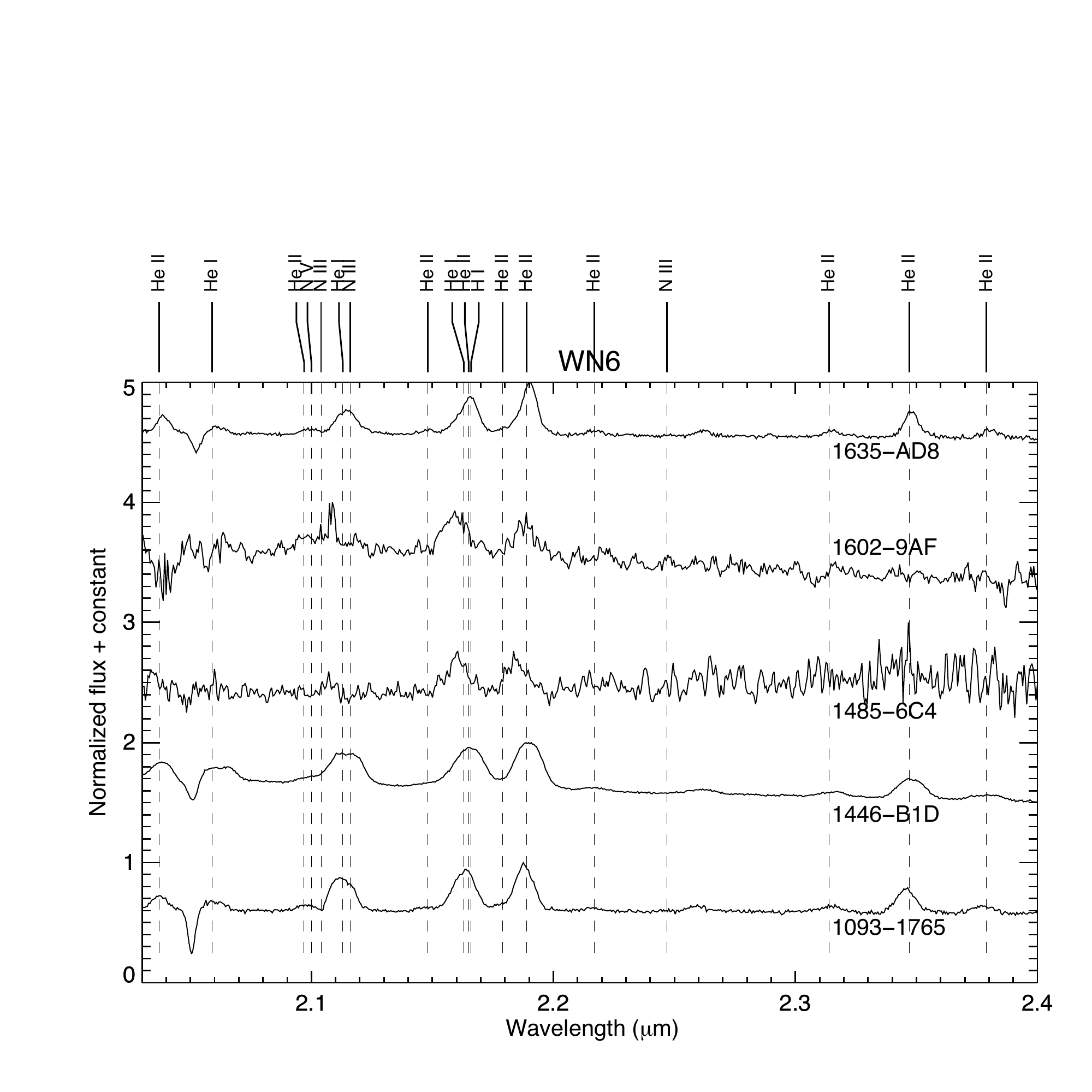}
  \includegraphics[width=\linewidth,clip=true,trim=0.25in 0.25in 0in 2in]{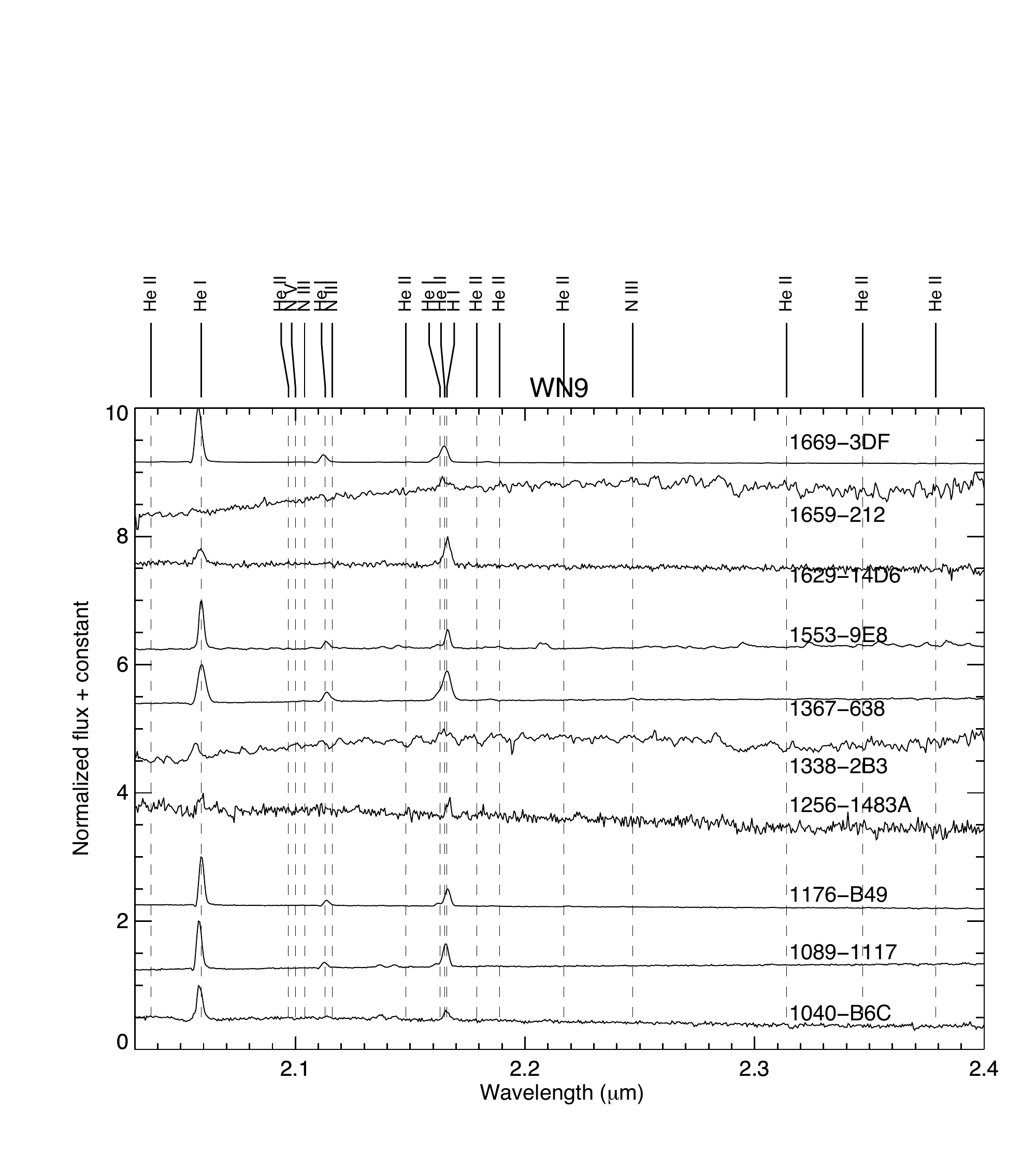}
\end{subfigure}%
\hspace*{\fill}%
\caption{\label{fig:wn} All new WN5 (top left), WN6 (top right), WN7 (middle left),  WN8 (bottom left), and WN9 (bottom right) objects classified in this work. Object 1089-1117 (cLBV transitioning to WNL) is included, as the spectral lines are similar to those of a WN9.}
\end{figure*}

\section{Spectral Follow-up \& Reduction}\label{sec:spec}
The first round of spectral follow-up for this candidate set was conducted over 7 nights at MDM and 12 half-nights at IRTF in 2011 June and July.

\subsection{IRTF}
At the 3m NASA Infrared Telescope Facility (IRTF), we obtained NIR spectra of 150 candidate WR stars, selected using the criteria above, with the SpeX spectrograph. Two of the nights were cloudy enough to prevent observations.  We operated in cross-dispersed mode with the 0.5" slit aligned and obtained an average resolving power of $\lambda/\Delta\lambda\sim1200$, over a wavelength range of  $0.8 - 2.4$~\micron.  

We first acquired each target in the guider camera, then took a single AB dither pattern, with exposure times varying from 30~s for our brightest targets to 200~s for our faintest.  Once we had confirmed the presence of emission lines we began a second  set of AB images so each WR candidate had four images obtained with an ABBA dither pattern along the slit. To minimize the overhead (slew and calibration target time) between sources, we chose nearby subsequent targets. 

After each several targets (typically 4-5), we observed an A0V star at a similar airmass for flux calibration and telluric correction. Internal flat-field and Ar arc lamp exposures were also acquired for pixel response and wavelength calibration, respectively. Additionally, we acquired spectra of almost all known spectral subtypes of Wolf--Rayet star. We reduced all data with \textsc{spextool} version 3.3 \citep{2004PASP..116..352V,2004PASP..116..362C} using standard settings.

\subsection{MDM 2011}
During a run of excellent weather over the 7 nights in 2011 June, we obtained 113 NIR spectra of candidate stars using TIFKAM in spectroscopic mode on the 2.4~m Hiltner telescope at MDM Observatory.  The weather conditions were excellent, with average seeing $\sim1.5''$.  We operated with the 100~\micron\ slit, the K blocking filter, and the J/K grism, providing wavelength coverage of $1.97-2.42$~\micron\ at a resolving power of $\lambda/\Delta\lambda\sim660$.  We performed a single AB dither pattern on each source once it had been placed on the slit in movie mode and a guide star acquired.  If on-the-fly extraction using \textsc{iraf} showed emission, a second AB dither was taken, giving each WR candidate an ABBA dither pattern along the slit.  Exposure times varied from 20~s for the brightest targets to 240~s for the faintest.  We also observed A0V stars at a variety of airmasses for flux calibrations and telluric corrections, acquiring internal flat-field exposures as well for calibrating pixel response.  Additionally, we obtained spectra of almost all known spectral subtypes of Wolf--Rayet star.

Spectra from this run were reduced with a combination of \textsc{iraf} packages and \textsc{idl} programs.  Trimming and flat-fielding were performed with \textsc{ccdproc} and \textsc{flatcombine}, and extraction was performed with \textsc{apall}, all in \textsc{iraf}.  Image arithmetic, wavelength calibration, and combining extracted spectra were done in \textsc{idl}; we used the \textsc{xtellcorgeneral} program included in the \textsc{spextool} package, also written in \textsc{idl}.

We discovered after the run that the NeAr arc lamp images we took at the telescope were faulty, and so wavelength calibration was performed using night sky lines taken from the unprocessed images.  The resulting wavelength calibration is in some cases mediocre, but there was a good enough match to perform telluric corrections, as well as to assign WR star types and subtypes.

\subsection{MDM 2012}
During early 2012, the original survey data were reduced again, using different methods to produce better images.  A new \textsc{idl} pipeline was constructed, creating flat and sky images by median-combining the first and last dither of each pointing for the entire month, using high-quality data images instead of relatively poor dome flats.  Then, during a 10-day observing run in the summer of 2012 (with 6 usable nights), we obtained 70 additional NIR spectra of candidate stars with TIFKAM at MDM, with the same instrument setup as in 2011.  

Reductions were performed entirely in \textsc{iraf}, primarily using the \textsc{kpnoslit} and \textsc{onedspec} packages.   We used a selection of long and short exposure flats to create a bad pixel mask with CCDMASK, and then performed trimming, bias-subtraction, and flat-fielding with \textsc{ccdproc} and \textsc{flatcombine}.  Once the initial preparation was complete, the spectra were extracted from the A-B images with \textsc{apall}.  Then we used \textsc{identify} to determine a wavelength solution for each spectrum from the 4 or 5 strong Ar lines, and \textsc{dispcor} to apply the solutions to the spectra.  All spectra for each individual object were combined using \textsc{scombine}, and then telluric correction was performed with \textsc{telluric}.

\begin{figure}
\includegraphics[width=\linewidth]{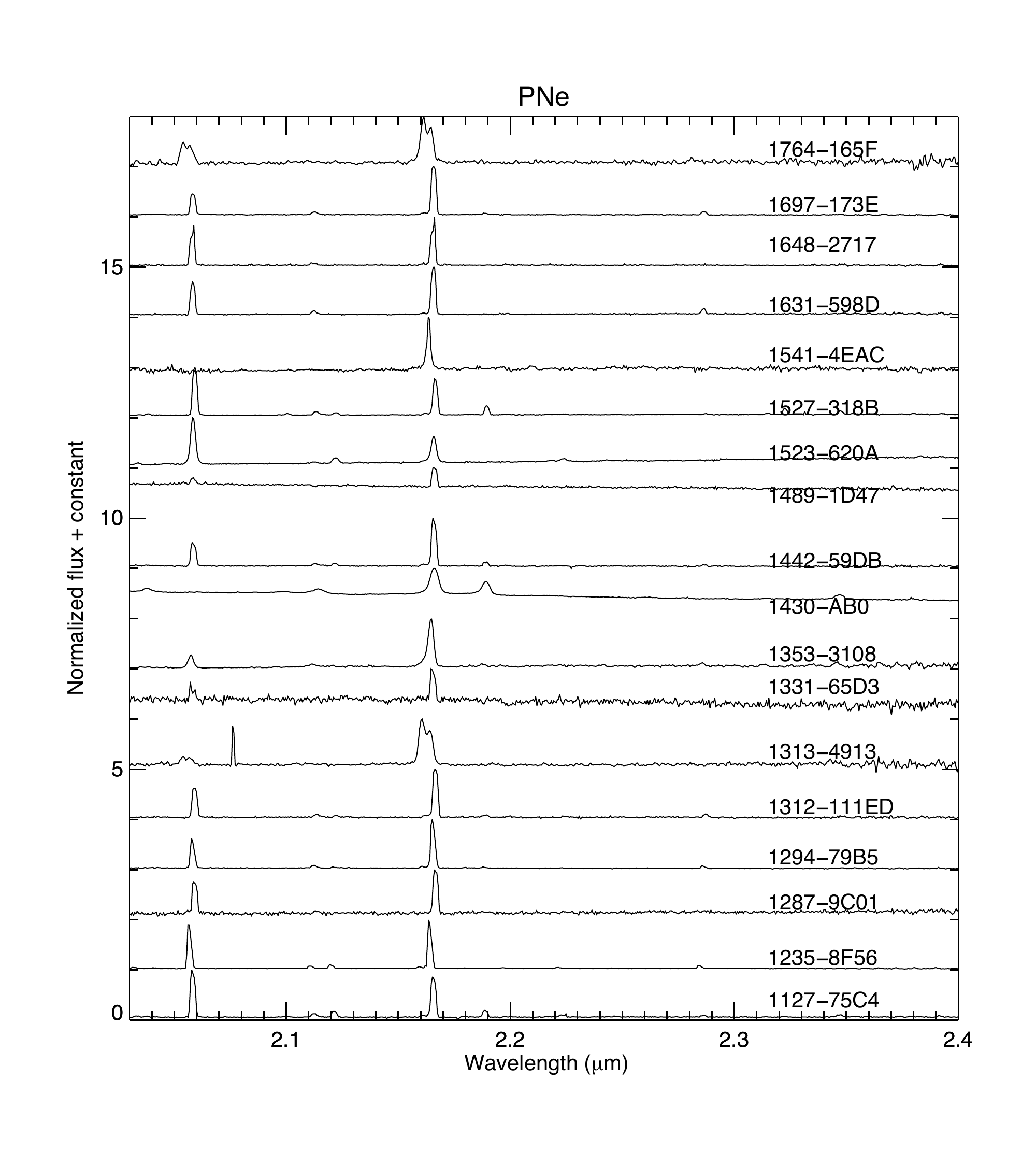}
\caption{\label{fig:pne}  All new PNe classified in this work. Note the doubled emission lines for 3 of the PN spectra. The selection criteria used to identify strong Wolf--Rayet candidates are also extremely effective at identifying new PNe; the search for new PNe is described in a paper in preparation.}
\end{figure}

\begin{figure}
\includegraphics[width=\linewidth]{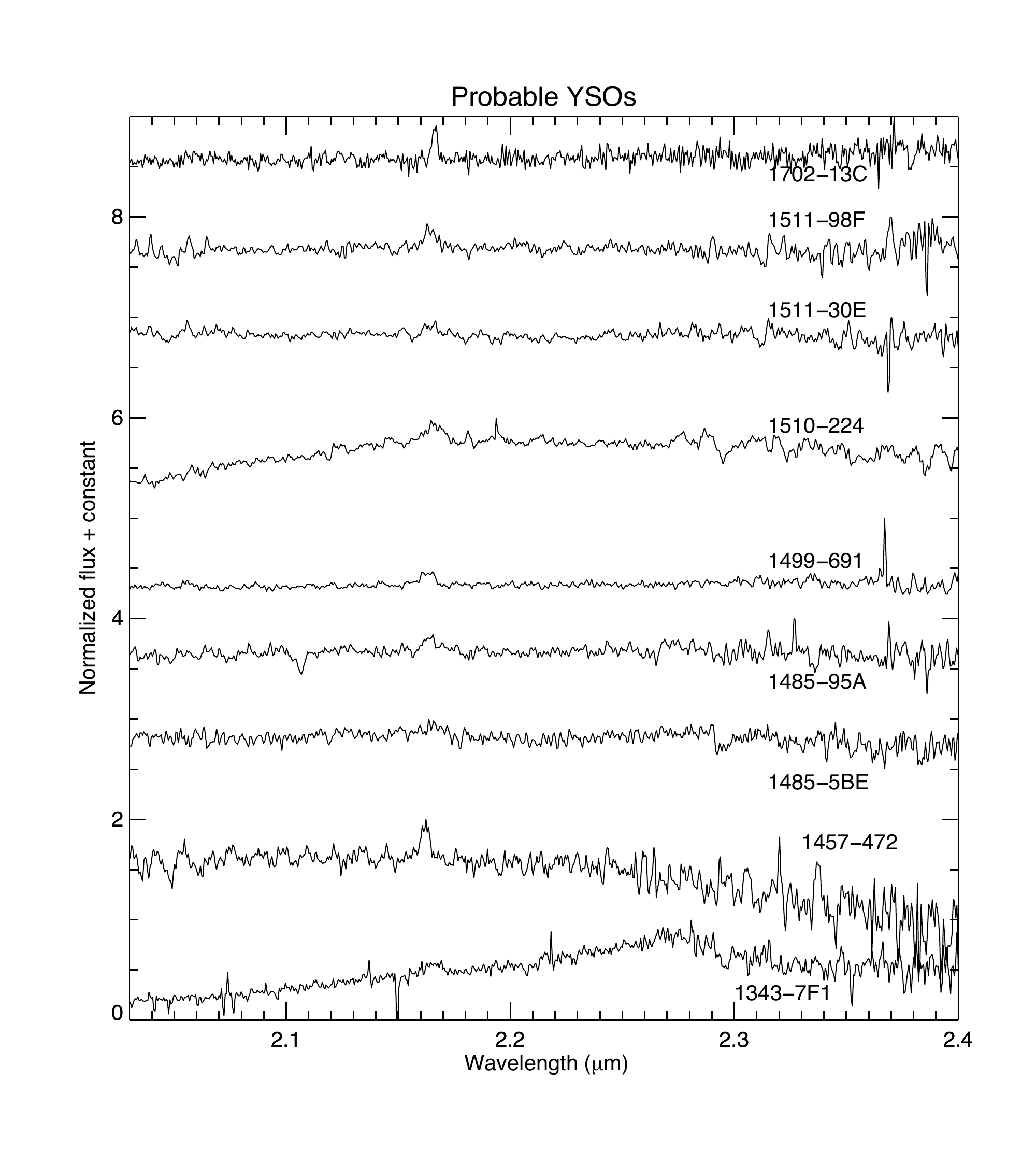}
\caption{\label{fig:yso}  Likely YSOs, as these spectra lack the CO bands redwards of 2.3~\micron\ which identify emitting red giants and supergiants.}
\end{figure}

\begin{figure}
\includegraphics[width=\linewidth,clip=true,trim=0.25in 0.25in 0in 1.5in]{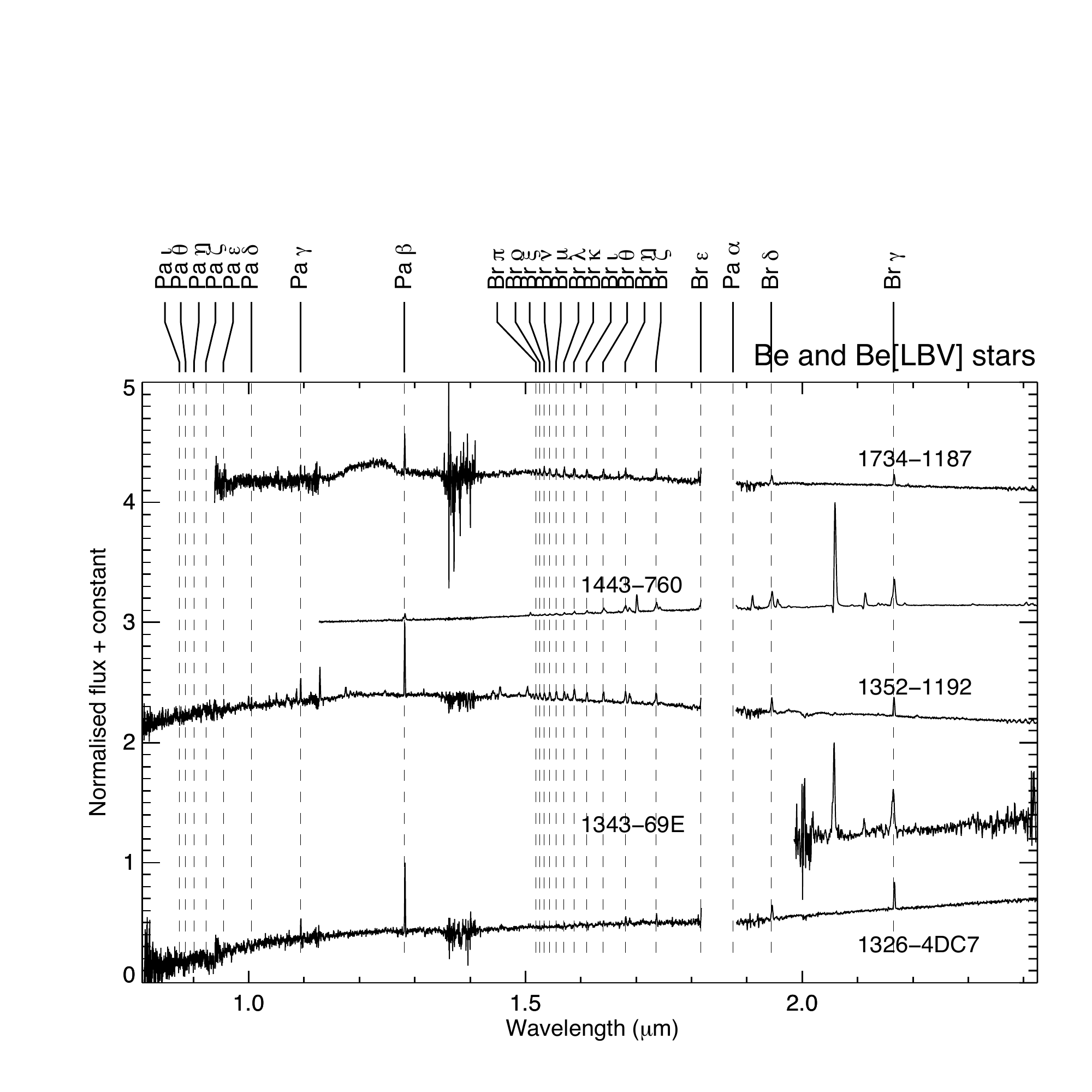}
\caption{\label{fig:be}  All new Be stars classified in this work. The full JHK spectrum is shown for these objects, displaying the prominent Hydrogen lines for those spectra for which we have J and H coverage. Be star interlopers are relatively common, as they are selected strongly by the BrGamma filter and can only be ruled out by obtaining spectra. Paschen and Brackett series emission lines are identified.}
\end{figure}

\begin{figure}
\includegraphics[width=\linewidth]{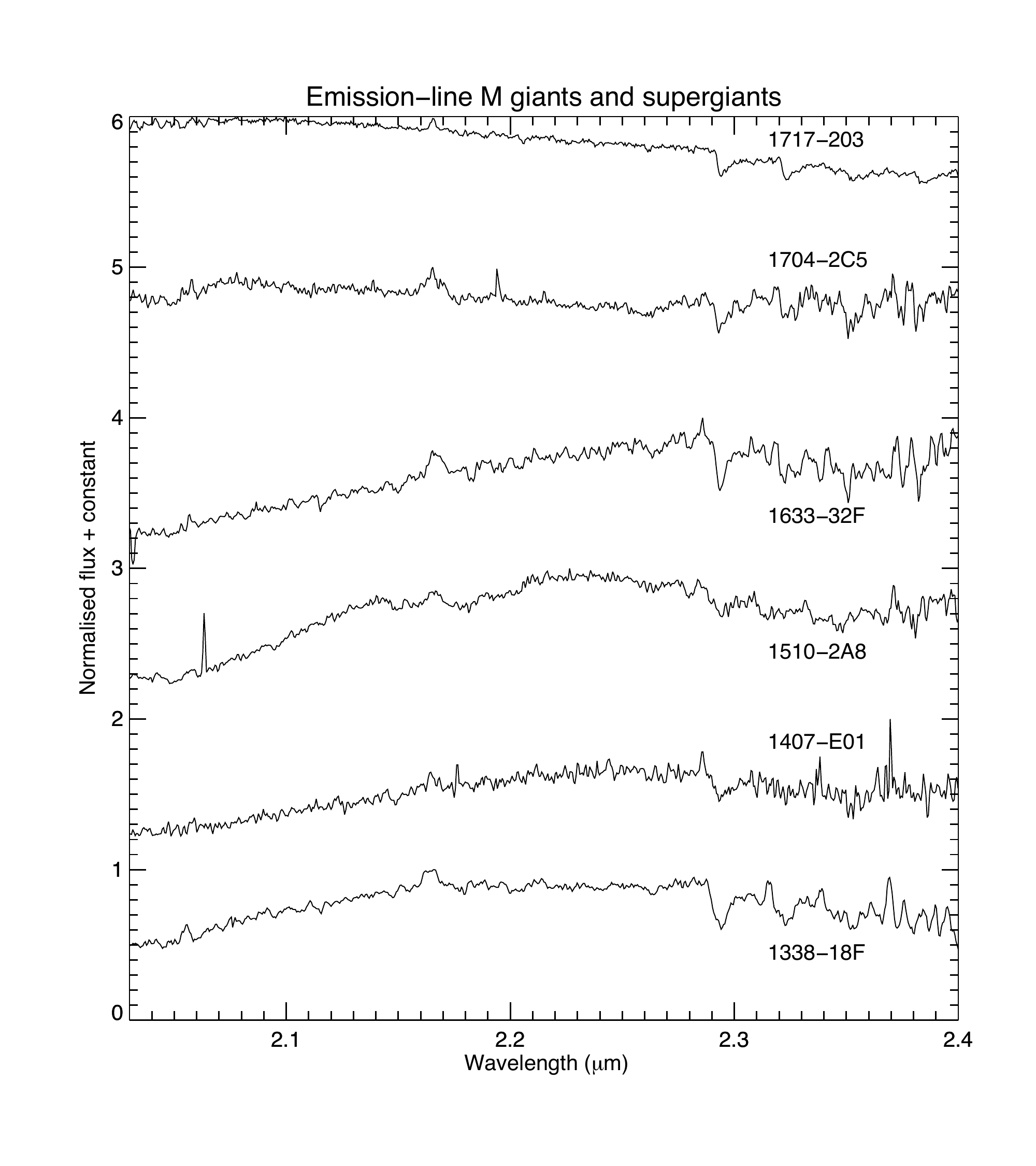}
\caption{\label{fig:me} Red giants or supergiants, with molecular CO bands redwards of 2.3~\micron, which showed \ion{H}{1} emission.}
\end{figure}

\begin{figure}
\includegraphics[width=\linewidth]{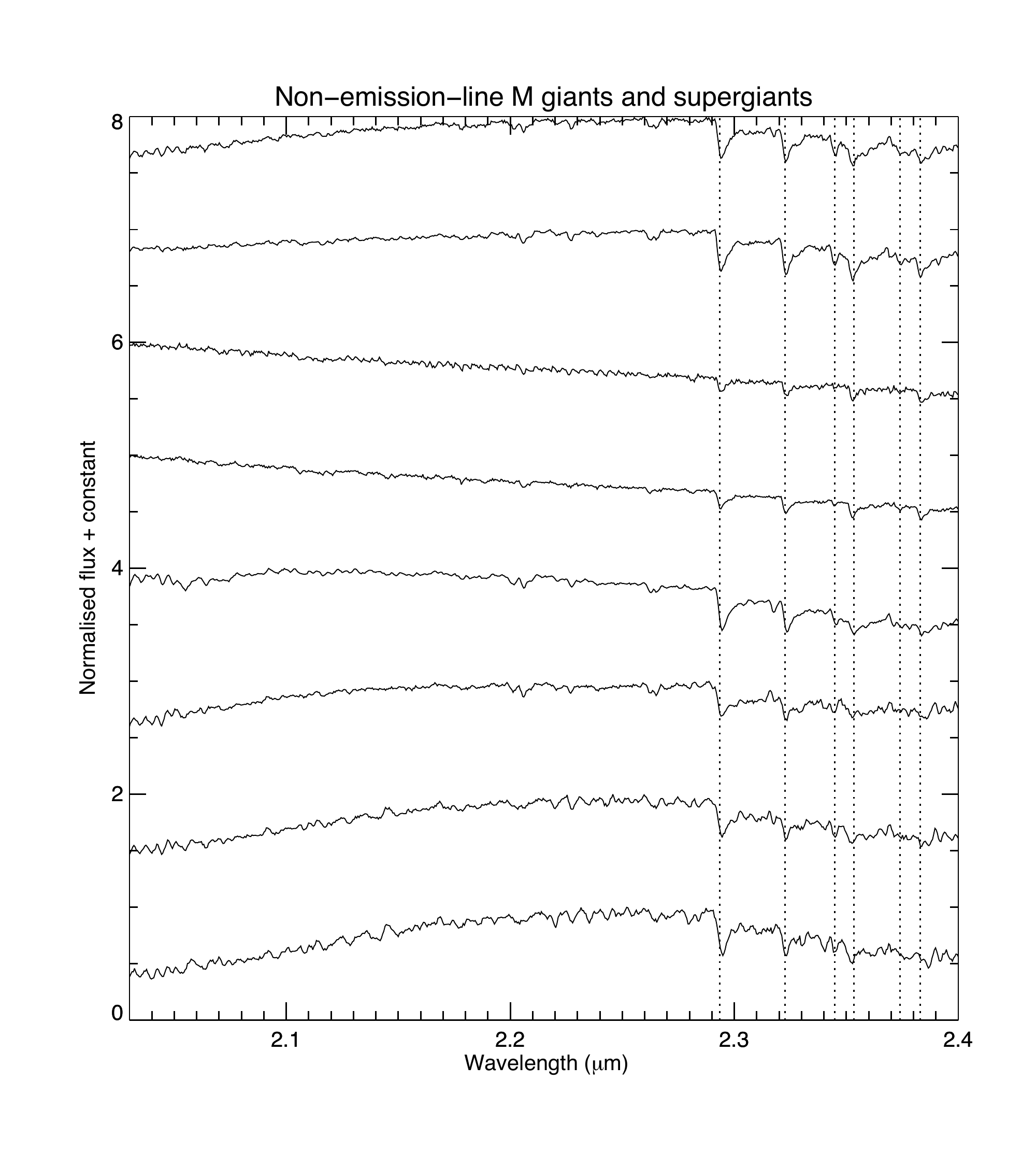}
\caption{\label{fig:duds}  A selection of non-emitting red giants or supergiants. These interlopers are selected simply because the slope of the red spectrum makes them brighter in \ion{C}{4} than in \ion{He}{1} or the nearest continuum filter. The dotted lines indicate the $\pres{^{12}}$C$\pres{^{16}}$O bandheads for $v=2-0$ (2.2935~\micron), $v=3-1$ (2.3227~\micron), $v=4-2$ (2.3535~\micron), and $v=5-3$ (2.3829~\micron), and the $\pres{^{13}}$C$\pres{^{16}}$O bandheads for $v=2-0$ (2.3448~\micron) and $v=3-1$ (2.3739~\micron).}
\end{figure}

\section{Results}\label{sec:res}
During 29 nights of observing we observed 333 candidates, finding 89 NIR emission sources (for a success rate of 27 per cent) in some of our survey's most crowded fields.  Of the emission-line sources, 49 were WR stars, 43 of which had never been previously identified, with 23 WC and 26 WN; finder charts for these new WR stars are included in appendix A.  WR types and subtypes were assigned by eye, comparing the relative strengths of nearby line pairs as in \citet{2006MNRAS.372.1407C}; classification would ideally be performed using EWs of the spectral lines, but in many cases they are difficult to obtain due to heavy blending which is easily compensated for by eye.  Figure~\ref{fig:newcuts} shows the positions of the new emission sources on the colour cut plots, and figures~\ref{fig:wc} and~\ref{fig:wn} show the spectra of the new WR stars, sorted by subtype.  Tables~\ref{tab:newwr} through~\ref{tab:dist} give the location, magnitude, subtype, and (for the confirmed WR stars) extinction for each NIR emission object classified in this paper. In addition, table~\ref{tab:lines} shows the wavelengths and transitions for the WR emission lines marked in figures~\ref{fig:wc} and~\ref{fig:wn}, and the measured equivalent widths are given in tables~\ref{tab:ewn1} thru~\ref{tab:ewc2}.

For a number of WR stars, the \textit{2MASS} magnitudes were not reliable (upper limits, poor photometric quality or PSF fitting, etc., as described in table~\ref{tab:wrmag}). In such cases, the $J$ and/or $CONT2$ magnitudes from the survey described in section~\ref{sec:obs} were used in place of the \textit{2MASS} $J$ and $K_s$ magnitudes, respectively. The alternate magnitudes were scaled (as were all other magnitudes from that survey, during the reduction process) to match those given in \textit{2MASS}. However, if the \textit{2MASS} $H$ magnitude could not be used, or the alternate $J$ or $CONT2$ magnitudes were unavailable, then extinction and distance were not calculated.

\begin{table*}
\caption{Spectroscopically-confirmed Wolf--Rayet stars. Objects are identified by field number and object number; this has no relation to RA or Dec. In addition, it is difficult to differentiate between features of WC$4-8$. A colon (:) indicates an uncertainty of up to $\pm2$ subtypes. \label{tab:newwr}}
\begin{center}
\begin{tabular}{lccccccccl}
\hline \noalign{\smallskip}
\textbf{Name} & \textbf{$\delta$ (J2000)} & \textbf{$\delta$ (J2000)} & \textbf{$l$} & \textbf{$b$} & \textbf{Type} & \textbf{Telescope} \\ 
\noalign{\smallskip} \hline
  1040-B6C & 16 04 03.76 &  -53 10 44.2 & -30.54 & -0.527 & WN9 & IRTF\\
 1089-1117 & 16 31 37.79 &  -48 14 55.3 & -23.98 & -0.038 & cLBV/WNL\fnm{1} & IRTF\\
 1093-1765 & 16 32 25.70 &  -47 50 46.1 & -23.60 &  0.139 & WN6 & IRTF\\
 1139-49EA & 16 54 08.46 &  -43 49 25.3 & -18.09 & -0.073 & WC6:: & IRTF\\
  1178-66B & 17 07 23.95 &  -39 19 54.4 & -13.02 &  0.738 & WC9 & IRTF\\
  1176-B49 & 17 12 34.87 &  -40 37 13.8 & -13.47 & -0.827 & WN9h & IRTF\\
 1198-6EC8 & 17 15 55.90 &  -37 19 12.0 & -10.41 &  0.575 & WC6:: & IRTF\\
1256-1483A & 17 40 59.36 &  -32 11 22.1 & -3.292 & -0.860 & WN9 & IRTF \\
 1319-3BC0 & 17 57 16.87 &  -25 23 13.8 &  4.376 & -0.416 & WC7: & IRTF \\
  1338-2B3 & 17 59 07.99 &  -22 36 43.0 &  6.991 &  0.605 & WN9 & MDM12 \\
  1343-284 & 18 03 28.37 &  -22 22 58.9 &  7.686 & -0.152 & WN8-9 & MDM12 \\
  1366-438 & 18 05 55.27 &  -19 29 44.1 & 10.483 &  0.765 & WN7-8\fnm{2} & MDM12 \\
  1367-638 & 18 09 06.22 &  -19 54 27.2 & 10.487 & -0.090 & WN9 & IRTF \\
  1381-19L & 18 12 02.41 &  -18 06 55.4 & 12.392 &  0.167 & WC9 & MDM11 \\
 1389-4AB6 & 18 14 14.09 &  -17 21 02.6 & 13.313 &  0.075 & WC7 & IRTF \\
 1389-1F5D & 18 14 17.37 &  -17 21 54.4 & 13.307 &  0.057 & WN8 & IRTF \\
  1446-B1D & 18 25 00.25 &  -10 33 23.5 & 20.536 &  0.983 & WN6\fnm{3} & IRTF \\
  1457-673 & 18 31 06.65 &  -09 48 01.4 & 21.904 &  0.004 & WC9d & MDM12 \\
  1485-6C4 & 18 36 55.53 &  -06 31 02.1 & 25.480 &  0.241 & WN6 & MDM12 \\
  1485-844 & 18 37 51.82 &  -06 31 19.1 & 25.583 &  0.032 & WN8 & MDM12 \\
 1495-1D8A & 18 39 40.60 &  -05 35 17.6 & 26.620 &  0.059 & WC8-9 & MDM12 \\
  1495-705 & 18 39 41.19 &  -05 57 36.3 & 26.290 & -0.113 & WN8 & MDM12 \\
  1514-AA0 & 18 41 06.79 &  -02 56 01.0 & 29.144 &  0.957 & WC8 & IRTF \\
 1509-2E64 & 18 42 26.61 &  -03 56 36.0 & 28.398 &  0.199 & WC9 & IRTF \\
 1525-2352 & 18 45 14.63 &  -02 05 05.7 & 30.370 &  0.427 & WC8: & IRTF \\
  1519-E43 & 18 45 49.88 &  -02 59 56.0 & 29.624 & -0.121 & WC7\fnm{4} & MDM12 \\
  1530-8FA & 18 46 00.97 &  -01 14 35.0 & 31.207 &  0.639 & WN5 & IRTF \\
  1541-3C8 & 18 50 02.75 &  -00 32 07.9 & 32.297 &  0.066 & WC8\fnm{4} & MDM11 \\
 1541-197C & 18 50 37.54 &  -00 01 21.1 & 32.819 &  0.171 & WC8 & MDM11 \\
  1544-FA4 & 18 51 33.09 &  -00 13 40.8 & 32.742 & -0.129 & WN5 & IRTF \\
  1553-9E8 & 18 52 33.12 &  +00 47 41.8 & 33.766 &  0.115 & WN9h & IRTF \\
 1547-1488 & 18 52 57.20 &  +00 02 54.1 & 33.148 & -0.315 & WN5 & MDM11 \\
 1553-15DF & 18 53 02.56 &  +01 10 22.7 & 34.159 &  0.178 & WC8 & IRTF \\
  1602-9AF & 19 02 42.32 &  +06 54 44.4 & 40.365 &  0.657 & WN6 & MDM11 \\
 1603-11AD & 19 04 20.14 &  +06 07 52.2 & 39.856 & -0.061 & WN5 & IRTF \\
 1609-1C95 & 19 06 10.68 &  +07 19 13.3 & 41.123 &  0.078 & WC9 & IRTF \\
 1626-4FC8 & 19 06 33.66 &  +09 07 20.8 & 42.767 &  0.822 & [WC6:]\fnm{5} & IRTF \\
 1629-14D6 & 19 10 06.40 &  +09 45 25.7 & 43.733 &  0.339 & WN9h & IRTF \\
  1627-A6D & 19 10 11.53 &  +08 58 39.6 & 43.051 & -0.040 &  WC7::\fnm{4} & IRTF \\
  1635-AD8 & 19 13 19.19 &  +09 55 29.0 & 44.248 & -0.285 & WN6 & IRTF \\
  1653-FFE & 19 14 40.73 &  +11 54 15.4 & 46.156 &  0.338 & WN5-6 & IRTF \\
  1651-BB4 & 19 15 37.26 &  +11 25 26.3 & 45.838 & -0.089 & WN5 & IRTF \\
 1647-1E70 & 19 15 52.52 &  +11 12 59.7 & 45.683 & -0.241 & WC8: & IRTF \\
  1659-212 & 19 17 22.20 &  +12 13 09.2 & 46.741 & -0.097 & WN9 & MDM12 \\
  1669-3DF & 19 18 31.35 &  +13 43 39.4 & 48.206 &  0.360 & WN9h & IRTF \\
 1660-1169 & 19 20 02.46 &  +12 08 20.3 & 46.975 & -0.712 & WC6: & IRTF \\
  1697-38F & 19 25 18.12 &  +17 02 15.9 & 51.895 &  0.477 & WC9 & IRTF \\
  1702-23L & 19 26 08.35 &  +17 46 23.1 & 52.637 &  0.651 & WC8 & MDM11 \\
  1695-2B7 & 19 27 17.98 &  +16 05 24.6 & 51.289 & -0.394 & WC9 & IRTF \\
\hline
\end{tabular} 
\end{center}
\begin{flushleft}
\fnt{1}{Originally identified in \cite{2010MNRAS.405.1047G} (MN42), and then re-classified as a cLBV transitioning to late WN in \cite{2012IAUS..282..267S}.}\\
\fnt{2}{Possibly an emitting M-giant.}\\
\fnt{3}{Originally identified in \citet{2007MNRAS.376..248H} (HDM 10).}\\
\fnt{4}{Originally identified in \citet{2012AJ....144..166S} (2w06, 2w10, 2w11).}\\
\fnt{5}{Originally classified as a [WC] in \cite{2010MNRAS.405.1047G} (MN102); the brackets indicate a central star of a Planetary Nebula, which displays WR-like emission.}
\end{flushleft}
\end{table*}

\begin{table*}
\caption{Other NIR emission sources. Objects are identified by field number and object number; this has no relation to RA or Dec. \label{tab:newem1}}
\begin{center}
\begin{tabular}{lccccccccl}
\hline \noalign{\smallskip}
\textbf{Name} & \textbf{$\delta$ (J2000)} & \textbf{$\delta$ (J2000)} & \textbf{$l$} & \textbf{$b$} & \textbf{Type\fnm{a}} & \textbf{Telescope} \\ 
\noalign{\smallskip} \hline
1127-75C4 & 17 29 37.54 & -35 13 43.8 & -7.126 & -0.515 & True PN\fnm{1} & IRTF\\
1235-8F56 & 17 31 50.69 & -34 10 44.3 & -5.999 & -0.319 & PN & IRTF\\
1294-79B5 & 17 46 01.68 & -27 26 01.3 & 1.33 & 0.706 & Possible PN\fnm{2} & IRTF \\
1287-9C01 & 17 47 14.65 & -28 26 48.8 & 0.603 & -0.05 & PN & IRTF \\
1313-4913 & 17 52 59.34 & -25 27 28.1 & 3.825 & 0.385 & Likely PN\fnm{2} & MDM12 \\
1326-4DC7 & 17 56 28.55 & -24 00 27.1 & 5.477 & 0.435 & Be & IRTF \\
1312-111ED & 17 56 41.57 & -26 31 09.5 & 3.33 & -0.869 & PN & IRTF \\
1338-18F & 17 58 57.24 & -22 20 24.6 & 7.206 & 0.776 & \fnm{a} & MDM12 \\
1331-65D3 & 17 59 42.43 & -23 51 43.7 & 5.972 & -0.132 & PN & IRTF \\
1343-69E & 18 02 22.35 &  -22 38 00.3 &  7.343 & -0.055 & B[e]/LBV\fnm{b} & MDM12 \\ 
1343-7F1 & 18 02 44.42 & -22 19 36.7 & 7.651 & 0.023 & \fnm{c} & MDM12 \\
1353-3108 & 18 04 08.45 & -20 57 05.8 & 9.009 & 0.416 & Likely PN\fnm{3} & MDM12 \\
1352-1192 & 18 06 40.77 & -21 40 17.4 & 8.67 & -0.452 & Be & IRTF \\
1407-E01 & 18 18 58.19 & -15 49 37.9 & 15.193 & -0.198 & \fnm{a} & MDM12 \\
1430-AB0 & 18 21 02.92 &  -12 27 45.8 & 18.397 &  0.945 & PN\fnm{d} & IRTF \\
1442-59DB & 18 24 07.91 & -11 06 42.6 & 19.945 & 0.913 & True PN\fnm{2} & IRTF \\
1443-760 & 18 28 33.39 &  -11 46 44.2 & 19.860 & -0.358 & B[e]/LBV\fnm{b} & IRTF \\
1457-472 & 18 29 40.38 & -09 31 21.5 & 21.986 & 0.447 & \fnm{c} & MDM12 \\
1485-5BE & 18 36 16.87 & -06 43 17.6 & 25.225 & 0.289 & \fnm{c} & MDM12 \\
1485-95A & 18 37 14.85 & -06 44 44.4 & 25.314 & 0.065 & \fnm{c} & MDM12 \\
1489-1D47 & 18 37 30.41 & -06 14 15.0 & 25.795 & 0.241 & PN & IRTF \\
1510-224 & 18 39 43.87 & -03 34 43.5 & 28.412 & 0.968 & \fnm{c} & MDM12 \\
1499-691 & 18 40 36.84 & -05 27 24.5 & 26.843 & -0.088 & \fnm{c} & MDM12 \\
1510-2A8 & 18 41 33.60 & -03 39 05.9 & 28.556 & 0.529 & \fnm{a} & MDM12 \\
1511-30E & 18 44 05.00 & -03 57 02.4 & 28.578 & -0.168 & \fnm{c} & MDM12 \\
1511-98F & 18 44 18.19 & -04 05 07.1 & 28.483 & -0.278 & \fnm{c} & MDM12 \\
1523-620A & 18 47 00.40 & -02 27 51.6 & 30.234 & -0.138 & Possible PN\fnm{4} & IRTF \\
1527-318B & 18 48 29.26 & -02 10 01.4 & 30.667 & -0.332 & Possible PN\fnm{4} & IRTF \\
1541-4EAC & 18 49 45.18 & -00 29 08.0 & 32.308 & 0.154 & PN & MDM11 \\
1633-32F & 19 10 37.97 & +09 47 34.5 & 43.824 & 0.240 & \fnm{a} & MDM12 \\
1631-598D & 19 10 56.59 & +09 28 36.6 & 43.579 & 0.026 & PN & IRTF \\
1648-2717 & 19 16 30.50 & +10 40 56.3 & 45.283 & -0.628 & PN & IRTF \\
1702-13C & 19 25 15.87 & +17 31 40.7 & 52.323 & 0.718 & \fnm{c} & IRTF \\
1697-173E & 19 25 53.53 & +16 53 31.5 & 51.834 & 0.284 & Likely PN\fnm{4} & IRTF \\
1717-203 & 19 30 57.41 & +19 29 51.8 & 54.698 & 0.473 & \fnm{a} & IRTF \\
1704-2C5 & 19 32 03.38 & +16 44 42.4 & 52.411 & -1.082 & \fnm{a} & MDM12 \\
1734-1187 & 19 32 28.69 & +21 11 30.4 & 56.354 & 0.976 & Be & IRTF \\
1764-165F & 19 45 32.87 & +23 28 10.4 & 59.823 & -0.536 & Likely PN\fnm{4} & MDM12 \\
\hline
\end{tabular}
\end{center}
\begin{flushleft}
\fnt{a}{These stars are emitting cool stars, M giants or supergiants \citep{2009ApJS..185..289R}.}\\
\fnt{b}{Originally identified in \cite{2011BSRSL..80..291W}.}\\
\fnt{c}{A number of emission sources seem likely to be YSOs \citep{1996AJ....112.2184G}.}\\
\fnt{d}{This object is missing the P Cygni absorption lines normally seen in a WN7h, and so is likely a high-ionisation nebula.}\\
\fnt{1}{Originally identified in \citet{2008MNRAS.384..525M}.}\\
\fnt{2}{Originally identified in \citet{1992secg.book.....A}.}\\
\fnt{3}{Originally identified in \citet{2003A&A...408.1029K}.}\\
\fnt{4}{Originally identified in \citet{2009A&A...501..539U}.}
\end{flushleft}
\end{table*}

\begin{table*}
\caption{Photometry for confirm ed WR stars.  $B$, $V$, and $R$ photometry is from the NOMAD catalogue; $J$, $H$, and $K_s$ photometry is from \textit{2MASS}; and $W1-4$ are from the \textit{WISE} All-Sky Source Catalog. Upper limits are marked with $>$, poor photometry quality (C or worse on the PH\_QUAL flag) with a colon (:), possible contamination by an asterisk (*), and poor PSF fitting with {}\textdagger. \label{tab:wrmag}}
\begin{center}
\begin{tabular}{lrrrrrrrrrr|rrr}
\hline\noalign{\smallskip}
\textbf{Name} & \textbf{$B$} & \textbf{$V$} & \textbf{$R$} & \textbf{$J$} & \textbf{$H$} & \textbf{$K_s$} & \textbf{$W1$} & \textbf{$W2$} & \textbf{$W3$} & \textbf{$W4$} & \textbf{$J-K_s$} & \textbf{$H-K_s$} & \textbf{$W1-W2$} \\
\noalign{\smallskip}\hline
1040-B6C &  --  &  --  &  --  & 13.13 & 11.90 & 11.17 & 10.41 & 10.15 & $>10.74$ & $>8.35$ & 1.96 & 0.73 & 0.26\\
1089-1117 &  --  &  --  &  --  & $>16.57$ & 12.60 & 10.28 & 8.53 & 7.53 & 6.58 & 1.62\textdagger & 6.29 & 2.31 & 1.01\\
1093-1765 &  --  &  --  &  --  & 15.15 & 12.93 & 11.57 & 10.46 & 9.98 & 8.71 & $>8.55$ & 3.58 & 1.35 & 0.49\\
1139-49EA &  --  &  --  &  --  & $>16.29$ & $>14.53$ & 13.09 &  --  &  --  &  --  &  -- & 3.20 & 1.44 & -- \\
1178-66B &  --  &  --  &  --  & 12.49 & 11.12 & 10.26 & 9.51 & 9.03 & 8.40 & 6.99 & 2.24 & 0.87 & 0.48\\
1176-B49 &  --  &  --  &  --  & 12.66 & 11.22 & 10.41 & 9.69 & 9.24 & 8.81 & $>7.86$ & 2.25 & 0.81 & 0.45\\
1198-6EC8 &  --  &  --  &  --  & $>16.15$ & $>14.68$ & 13.47 &  --  &  --  &  --  &  -- & 2.69 & 1.21 & -- \\
1256-1483A &  --  &  --  &  --  &  13.98 &  12.46  &  11.86 & 11.07 & 10.81 & 4.89 & 0.84 & 2.12 & 0.60 & 0.27\\
1319-3BC0 &  --  &  --  &  --  & 14.89: & $>12.72$ & $>11.46$ &  10.89  &  10.39  &  $>8.97$  &  $>7.58$  & 3.42 & 1.26 & 0.50\\
1338-2B3 & -- & -- & -- & 12.61 & $>10.48$ &  $>9.21$ &  -- &  -- &  -- &  -- & 3.40 & 1.27 & --\\
1343-284 & 17.39 & 15.60 & 14.72 & 10.47 &  9.60 &  9.02 &  8.40 &  8.08 &  7.68 &  $>5.82$ & 1.46 & 0.58 & 0.32\\
1366-438 & -- & -- & -- & 12.91 & $>10.58$ & $>9.20$ &  7.78 & 7.26 &  5.98 &  4.65 & 3.71 & 1.38 & 0.53\\
1367-638 &  --  &  --  &  --  & 15.96: & 12.43 & 10.40 & -- & -- & -- & -- & 5.57 & 2.03 & --\\
1381-19L & 15.48  & 13.86  & 13.78 & 9.66 & 8.26 & 7.80* & 7.34: &  6.62  & 5.90 & 4.89 & 1.87 & 0.83 & 0.72\\
1389-4AB6 &  --  &  --  & 20.74 & 16.13 & 14.26 & 12.24 & 10.22 & 9.36 & 8.37 & 6.88 & 3.89 & 2.03 & 0.86\\
1389-1F5D &  --  &  --  &  & $>17.57$ & 13.28 & 11.05 & 9.67 & 8.79 & 8.91 & $>5.78$ & 6.52 & 2.23 & 0.88\\
1446-B1D &  --  &  --  & 17.24 & 12.21 & 11.24 & 10.61 & 9.68 & 9.31 & 8.30 & 6.95 & 1.60 & 0.63 & 0.37\\
1457-673 & -- & -- & -- & 14.85 & $>11.44$ &  $>9.35$ &  7.55 &  6.66 &  5.28 &  3.62 & 5.50 & 2.09 & 0.89\\
1485-6C4 & -- & -- & 16.66 & 12.04 & 10.81 & 10.02 &  9.23 &  8.90 &  8.06 &  $>5.30$ & 2.02 & 0.79 & 0.33\\
1485-844 & -- & -- & -- & 14.72 & 11.30 &  9.49 &  8.07 &  7.40 &  6.84 &  5.30 & 5.23 & 1.81 & 0.68\\
1495-1D8A & -- & -- & -- & $>14.64$ & $>12.36$ & 11.72 & 9.72\textdagger & 9.22\textdagger & $>9.87$ & $>6.75$ & 2.92 & 0.65 & 0.50\\
1495-705 & -- & -- & -- & 14.97 & 11.38 &  9.16 & 7.31 &  5.92 &  4.75 &  3.77 & 5.81 & 2.22 & 1.39\\
1514-AA0 &  --  &  --  &  --  & 12.92 & 11.24 & 10.54 & 10.06 & 10.13 & 10.15 & $>7.27$ & 2.38 & 0.71 & -0.07\\
1509-2E64 &  --  &  --  &  --  & $>15.78$ & 14.87: & 12.45 & 9.83\textdagger & 9.01\textdagger & 8.33 & 6.36: & 3.33 & 2.42 & 0.82\\
1525-2352 &  --  &  --  &  --  & $>17.93$ & 14.75 & 12.44 & 10.34\textdagger & 9.53 & $>9.01$ & $>6.23$ & 5.48 & 2.30 & 0.81\\
1519-E43 & -- & -- & -- & $>13.65$ & 13.04 & 11.26 &  9.59\textdagger &  9.01 &  9.66: &  $>7.09$ & 2.39 & 1.78 & 0.57\\
1530-8FA &  --  &  --  &  --  & 12.76 & 11.43 & 10.64 & 9.03 & 8.72 & 9.90 & $>7.82$ & 2.12 & 0.78 & 0.31\\
1541-3C8 &  -- & -- & -- & $>16.98$ & 13.86 & 12.15 & 10.73 & 10.06 & $>8.54$ & $>6.28$ & 4.82 & 1.70 & 0.67\\
1541-197C &  -- & -- & -- & $>14.74$ & 13.33 & $>11.64$ & 10.05 & 9.30 & 8.73 & 4.45\textdagger & 3.11 & 1.69 & 0.75\\
1544-FA4 &  --  &  --  & -- &  $>13.50$ & 12.24\textdagger & 10.83\textdagger  &  9.61  &  9.02  &  8.37  &  $>7.17$  & 2.68 & 1.41 & 0.59\\
1553-9E8 &  --  &  --  &  --  & 15.76 & 12.86 & 10.98 & 9.54 & 8.54 & 8.47 & $>6.35$ & 4.78 & 1.89 & 1.00\\
1547-1488 & --  &  --  &  --  & 13.96 & 12.25 & 11.16 & 9.99 & 9.51 & 8.62 & 7.39: & 2.80 & 1.10 & 0.48\\
1553-15DF &  --  &  --  &  --  & $>17.78$ & 14.80 & 12.07 & 9.70 & 8.62 & $>8.28$ & 4.03 & 5.71 & 2.74 & 1.07\\
1602-9AF &  --  &  --  & -- & 13.11 & 11.62 & 11.05 & 10.59 & 10.67 & 9.60 & 7.27 & 2.06 & 0.57 & -0.10\\
1603-11AD &  --  &  --  &  --  & 16.13 & 13.64 & 12.16 & 10.43\textdagger & 9.80 & 8.76 & 6.93 & 3.98 & 1.48 & 0.63\\
1609-1C95 &  --  &  --  &  --  & $>18.41$ & 15.07 & 11.93 & 9.91 & 8.35 & 8.58 & $>8.19$ & 6.48 & 3.14 & 1.57\\
1626-4FC8 &  --  &  --  & 17.40 & 15.59 & 14.86 & 13.89 & 12.77 & 11.79 & 5.71\textdagger & 2.09\textdagger & 1.71 & 0.97 & 0.99\\
1629-14D6 &  --  &  --  &  --  & 14.73 & 13.44 & 12.63 & 11.75\textdagger & 11.40 & $>10.42$ & $>7.93$ & 2.10 & 0.82 & 0.35\\
1627-A6D &  --  &  --  &  --  & $>15.78$ & 13.54 & 11.75 & 10.51 & 9.80 & $>10.92$ & $>6.77$ & 4.03 & 1.78 & 0.70\\
1635-AD8 &  --  &  --  &  --  & 15.20 & 12.87 & 11.48 & 10.27 & 9.63 & 9.26 & 8.52 & 3.72 & 1.40 & 0.65\\
1653-FFE &  --  &  --  &  --  & $>15.22$ & $>13.00$ & 11.66 & 10.37 & 9.78 & 9.17 & 8.61: & 3.56 & 1.34 & 0.59\\
1651-BB4 &  --  &  --  &  --  & 15.83 & $>13.21$ & $>11.78$ & 10.30 & 9.53 & 8.60 & 5.18 & 4.05 & 1.43 & 0.77\\
1647-1E70 &  --  &  --  &  --  & $>17.53$ & 15.10 & 12.57 & 10.74\textdagger & 9.66 & 8.65 & 4.93 & 4.97 & 2.53 & 1.08\\
1659-212 & -- & -- & -- & 13.44 & 10.89 &  9.53 &  8.56 &  8.00 &  7.06 &  6.00 & 3.91 & 1.36 & 0.56\\
1669-3DF &  --  &  --  &  --  & 12.79 & 11.03 & 9.97 & 8.81 & 8.22 & 7.61 & 6.57 & 2.82 & 1.07 & 0.59\\
1660-1169 &  --  &  --  &  --  & 14.65 & 13.29 & 12.08 & 11.65 & 11.22 & $>11.15$ & $>8.72$ & 2.58 & 1.22 & 0.43\\
1697-38F &  --  &  --  & -- & 12.97\textdagger & 11.18\textdagger & 9.92 & 8.54 & 7.87 & 7.58 & 6.57 & 3.04 & 1.25 & 0.67\\
1702-23L &  20.44 &  17.29  &  16.15  &  11.86 &  10.97  &  10.21  &  9.66  &  9.26  &  8.60  &  7.62 & 1.66 & 0.76 & 0.40\\
1695-2B7 &  --  &  --  &  --  & 13.01 & 11.09 & 9.64 & 8.40 & 7.70 & 7.27 & 7.02 & 3.37 & 1.45 & 0.71\\
\hline
\end{tabular}
\end{center}
\end{table*}

\begin{table*}
\caption{Photometry for other NIR emission sources.  $B$, $V$, and $R$ photometry is from the NOMAD catalogue; $J$, $H$, and $K_s$ photometry is from \textit{2MASS}; and $W1-4$ are from the \textit{WISE} All-Sky Source Catalog. Upper limits are marked with $>$, poor photometry quality (C or worse on the PH\_QUAL flag) with a colon (:), possible contamination by an asterisk (*), and poor PSF fitting with {}\textdagger.\label{tab:newem2}}
\begin{center}
\begin{tabular}{lrrrrrrrrrr|rrr}
\hline\noalign{\smallskip}
\textbf{Name} & \textbf{$B$} & \textbf{$V$} & \textbf{$R$} & \textbf{$J$} & \textbf{$H$} & \textbf{$K_s$} & \textbf{$W1$} & \textbf{$W2$} & \textbf{$W3$} & \textbf{$W4$} & \textbf{$J-K_s$} & \textbf{$H-K_s$} & \textbf{$W1-W2$} \\
\noalign{\smallskip}\hline
1127-75C4 & -- & -- & 16.08 & 14.99 & 13.92 & 12.24 & -- &  & -- & -- & 2.75 & 1.68 & --\\
1235-8F56 & -- & -- & -- & $>16.48$ & $>13.85$ & 12.7 & -- &  & -- & -- & 3.78 & 1.15 & \\
1294-79B5 & -- & -- & -- & $>13.44$ & 12.44 & 11.57 & 10.18\textdagger & 9.63 & 4.33 & 0.54 & 1.87 & 0.87 & 0.55\\
1287-9C01 & 18.42 & 17.24 & 16.97 & $>14.81$ & $>14.00$ & 12.60 & 8.36\textdagger & 5.02\textdagger & $>0.81$ & $>1.68$ & 2.74 & 2.17 & 3.34\\
1313-4913 & -- & -- & 16.01 & 13.45 & 12.79 & 11.8 & 9.9\textdagger & 9.33\textdagger & 4.08 & -0.36 & 1.65 & 0.99 & 0.57\\
1326-4DC7 & -- & -- & -- & 13.91 & 12.68 & 11.4 & 9.12 & 8.01 & 5.46 & 4.11 & 2.51 & 1.28 & 1.11\\
1312-111ED & -- & -- & -- &  & -- &  & -- &  & -- & -- &  --&  --& --\\
1338-18F & -- & -- & -- & 14.32 & 11.48 & 9.88 & 8.37 & 7.69 & 7.19 & 6.55 & 4.44 & 1.60 & 0.68\\
1331-65D3 & -- & -- & -- & 15.69 & $>13.43$ & $>12.51$ & 11.82 & 12.06 & 6.06 & 1.89 & 3.18 & 0.92 & -0.24\\
1343-69E & -- & -- & -- & 13.60 & 11.07 & 9.59 & 8.46\textdagger & 7.72 & 6.67 & 3.64 & 4.01 & 1.48 & 0.73\\
1343-7F1 & -- & -- & -- & $>13.99$ & 11.57\textdagger & 9.47 & 8.21 & 7.38 & 6.5 & 4.72 & 4.52 & 2.10 & 0.83\\
1353-3108 & -- & -- & -- & 13.82 & 12.8 & 11.40 & -- &  & -- & -- & 2.42 & 1.40 & --\\
1352-1192 & -- & -- & 18.13 & 12.99 & 12.05 & 11.41 & 10.56 & 10.30 & 10.29: & $>7.40$ & 1.58 & 0.64 & 0.26\\
1407-E01 & 19.16 & -- & 16.19 & $>13.72$ & 11.61 & 9.98 & 8.7 & 8.18 & 7.55 & $>7.47$ & 3.74 & 1.63 & 0.52\\
1430-AB0 & 17.07 & 15.21 & 16.90 & 11.70\textdagger & 11.11\textdagger & 10.64 & 9.92 & 9.63 & 9.12 & >$8.19$ & 1.06 & 0.48 & 0.30\\
1442-59DB & -- & -- & -- & 14.35 & 13.85 & 12.63 & 10.17\textdagger & 9.62\textdagger & 4.79 & 1.69 & 1.72 & 1.22 & 0.55\\
1443-760 & 18.40 & -- & 14.38 & 14.35 & 11.41 & 9.7 & 8.33\textdagger & 7.42\textdagger & 6.72 & 2.93\textdagger & 4.65 & 1.71 & 0.91\\
1457-472 & -- & -- & -- & 10.59 & 10.19 & 9.84 & 9.14\textdagger & 8.82\textdagger & 7.7 & 5.85 & 0.75 & 0.35 & 0.32\\
1485-5BE & 13.47 & 13.10 & 12.75 & 11.01 & 10.26 & 9.92 & 9.39 & 8.36 & 6.5\textdagger & 2.97\textdagger & 1.09 & 0.34 & 1.03\\
1485-95A & 14.98 & 13.88 & 12.70 & $>11.76$ & $>10.80$ & 10.34 & 9.77 & 9.59 & $>9.42$ & $>6.31$ & 1.42 & 0.46 & 0.18\\
1489-1D47 & 17.34 & 16.14 & 14.80 & 14.08 & 12.12 & 10.62\textdagger & 7.55\textdagger & 6.52\textdagger & 1.21 & -2.04\textdagger & 3.46 & 1.50 & 1.03\\
1510-224 & -- & -- & -- & 10.1 & 8.43 & 7.56 & 6.95 & 6.53 & 4.93 & 3.72 & 2.54 & 0.87 & 0.42\\
1499-691 & -- & -- & -- & 10.86 & 9.94 & 9.29 & 8.67 & 8.41 & 8.57 & $>6.96$ & 1.57 & 0.65 & 0.26\\
1510-2A8 & 18.79 & 17.69 & 15.52 & 11.1 & 9.29 & 8.25 & 7.32 & 6.85 & 6.59 & 6.37 & 2.85 & 1.04 & 0.47\\
1511-30E & -- & -- & 18.08 & 9.98 & 9.63 & 9.55 & 9.33 & 9.31 & 8.84: & $>5.54$ & 0.43 & 0.08 & 0.02\\
1511-98F & 12.16 & 11.73 & 11.44 & 11.13 & 10.84 & 10.66 & 10.38 & 10.38 & $>9.05$ & $>6.70$ & 0.47 & 0.18 & 0.00\\
1523-620A & 13.62 & 13.41 & 12.27 & $>16.29$ & $>14.31$ & 13.22: & $>9.54$ & 7.84 & 3.08 & 0.03 & 3.07 & 1.09 & 1.70\\
1527-318B & -- & -- & -- & $>16.82$ & 14.99: & 12.89 & 9.73 & 7.3 & 2.15 & -1.2 & 3.93 & 2.10 & 2.43\\
1541-4EAC & -- & -- & -- & 15.05 & 13.87 & 12.8 & -- &  & -- & -- & 2.25 & 1.07 & --\\
1633-32F & -- & -- & 18.52 & 14.91 & 12.33 & 10.32 & 5.9 & 4.11 & 1.51 & 0.41 & 4.59 & 2.01 & 1.79\\
1631-598D & -- & -- & -- & $>18.25$ & $>16.65$ & 14.08 & 12.22 & 10.75 & 5.39 & 2.44 & 4.17 & 2.57 & 1.47\\
1648-2717 & -- & -- & -- & $>16.44$ & 14.88 & 13.09 & 11.55 & 10.45 & 4.83 & 0.99 & 3.35 & 1.79 & 1.10\\
1702-13C & -- & -- & -- & 11.5 & 9.72 & 8.99 & 8.54 & 8.62 & 8.68 & $>7.63$ & 2.51 & 0.73 & -0.08\\
1697-173E & -- & 17.72 & 18.69 & 14.18 & 13.23 & 11.98 & 10.73 & 9.68 & 4.09 & 0.33 & 2.20 & 1.25 & 1.05\\
1717-203 & -- & -- & 17.4 & 11.95 & 10.21 & 9.23 & 8.04 & 7.57 & 6.83 & 6.01 & 2.72 & 0.98 & 0.47\\
1704-2C5 & -- & -- & -- & 10.95 & 10.37 & 9.89 & 9.13 & 8.81 & 7.78 & 4.72 & 1.06 & 0.48 & 0.32\\
1734-1187 & 16.20 & 14.87 & 14.27 & 14.15 & 13.12 & 12.42 & 11.74 & 11.39 & 10.67 & 8.15 & 1.73 & 0.70 & 0.35\\
1764-165F & -- & -- & -- & 14.42 & 13.7 & 12.57 & 10.43\textdagger & 9.78\textdagger & 4.66 & 0.82 & 1.85 & 1.13 & 0.65\\
\hline
\end{tabular}
\end{center}
\end{table*}

\begin{table*}
\caption{Extinction and distances for confirmed WR stars. It is difficult to differentiate between features of WC$4-8$; colon (:) indicates an uncertainty of up to $\pm2$ subtypes. $K_s$-band extinction was calculated from \textit{2MASS} colours and subtype values provided in \citet{2006MNRAS.372.1407C}, while M$_{K_s}$ values are derived for spectral subtypes from \citet{2015MNRAS.447.2322R}. Distances ($d$) and Galactocentric radii (R$_G$) are in kpc, with typical uncertainties of $\sim25$ per cent.\label{tab:dist}}
\begin{center}
\begin{tabular}{lccccccccccc}
\hline \noalign{\smallskip}
\textbf{Name} & \textbf{Subtype} & \textbf{$J$} & \textbf{$H$} & \textbf{$K_s$} & \textbf{A$^{J-K_s}_{K_s}$} & \textbf{A$^{H-K_s}_{K_s}$} & \textbf{A$_{K_s}$} & \textbf{M$_{K_s}$} & \textbf{DM} & \textbf{$d$} & \textbf{R$_G$}  \\ 
\noalign{\smallskip} \hline
1040-B6C &  WN9 &13.13 & 11.90 & 11.17 & 1.32 & 1.32 & 1.3 & -6.32 & 16.2 & 17.2 & 10.7 \\
1089-1117 &  cLBV/WNL &16.82 & 12.60 & 10.28 & 4.21 & 4.38 & 4.3 & -6.32 & 12.3 & 2.9 & 6.0 \\
1093-1765 &  WN6 &15.15 & 12.93 & 11.57 & 2.46 & 2.40 & 2.4 & -4.94 & 14.1 & 6.6 & 3.6 \\
1139-49EA &  WC6:: &16.25 & -- & 13.09 & 2.12 & -- & -- & -4.66 & -- & -- & -- \\
1178-66B & WC9 &12.49 & 11.12 & 10.26 & 1.57 & 1.50 & 1.5 & -4.57 & 13.3 & 4.6 & 4.2 \\
1176-B49 & WN9h &12.66 & 11.22 & 10.41 & 1.47 & 1.51 & 1.5 & -6.34 & 15.3 & 11.3 & 3.6 \\
1198-6EC8 & WC6:: & -- & -- & 13.47 & -- & -- & -- & -4.66 & -- & -- & -- \\
1256-1483A & WN9 & 13.98 & 12.46 & 11.86 & 1.42 & 1.09 & 1.26 & -6.32 & 16.9 & 24.2 & 15.7\\
1319-3BC0 & WC7: &14.55 & -- & 12.18 & -- & 1.59 & -- & -4.84 & -- & -- & --  \\
1338-2B3 & WN9 &12.61 & -- & 8.79 & -- & 2.56 & -- & -6.32 & -- & -- & --  \\
1343-284 & WN8-9 &10.47 & 9.60 & 9.02 & 1.06 & 0.98 & 1.0 & -5.82 & 13.8 & 5.8 & 2.9 \\
1366-438 & WN7-8 &12.91 & -- & 8.77 & -- & 2.78 & -- & -5.49 & -- & -- & --  \\
1367-638 & WN9 &16.15 & 12.43 & 10.40 & 3.70 & 3.86 & 3.8 & -6.32 & 12.9 & 3.9 & 4.7 \\
1381-19L & WC9 &9.66 & 8.62 & 8.69 & -0.13 & 0.65 & 0.3 & -4.57 & 13.0 & 4.0 & 7.0 \\
1389-4AB6 & WC7 &16.13 & 14.26 & 12.24 & 3.69 & 2.61 & 3.2 & -4.84 & 13.9 & 6.1 & 2.9 \\
1389-1F5D & WN8 &-- & 13.28 & 11.05 & 4.05 & -- & -- & -5.82 & -- & -- & --  \\
1446-B1D & WN6 &12.21 & 11.24 & 10.61 & 1.14 & 1.07 & 1.1 & -4.94 & 14.5 & 7.8 & 3.0 \\
1457-673 & WC9d &14.52 & 11.44 & 9.35 & 3.81 & 3.46 & 3.6 &-4.57 & 10.3 & 1.1 & 7.5 \\
1485-6C4 & WN6 &12.04 & 10.81 & 10.02 & 1.44 & 1.35 & 1.4 & -4.94 & 13.6 & 5.2 & 4.4 \\
1485-844 & WN8 &14.72 & 11.30 & 9.49 & 3.29 & 3.50 & 3.4 & -5.82 & 12.0 & 2.4 & 6.4 \\
1495-1D8A & WC8-9 &-- & -- & 11.72 & -- & -- & -- & -5.04 & -- & -- & --  \\
1495-705 & WN8 &14.97 & 11.38 & 9.16 & 4.03 & 3.89 & 4.0 & -5.82 & 11.0 & 1.6 & 7.1 \\
1514-AA0 & WC8 &12.92 & 11.24 & 10.54 & 1.29 & 1.60 & 1.4 & -5.04 & 14.1 & 6.7 & 4.2 \\
1509-2E64 & WC9 & -- & -- & 12.45 & -- & -- & -- & -4.57 & -- & -- & --  \\
1525-2352 & WC8: & -- & 14.75 & 12.44 & 4.19 & -- & -- & -5.04 & -- & -- &  -- \\
1519-E43 & WC7 & -- & 13.04 & 11.26 & 3.23 & -- & -- & -4.84 & -- & -- &  -- \\
1530-8FA & WN5 &12.76 & 11.43 & 10.64 & 1.43 & 1.42 & 1.4 & -3.86 & 13.1 & 4.1 & 5.4 \\
1541-3C8 & WC8 & -- & 13.86 & 12.15 & 3.10 & -- & -- & -5.04 & -- & -- &  -- \\
1541-197C & WC8 &15.14 & 13.28 & 11.62 & 3.03 & 2.36 & 2.7 & -5.04 & 14.0 & 6.2 & 4.7 \\
1544-FA4 & WN5 &13.74 & -- & 10.83 & 1.95 & -- & -- & -3.86 & -- & -- & -- \\
1553-9E8 & WN9h &15.76 & 12.86 & 10.98 & 3.44 & 3.21 & 3.3 & -6.34 & 14.0 & 6.3 & 4.8 \\
1547-1488 & WN5 &13.96 & 12.25 & 11.16 & 1.99 & 1.87 & 1.9 & -3.86 & 13.1 & 4.1 & 5.5 \\
1553-15DF & WC8 & -- & 14.80 & 12.07 & 4.98 & -- & -- & -5.04 & -- & -- &  -- \\
1602-9AF & WN6 &13.11 & 11.62 & 11.05 & 1.04 & 1.38 & 1.2 & -4.94 & 14.8 & 9.1 & 6.1 \\
1603-11AD & WN5 &16.13 & 13.64 & 12.17 & 2.69 & 2.66 & 2.7 & -3.86 & 13.3 & 4.7 & 5.8 \\
1609-1C95 & WC9 &-- & 15.07 & 11.93 & 5.72 & -- & -- & -4.57 & -- & -- &  -- \\
1626-4FC8 & [WC6:] &15.59 & 14.86 & 13.89 & 1.77 & 1.14 & 1.5 & -- & -- & -- & -- \\
1629-14D6 & WN9h &14.73 & 13.44 & 12.63 & 1.49 & 1.41 & 1.5 & -6.34 & 17.5 & 31.9 & 26.4 \\
1627-A6D & WC7:: &15.85 & 13.54 & 11.75 & 3.25 & 2.75 & 3.0 & -4.84 & 13.6 & 5.2 & 5.9 \\
1635-AD8 & WN6 &15.20 & 12.87 & 11.48 & 2.54 & 2.49 & 2.5 & -4.94 & 13.9 & 6.0 & 5.9 \\
1653-FFE & WN5-6 &15.15 & -- & 11.66 & -- & 2.34 & -- & -3.86 & -- & -- & --  \\
1651-BB4 & WN5 &15.83 & -- & 11.73 & -- & 2.74 & -- & -3.86 & -- & -- & --  \\
1647-1E70 & WC8: &-- & 15.10 & 12.57 & 4.60 & -- & -- & -5.04 & -- & -- & --  \\
1659-212 & WN9 &13.44 & 10.89 & 9.53 & 2.48 & 2.62 & 2.6 & -6.32 & 13.3 & 4.6 & 6.3 \\
1669-3DF & WN9h &12.79 & 11.04 & 9.97 & 1.95 & 1.89 & 1.9 & -6.34 & 14.4 & 7.5 & 6.6 \\
1660-1169 & WC6: &14.65 & 13.29 & 12.08 & 2.21 & 1.73 & 2.0 & -4.66 & 14.8 & 9.0 & 7.0 \\
1697-38F & WC9 &12.97 & -- & 9.92 & 2.28 & -- & -- & -4.57 & -- & -- & -- \\
1702-23L & WC8 &11.86 & 11.00 & 10.21 & 1.44 & 1.11 & 1.3 & -5.04 & 14.0 & 6.2 & 6.8 \\
1695-2B7 & WC9 &13.01 & 11.09 & 9.64 & 2.64 & 2.26 & 2.5 & -4.57 & 11.8 & 2.3 & 7.3 \\
\hline
\end{tabular}
\end{center}
\end{table*}

\begin{table*}
\caption{Lines in the NIR Wolf--Rayet spectrum, a subset of \citet[table 2]{1997ApJ...486..420F}. These lines are marked in the spectra shown in figures~\ref{fig:wc} and~\ref{fig:wn}.\label{tab:lines}}
\begin{center}
\begin{tabular}{clcl}
\hline \noalign{\smallskip}
\textbf{Wavelength (\micron)} & \textbf{Transition} & \textbf{Wavelength (\micron)} & \textbf{Transition} \\
\noalign{\smallskip} \hline
2.037 & \ion{He}{2} $15-8$ & 2.148 & \ion{He}{2} $24-9$\\
2.059 & \ion{He}{1} $\mathrm{2s\pres{^1}S-2p\pres{^1}P^0}$ & 2.162 & \ion{He}{1} $\mathrm{7\pres{^1}L-4\pres{^1}D}$ \\
2.071 & \ion{C}{4} $\mathrm{3p\pres{^2}P_{1/2}^0-3d\pres{^2}D_{3/2}}$ & 2.165 & \ion{He}{2} $14-8$\\
2.080 & \ion{C}{4} $\mathrm{3p\pres{^2}P_{3/2}^0-3d\pres{^2}D_{5/2}}$ & 2.166 & \ion{H}{1} $7-4$\\
2.084 & \ion{C}{4} $\mathrm{3p\pres{^2}P_{3/2}^0-3d\pres{^2}D_{3/2}}$ & 2.179 & \ion{He}{2} $23-9$ \\
2.097 & \ion{He}{2} $26-9$ & 2.189 & \ion{He}{2} $10-7$ \\
2.100 & \ion{N}{5} $11-10$ & 2.217 & \ion{He}{2} $22-9$\\
2.104 & \ion{C}{3}/\ion{N}{3} $8-7$ & 2.247 & \ion{N}{3} $\mathrm{5s\pres{^2}S_{1/2}-5p\pres{^2}P^0_{3/2}}$ \\
2.108 & \ion{C}{3} $\mathrm{5s\pres{^1}S-5p\pres{^1}P^0}$ & 2.278 & \ion{C}{4}/\ion{N}{4} $15-12$ \\
2.113 & \ion{He}{1} $\mathrm{4s\pres{^3}S-3p\pres{^3}P}$ & 2.314 & \ion{He}{2} $20-19$ \\
2.115 & \ion{C}{3}/\ion{N}{3} $8-7$ & 2.320 & \ion{C}{4} $\mathrm{13d\pres{^2}D-11p\pres{^2}P^0}$ \\
2.116 & \ion{C}{3}/\ion{N}{3} $8-7$ & 2.335 & \ion{C}{4} $\mathrm{9p\pres{^2}P_{1/2}^0-10d\pres{^2}D}$, $\mathrm{9p\pres{^2}P_{3/2}^0-10d\pres{^2}D}$ \\
2.122 & \ion{C}{3} $\mathrm{4p\pres{^1}P^0-4d\pres{^1}D}$ & 2.347 & \ion{He}{2} $13-8$ \\
2.139 & \ion{C}{4} $\mathrm{9s\pres{^2}S-10p\pres{^2}P^0}$ & 2.379 & \ion{He}{2} $19-9$ \\
\hline
\end{tabular}
\end{center}
\end{table*}

\begin{table*}
\caption{Measured equivalent widths for confirmed WN stars between 2.037 and 2.115~\micron. Lines were fit as gaussian profiles using \textsc{pan}\fnm{1} and then summed to find the equivalent width. Atomic transitions for each line are given in table~\ref{tab:lines}. The three numbers in each entry are, in order: line centre from the gaussian fit; equivalent width; line FWHM.\label{tab:ewn1}}
\begin{center}
\begin{tabular}{llcccc}
\hline \noalign{\smallskip}
\textbf{Name} & \textbf{Subtype} & \ion{He}{2} 2.037~\micron & \ion{He}{1} 2.059~\micron & \ion{He}{2} 2.097~\micron  & \ion{C}{3} 2.108~\micron \\
 & & & & \ion{N}{5} 2.100~\micron & \ion{He}{1} 2.113~\micron \\
 & & & & \ion{C}{3} 2.104~\micron & \ion{C}{3} 2.115~\micron \\
 & & & & \ion{N}{3} 2.104~\micron & \ion{N}{3} 2.115~\micron \\
\noalign{\smallskip} \hline
1040-B6C & WN9 & -- & 2.0583 25.21 27.6 & -- & -- \\
1089-1117 & cLBV/WNL & -- & 2.0581 101.85 20.6 & -- & 2.1127 9.01 28.0 \\
1093-1765 & WN6 & -- & -- & -- & 2.1126 40.27 106.3 \\
1176-B49 & WN9h & -- & 2.0591 61.87 24.7 & -- & 2.1137 7.27 26.2 \\
1256-1438A & WN9 & -- & 2.0592 5.45 24.4 & -- & -- \\
1338-2B3 & WN9 & -- & 2.0554 11.90 29.3 & -- & -- \\
1343-284 & WN8-9 & -- & 2.0549 16.24 42.0 & -- & -- \\
1366-438 & WN7-8 & -- & 2.0540 6.27 16.1 & -- & -- \\
1367-638 & WN9 & -- & 2.0592 57.71 39.0 & 2.1033 2.72 61.0 & 2.1138 13.03 37.8 \\
1389-1F5D & WN8 & -- & 2.0601 67.62 55.6 & 2.1026 1.94 26.0 & 2.1142 32.98 51.6 \\
1446-B1D & WN6 & 2.0377 13.76 84.8 & 2.0638 15.17 108.5 & 2.0974 2.07 39.6 & 2.1140 49.41 154.8 \\
1485-6C4 & WN6 & -- & -- & 2.1060 10.50 44.9 & -- \\
1485-844 & WN8 & -- & 2.0574 30.01 41.4 & -- & 2.1123 5.15 40.8 \\
1495-705 & WN8 & -- & 2.0577 38.48 35.0 & -- & -- \\
1530-8FA & WN5 & 2.0385 18.19 87.5 & 2.0604 2.10 46.0 & 2.0993 4.19 43.1 & 2.1145 26.21 131.1 \\
1544-FA4 & WN5 & 2.0356 22.29 118.1 & -- & -- & 2.1117 39.51 282.1 \\
1553-9E8 & WN9h & -- & 2.0591 77.32 25.1 & -- & 2.1136 13.25 30.8 \\
1547-1488 & WN5 & -- & -- & 2.0964 91.42 252.9 & 2.1095 26.70 68.1 \\
1602-9AF & WN6 & -- & -- & -- & 2.1069 51.82 182.3 \\
1603-11AD & WN5 & 2.0373 14.96 87.9 & -- & 2.0923 1.36 43.9 & 2.1142 23.05 131.8 \\
1629-14D6 & WN9h & -- & 2.0582 14.75 43.9 & -- & -- \\
1635-AD8 & WN6 & 2.0388 17.52 75.1 & 2.0611 5.24 43.9 & 2.0967 5.17 87.9 & 2.1146 33.55 109.5 \\
1653-FFE & WN5-6 & 2.0375 19.58 88.0 & 2.0600 2.97 87.9 & 2.0961 5.01 65.9 & 2.1130 27.61 104.8 \\
1651-BB4 & WN5 & 2.0346 13.56 109.9 & 2.0561 3.53 65.9 & -- & 2.1114 27.17 153.8 \\
1659-212 & WN9 & -- & 2.0543 3.81 22.2 & -- & -- \\
1669-3DF & WN9h & -- & 2.0579 136.96 31.0 & -- & 2.1124 19.10 30.9 \\
\hline
\end{tabular}
\end{center}
\begin{flushleft}
\fnt{1}{http://ifs.wikidot.com/pan}
\end{flushleft}
\end{table*}

\begin{table*}
\caption{Measured equivalent widths for confirmed WN stars between 2.148 and 2.217~\micron, using \textsc{pan} to fit gaussian profiles as in table~\ref{tab:ewn1}.  The three numbers in each entry are, in order: line centre from the gaussian fit; equivalent width; line FWHM.\label{tab:ewn2}}
\begin{center}
\begin{tabular}{llcccc}
\hline \noalign{\smallskip}
\textbf{Name} & \textbf{Subtype} & \ion{He}{2} 2.148~\micron & \ion{He}{1} 2.162~\micron & \ion{He}{2} 2.189~\micron & \ion{He}{2} 2.217~\micron \\
 & & & \ion{He}{2} 2.165~\micron & & \\
 & & & \ion{H}{1} 2.166~\micron & & \\
\noalign{\smallskip} \hline
1040-B6C & WN9 & -- & 2.1655 3.06 9.4 & -- & -- \\
1089-1117 & cLBV/WNL & 2.1431 4.0 35.2 & -- & -- & -- \\
1093-1765 & WN6 & -- & 2.1633 51.04 107.1 & 2.1879 49.25 92.4 & -- \\
1176-B49 & WN9h & -- & 2.1623 3.57 23.5 & -- & -- \\
 & & & 2.1663 25.52 27.5 & & \\
1256-1438A & WN9 & -- & 2.1671 6.84 24.4 & -- & -- \\
1338-2B3 & WN9 & -- & 2.1637 7.06 45.6 & -- & -- \\
1343-284 & WN8-9 & -- & 2.1628 25.38 57.1 & -- & -- \\
1366-438 & WN7-8 & -- & 2.1628 16.96 52.2 & 2.1833 4.84 48.9 & -- \\
1367-638 & WN9 & -- & 2.1657 48.98 49.4 & 2.1848 1.71 46.7 & -- \\
1389-1F5D & WN8 & -- & 2.1648 56.36 76.9 & 2.1904 6.87 37.0 & -- \\
1446-B1D & WN6 & 2.1471 2.04 62.7 & 2.1653 60.77 146.8 & 2.1898 65.02 127.4 & 2.2165 3.80 89.1 \\
1485-6C4 & WN6 & -- & 2.1592 57.82 90.1 & 2.1834 49.17 87.7 & -- \\
1485-844 & WN8 & -- & 2.1638 37.66 54.7 & -- & -- \\
1495-705 & WN8 & -- & 2.1640 43.66 45.0 & -- & 2.2057 7.54 42.4 \\
1530-8FA & WN5 & 2.1485 2.97 71.7 & 2.1655 33.30 104.3 & 2.1898 62.55 100.3 & -- \\
1544-FA4 & WN5 & -- & 2.1643 51.63 219.4 & 2.1896 123.53 198.4 & -- \\
1553-9E8 & WN9h & 2.1457 9.62 66.9 & 2.1638 17.09 89.9 & 2.1859 6.56 86.8 & 2.2078 13.22 41.0 \\
 & & & 2.1664 21.19 20.7 & & \\
1547-1488 & WN5 & -- & 2.1627 39.94 88.6 & 2.1875 120.39 134.6 & -- \\
1602-9AF & WN6 & -- & 2.1594 73.61 130.0 & 2.1881 49.74 103.3 & -- \\
1603-11AD & WN5 & -- & 2.1652 36.48 109.9 & 2.1884 104.70 153.8 & -- \\
1629-14D6 & WN9h & -- & 2.1662 16.92 22.0 & -- & -- \\
1635-AD8 & WN6 & 2.1472 3.81 65.9 & 2.1651 45.85 100.6 & 2.1899 54.25 84.5 & 2.2164 2.35 43.9 \\
1653-FFE & WN5-6 & 2.1461 3.20 87.9 & 2.1643 46.18 127.2 & 2.1884 84.10 122.5 & 2.2175 5.60 109.9 \\
1651-BB4 & WN5 & -- & 2.1631 45.21 138.6 & 2.1882 121.68 203.4 & -- \\
1659-212 & WN9 & -- & 2.1636 8.83 46.9 & -- & -- \\
1669-3DF & WN9h & -- & 2.1645 57.67 45.7 & -- & -- \\
\hline
\end{tabular}
\end{center}
\end{table*}

\begin{table*}
\caption{Measured equivalent widths for confirmed WC stars between 2.037 and 2.115~\micron, using \textsc{pan} to fit gaussian profiles as in table~\ref{tab:ewn1}.  The three numbers in each entry are, in order: line centre from the gaussian fit; equivalent width; line FWHM.\label{tab:ewc1}}
\begin{center}
\begin{tabular}{llcccc}
\hline \noalign{\smallskip}
\textbf{Name} & \textbf{Subtype} & \ion{He}{2} 2.037~\micron & \ion{He}{1} 2.059~\micron & \ion{C}{4} 2.071~\micron  & \ion{C}{3} 2.104~\micron \\
 & & & & \ion{C}{4} 2.080~\micron & \ion{C}{3} 2.115~\micron \\
 & & & & \ion{C}{4} 2.084~\micron & \\
\noalign{\smallskip} \hline
1139-49EA &  WC6:: & -- & -- & 2.0754 688.74 203.1 & 2.1116 105.67 159.2 \\
1178-66B & WC9 & -- & 2.0603 76.96 73.1 & 2.0786 6.38 41.3 & 2.1086 12.00 55.0 \\
 & & & &  2.0854 7.74 55.8 & 2.1141 28.18 54.6 \\
1198-6EC8 & WC6:: & -- & -- & 2.0751 1856.57 267.1 & 2.1132 269.48 207.5 \\
1319-3BC0 & WC7: & -- & -- & 2.0750 595.53 241.5 & 2.1123 110.36 237.8 \\
1381-19L & WC9 & -- & 2.0553 88.10 63.3 & 2.0693 170.15 193.5 & 2.1068 123.28 224.5 \\
1389-4AB6 & WC7 & -- & -- & 2.0767 309.24 244.1 & 2.1164 61.56 170.9 \\
1457-673\fnm{a} & WC9d & -- & 2.0536 21.86 55.3 & -- & -- \\
1495-1D8A & WC8-9 & -- & 2.0590 92.39 98.7 & 2.0793 78.20 114.0 & 2.1109 114.26 91.9 \\
1514-AA0 & WC8 & -- & -- & 2.0765 339.28 191.6 & 2.1129 131.97 145.3 \\
1509-2E64 & WC9 & 2.0385 6.25 46.7 & 2.0622 196.58 130.6 & 2.0807 153.45 131.8 & 2.1131 166.77 123.0 \\
1525-2352 & WC8: & -- & -- & 2.0779 390.00 175.8 & 2.1136 119.99 109.9 \\
1519-E43 & WC7 & -- & -- & 2.0755 448.47 213.1 & 2.1115 178.18 161.9 \\
1541-3C8 & WC8 & -- & -- & 2.0822 217.71 145.9 & 2.1159 177.26 151.7 \\
1541-197C\fnm{b} & WC8 & -- & -- & 2.0723 270.80 188.4 & 2.1087 157.25 155.7 \\
1553-15DF & WC8 & -- & -- & 2.0692 33.69 100.5 & 2.1125 41.06 115.7 \\
 & & & & 2.0800 55.07 104.4 & \\
1609-1C95 & WC9 & -- & 2.0604 46.12 80.4 & 2.0694 6.09 34.2 & 2.1126 39.84 93.2 \\
 & & & & 2.0785 12.22 48.5 & \\
 & & & & 2.0851 11.35 59.6 & \\
1626-4FC8 & [WC6:] & -- & -- & 2.0696 94.76 65.9 & 2.1142 114.74 109.9 \\
 & &  & & 2.0798 353.35 109.9 & \\
1627-A6D & WC7:: & -- & -- & 2.0760 693.81 250.4 & 2.1131 171.41 282.6 \\
1647-1E70 & WC8: & -- & -- & 2.0764 605.00 192.3 & 2.1123 163.74 131.8 \\
1660-1169 & WC6: & -- & -- & 2.0713 1059.04 179.9 & 2.1107 309.08 283.3 \\
 & & & & 2.0857 373.04 126.7 & \\
1697-38F & WC9 & 2.0387 3.48 89.8 & 2.0614 12.69 82.2 & 2.0683 10.22 48.4 & 2.1122 61.58 113.5 \\
 & & & & 2.0799 62.47 139.0 & \\
1702-23L & WC8 & -- & -- & 2.0761 678.75 204.2 & 2.1114 575.71 163.1 \\
1695-2B7\fnm{c} & WC9 & 2.0383 4.12 71.2 & 2.0626 14.67 83.4 & 2.0687 19.80 50.0 & 2.1126 91.34 113.1 \\
  & & & & 2.0799 106.61 128.7 & \\
\hline
\end{tabular}
\end{center}
\end{table*}
 
\begin{table*}
\caption{Measured equivalent widths for confirmed WC stars between 2.139 and 2.217~\micron, using \textsc{pan} to fit gaussian profiles as in table~\ref{tab:ewn1}.  The three numbers in each entry are, in order: line centre from the gaussian fit; equivalent width; line FWHM.\label{tab:ewc2}}
\begin{center}
\begin{tabular}{llcccc}
\hline \noalign{\smallskip}
\textbf{Name} & \textbf{Subtype} & \ion{C}{4} 2.139~\micron & \ion{He}{1} 2.162~\micron & \ion{He}{2} 2.189~\micron & \ion{He}{2} 2.217~\micron \\
 & & & \ion{He}{2} 2.165~\micron & & \\
\noalign{\smallskip} \hline
1139-49EA &  WC6:: & -- & -- & 2.1896 33.45 95.5 & -- \\
1178-66B & WC9 & 2.1392 0.98 26.8 & 2.1643 32.19 80.3 & -- & --\\
1198-6EC8 & WC6:: & -- & -- & 2.1880 54.71 134.3 & -- \\
1319-3BC0 & WC7: & -- & 2.1687 28.39 170.2 & 2.1890 43.11 155.7 & -- \\
1381-19L & WC9 & -- & 2.1646 40.83 117.9 & 2.1867 43.35 160.5 & 2.2189 61.83 156.4 \\
1389-4AB6 & WC7 & -- & 2.1575 3.23 51.2 & 2.1842 29.36 170.9 & -- \\
1457-673\fnm{a} & WC9d & -- & 2.1625 21.57 65.2 & -- & -- \\
1495-1D8A & WC8-9 & 2.1366 11.39 39.0 & 2.1616 60.14 66.2 & 2.1829 32.05 152.2 & -- \\
1514-AA0 & WC8 & 2.1405 8.23 87.9 & 2.1648 58.78 148.9 & 2.1889 51.93 117.5 & -- \\
1509-2E64 & WC9 & 2.1397 14.98 109.9 & 2.1647 59.88 117.9 & 2.1879 44.72 115.0 & 2.2230 10.25 109.9 \\
1525-2352 & WC8: & 2.1374 3.58 65.9 & 2.1700 6.90 65.9 & 2.1897 21.92 109.9 & -- \\
1519-E43 & WC7 & -- & 2.1628 96.08 181.4 & 2.1876 70.14 143.9 & -- \\
1541-3C8 & WC8 & -- & -- & 2.1925 34.46 113.8 & 2.2147 96.09 282.5 \\
1541-197C\fnm{b} & WC8 & 2.1318 44.09 154.9 & 2.1637 11.95 61.6 & 2.1865 8.98 70.7 & -- \\
1553-15DF & WC8 & -- & 2.1648 12.28 93.3 & 2.1859 14.98 130.2 & -- \\
1609-1C95 & WC9 & -- & 2.1646 11.22 67.7 & 2.1841 15.61 132.3 & -- \\
1626-4FC8 & [WC6:] & -- & 2.1661 16.61 43.9 & 2.1900 49.48 87.9 & -- \\
1627-A6D & WC7:: & -- & 2.1591 19.65 144.7 & 2.1850 82.65 267.4 & -- \\
1647-1E70 & WC8: & 2.1362 5.07 65.9 & 2.1670 15.77 87.9 & 2.1887 32.37 109.9 & -- \\
1660-1169 & WC6: & -- & -- & 2.1855 60.61 219.5 & -- \\
1697-38F & WC9 & 2.1396 6.78 87.9 & 2.1638 17.12 106.4 & 2.1881 11.09 78.8 & 2.2231 2.63 68.9 \\
1702-23L & WC8 & -- & 2.1617 251.42 94.9 & 2.1848 958.74 302.8 & -- \\
1695-2B7\fnm{c} & WC9 & 2.1394 6.73 65.0 & 2.1641 21.24 97.2 & 2.1894 12.09 62.8 & -- \\
\hline
\end{tabular}
\end{center}
\begin{flushleft}
\fnt{a}{This spectrum also contains the \ion{C}{3} line at 2.122~\micron\ (2.1264 4.18 29.0).}\\
\fnt{b}{This spectrum also contains the \ion{He}{2} line at 2.148~\micron\ (2.1501 29.00 108.7).}\\
\fnt{c}{This spectrum also contains the \ion{He}{2} line at 2.179~\micron\ (2.1822 11.86 97.0).}\\
\end{flushleft}
\end{table*}

One of the confirmed WR stars is particularly notable: 1627-A6D, a WC7:: which is associated with a \textit{Chandra} x-ray source (CXO J191011.5+085839); no other WRs described here were successfully matched to \textit{Chandra} sources. 1627-A6D was originally identified in \citet{2012AJ....144..166S}, also as a WC7. Using the relation $A_K/A_V = 0.112$ from \citet{1985ApJ...288..618R} with the calculated $A_{Ks}$ of 3.0 (table~\ref{tab:dist}), we derive a $V$-band extinction of 26.8 for $R=3.11$ extinction. Then, we derive the hydrogen column density as in \citet{1995A&A...293..889P}, arriving at $N_h = 4.82\times10^{22}$~cm$^{-2}$. This source has an ACIS--broad x-ray flux of $3.1\times10^{-14}$~erg s$^{-1}$ cm$^{-2 }$, and assuming a characteristic temperature of 1~keV, an unabsorbed flux of $4.1\times10^{-13}$~erg s$^{-1}$ cm$^{-2}$ (calculated via WebPIMMS\footnote{https://heasarc.gsfc.nasa.gov/cgi-bin/Tools/w3pimms/w3pimms.pl}). Using the calculated distance shown in table~\ref{tab:dist}, we find an x-ray luminosity of $1.7\times10^{33}$~erg s$^{-1}$, which is consistent with other WN x-ray luminosities, but two orders of magnitude higher than the WCs in \citet{2006Ap&SS.304...97S}. For a characteristic temperature of 3~keV, this becomes an unabsorbed flux of $1.0\times10^{-13}$~erg s$^{-1}$ cm$^{-2}$ and an x-ray luminosity of $3.4\times10^{32}$~erg s$^{-1}$; if the characteristic temperature is 10~keV, the unabsorbed flux would be $7.6\times10^{-14}$~erg s$^{-1}$ cm$^{-2}$, for an x-ray luminosity of $2.5\times10^{32}$~erg s$^{-1}$.  As there are no known single WC stars which show x-ray emission, it is statistically most likely that 1627-A6D is a colliding-wind binary, but without more extensive x-ray spectroscopy, we cannot discount the possibility of a WC7 + compact companion.

In addition to new WR stars, we classified a number of other emission-line objects which were selected using our tools.  The 17 planetary nebulae (spectra shown in figure~\ref{fig:pne}) display strong emission in the \ion{He}{1} 2.06~\micron\ and Br $\gamma$/\ion{He}{1} 2.17~\micron\ filters, with little to no continuum.  PNe are quite easy to identify, particularly using \textit{2MASS} and \textit{WISE} colour criteria, as described in \citet{2014AJ....147..115F} and a paper in preparation. We also observed nine emission-line sources which are likely to be emitting YSOs (cf. \citealt{1996AJ....112.2184G}) due to the lack of CO bands redwards of 2.3~\micron; these spectra are shown in figure~\ref{fig:yso}. 

Another source of interlopers in the WR candidate set is Be stars \citep[see also][]{2014AJ....147..115F,2010ApJ...710..706M}; the 5 Be spectra are shown in figure~\ref{fig:be}.  These spectra show relatively weak emission in the 2.17~\micron\ filter, with strong continuum emission and a large number of hydrogen lines, especially in the H-band. Finally, the largest category of contaminants amongst our candidates are M giants and supergiants, often with Br-$\gamma$ emission, as shown in figures~\ref{fig:me} and~\ref{fig:duds} (cf. \citealt{2009ApJS..185..289R}). Contaminants with no emission were generally chosen using criteria from the \ion{He}{2} and Br-$\gamma$ filters early on in the selection process; due to the shape of the spectra, these stars appeared brighter in those filters than in at least one of the continuum filters. Once we discovered this trend, that criterion was no longer used.

\section{Galactic WR Distribution}\label{sec:dist}
\begin{figure*}
\begin{minipage}[c]{0.72\linewidth}
\includegraphics[width=\linewidth]{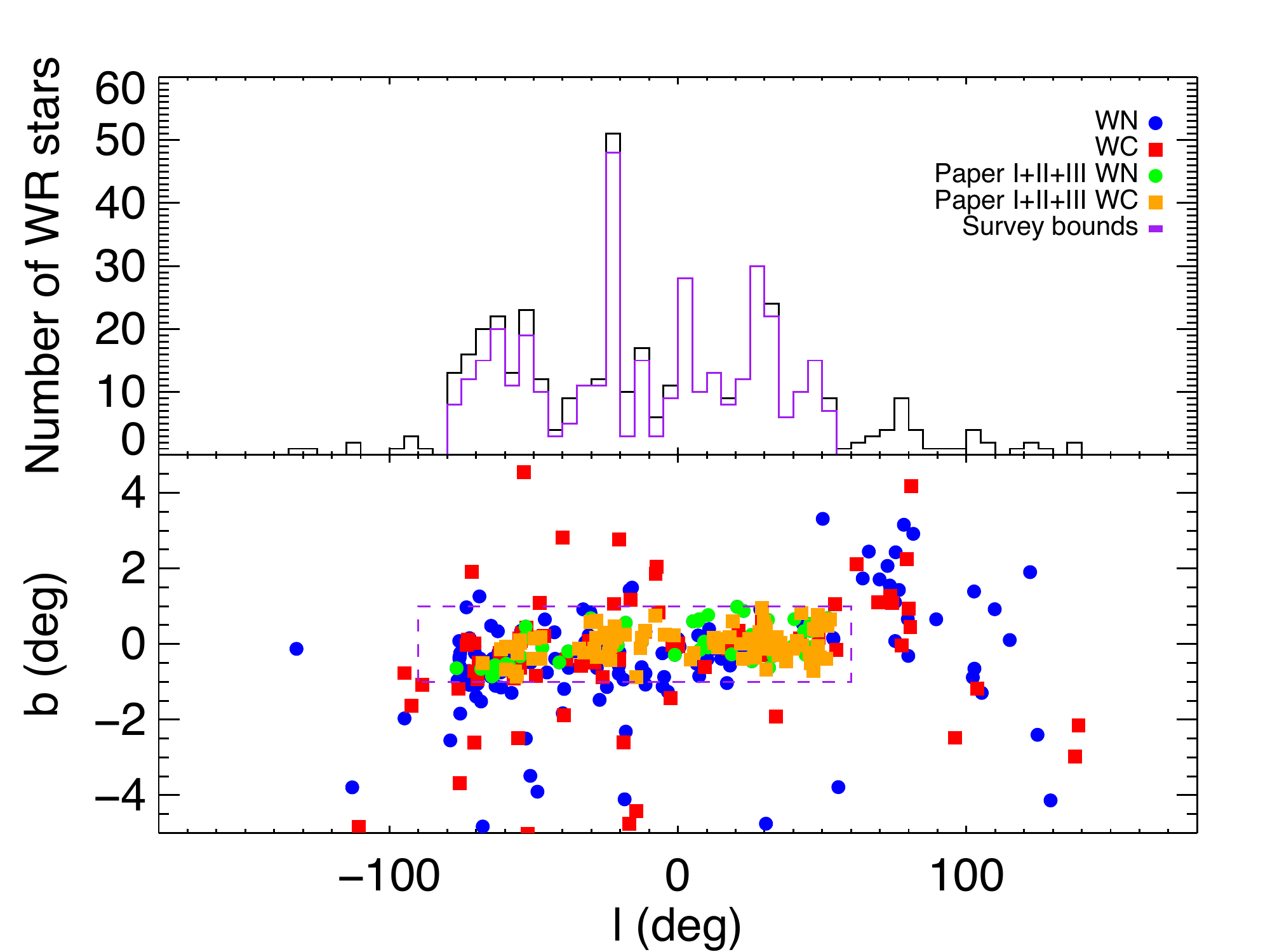}
\end{minipage}\hfill
\begin{minipage}[c]{0.28\linewidth}
\caption{\label{fig:proj} Top: A histogram of WR stars per 5\degr of longitude, with those WRs lying within the survey bounds overplotted in cyan.  80 per cent of all current WR stars lie within the bounds of the survey.  The histogram shows clear spikes where the telescope looks along the Milky Way's spiral arms, as predicted by simulations presented in the appendix of Paper I. Bottom: The distribution of WR stars projected on the sky, as a function of Galactic latitude and longitude.  Blue circles (WN) and red squares (WC) are WR stars from the literature, while green circles (WN) and orange squares (WC) indicate the contribution by this survey, as detailed in Papers I, II, and this work.  The cyan box shows the survey extent; the great majority of WR stars in the Galaxy lie in the Galactic plane, within 60-90\degr of the Galactic centre.}
\end{minipage}
\end{figure*}

We expect to find WR stars in areas with recent star formation, which in spiral galaxies like the Milky Way will be in the plane of the Galaxy, concentrated most closely along the spiral arms.  Figure~\ref{fig:proj} shows a plot of the distribution of confirmed WR stars as projected on to the sky, using absolute-magnitude calibrations for the various WR types from \citet{2015MNRAS.447.2322R}, and coordinates from the Galactic Wolf--Rayet Catalogue associated with the same publication\footnote{http://pacrowther.staff.shef.ac.uk/WRcat/index.php}.  As expected, the WR stars cluster strongly within 1\degr of $b=0$; 80 per cent of all confirmed WR stars lie within the bounds of the survey described in section~\ref{sec:obs}, despite being limited to $\pm1\degr$ in Galactic latitude.  Also included in figure~\ref{fig:proj} is a histogram showing the distribution of WR stars in Galactic longitude.  The plot shows heavy clustering at $l=\pm30\degr$, where the telescope is pointed along the Carina or Norma arms, as predicted in simulations of WR distribution based on local surface densities, described in the appendix to Paper I. 

Figure~\ref{fig:galp} shows the distribution of WR stars on the Galactic plane, overlaid on an artist's representation of the Milky Way so as to highlight the spiral arms.  This work (in combination with Papers I and II) more than doubles the number of confirmed WR stars on the far side of the Galaxy, including the most distant WR stars yet identified.  The majority of new WR stars classified here roughly trace the northern curve of the Carina arm, where one would expect the most massive star formation.

\subsection{Completeness and the Survey in Context}\label{sec:compl}
\begin{figure*}
\centering
\hspace*{\fill}%
\begin{subfigure}{.48\textwidth}
  \centering
  \includegraphics[width=\linewidth]{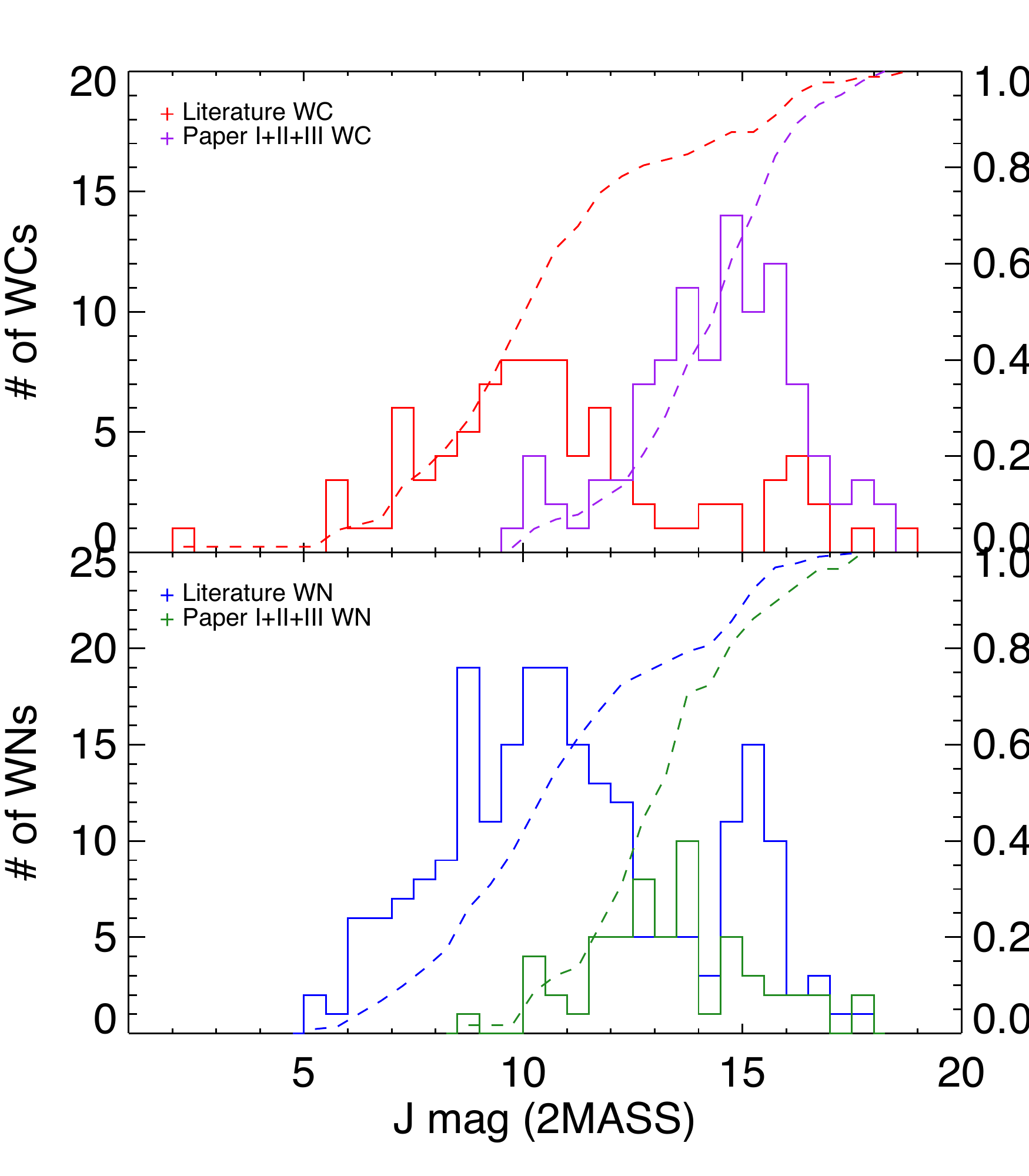}
\end{subfigure}\hfill%
\begin{subfigure}{.48\textwidth}
  \centering
  \includegraphics[width=\linewidth]{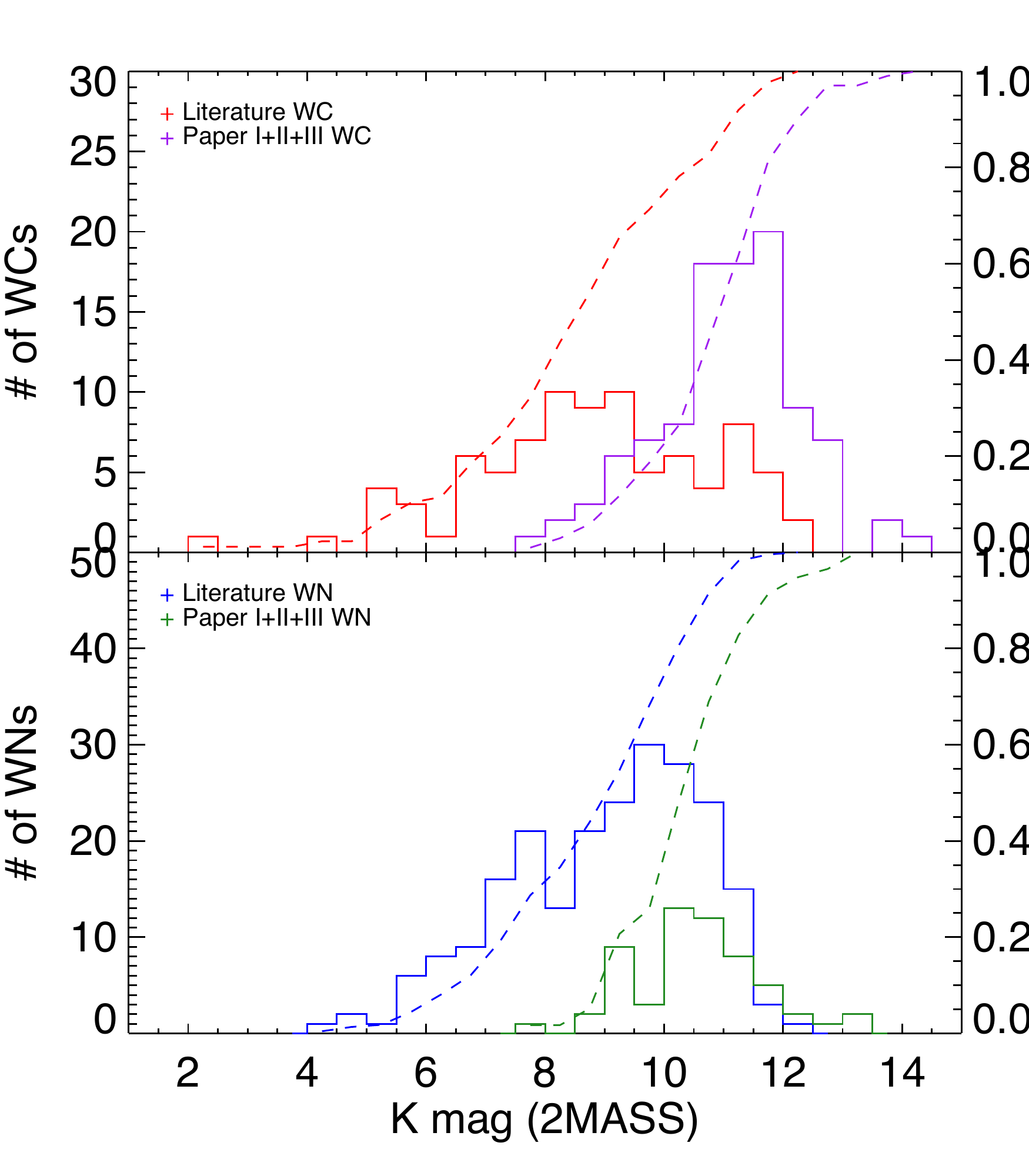}
\end{subfigure}%
\hspace*{\fill}%
\caption{\label{fig:hist_subtype}Histogram of $J-$band (left) and $K-$band (right) \textit{2MASS} magnitudes for WN and WC stars, identified in the literature or from the Shara et al. survey. Also shown is the cumulative fraction identified, as a function of magnitude in each case. The survey detailed in Papers I and II and in this paper has contributed WR stars in regions of magnitude-space which are poorly probed by the literature; the survey has identified far more faint WCs in particular than other surveys.}
\end{figure*}

\begin{figure*}
\includegraphics[width=\linewidth]{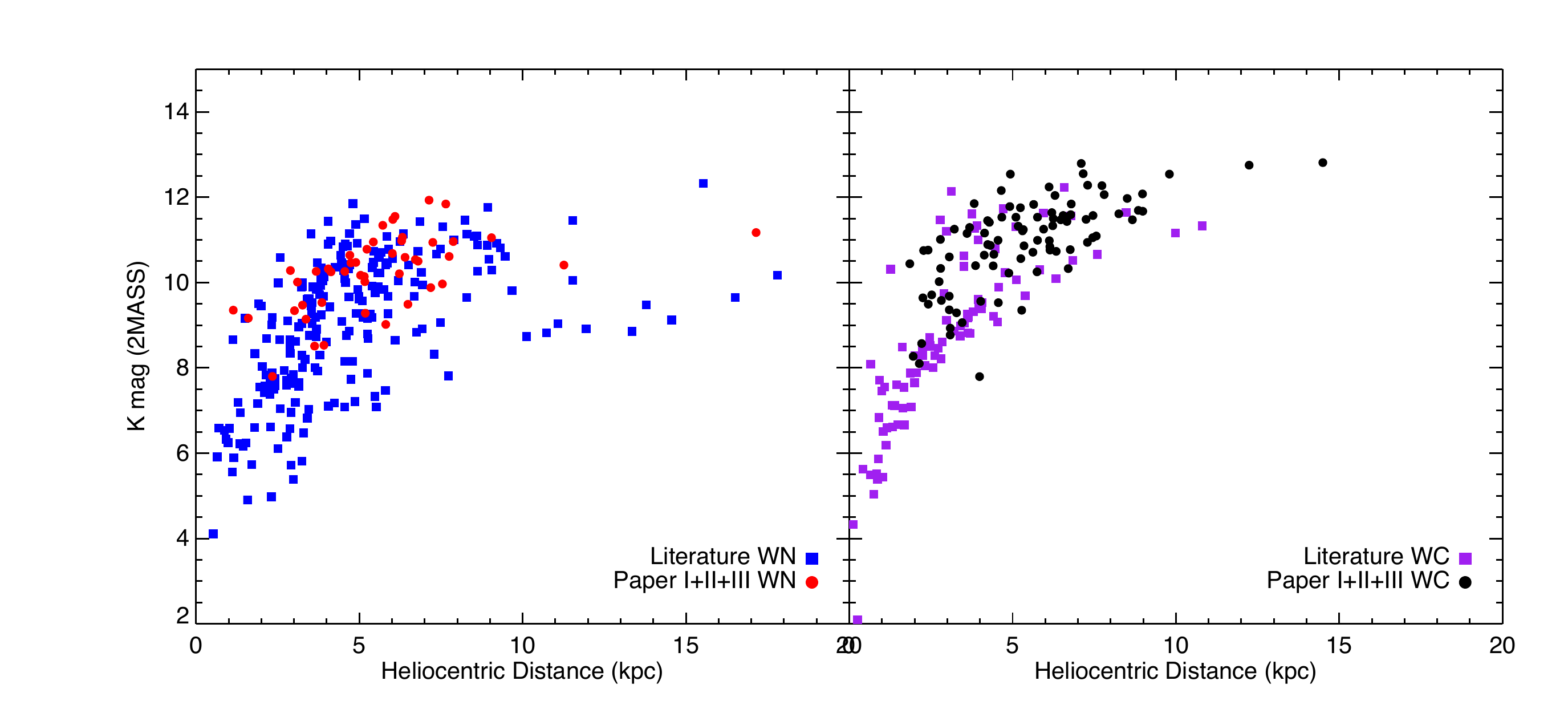}
\caption{\label{fig:distot} $K_s$-band magnitude (\textit{2MASS}) as a function of heliocentric distance for WCs and WNs. The Shara et al. survey has identified the most distant and the faintest WNs and (especially) WCs in the Milky Way.}
\end{figure*}

The image subtraction methods detailed in section~\ref{sec:obs}, in addition to the candidate selection methods described in Papers I and II, have allowed us to identify the faintest, furthest WR stars in the Galaxy. Figure~\ref{fig:hist_subtype} shows histograms and cumulative fractions of the $J-$ and $K_s-$magnitudes, respectively, for WR stars identified by the Shara et al. survey and in the literature. There is a clear divergence in the range of typical magnitudes between the two populations; the WR stars identified by this survey are significantly fainter than those in the literature, particularly for WC stars, which can be far more difficult to identify if the emission lines are diluted by continuous NIR emission by the star's shroud of dust (a characteristic of WC9ds, particularly near the Galactic centre). In addition, figure~\ref{fig:distot} shows that these faint WR stars lie at greater distances from the Solar System than any WR stars previously identified. The Shara et al. survey has more than doubled the number of confirmed WR stars on the far side of the Milky Way.

\begin{figure*}
\begin{minipage}[c]{0.48\linewidth}
\includegraphics[width=\linewidth]{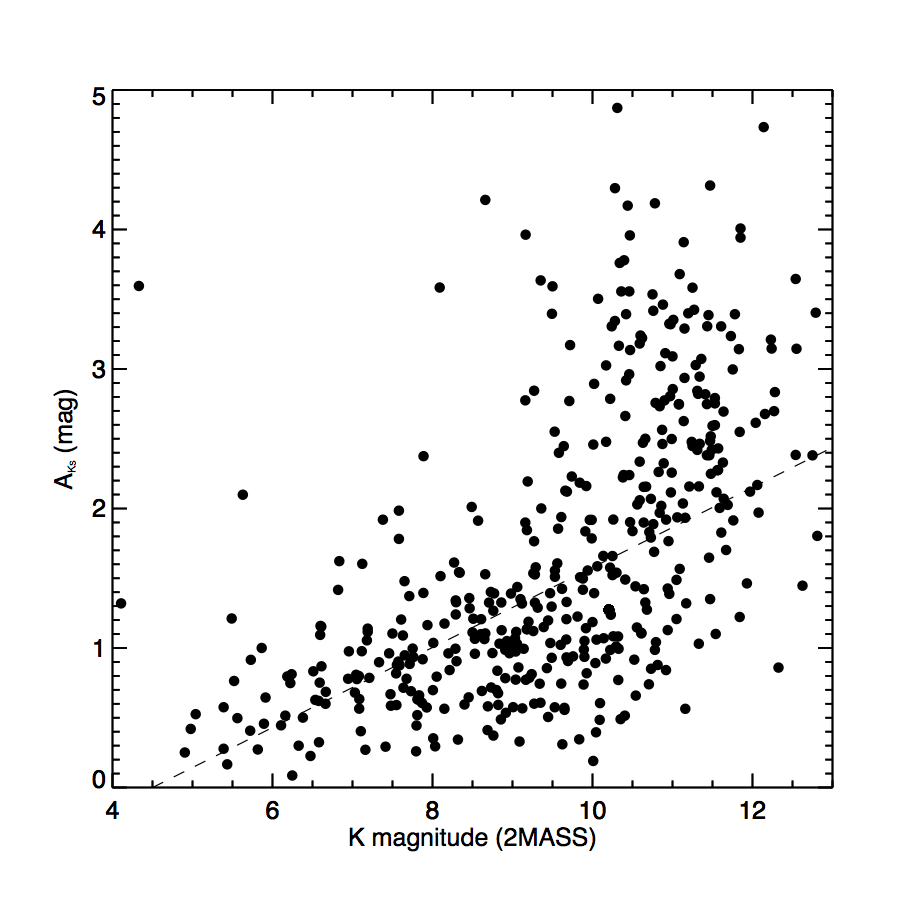}
\end{minipage}
\begin{minipage}[c]{0.48\linewidth}
\includegraphics[width=\linewidth]{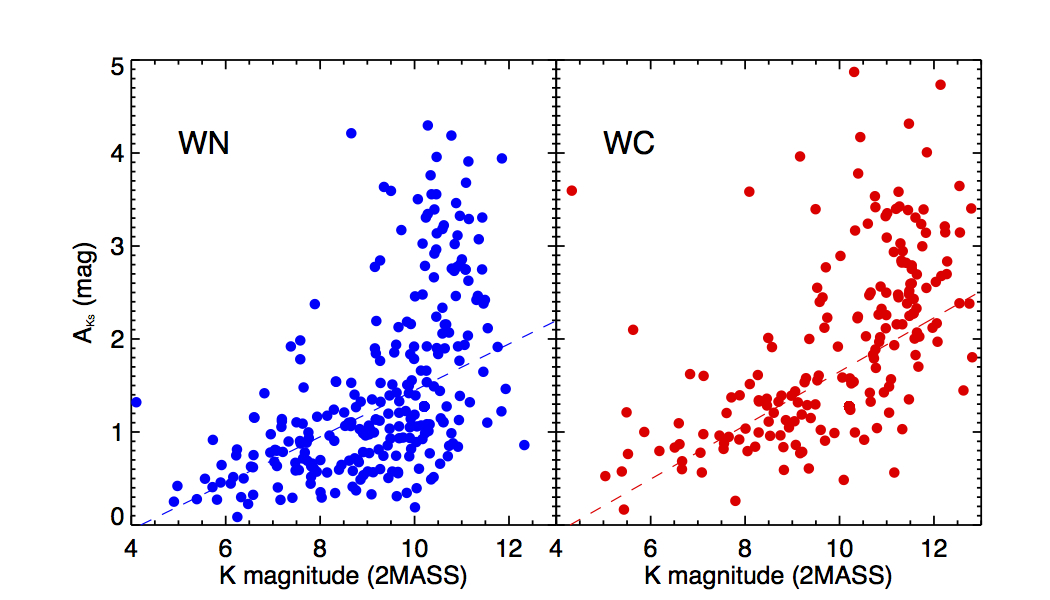}
\caption{\label{fig:ak}  $K_s$-band absorption vs magnitude, using \textit{2MASS} magnitudes of WR stars from this paper and the literature. Above shows WNs (left) and WCs (right) only, and left shows the total population. In each case a Least-Absolute-Deviation fit to the points has been included, showing the expected trend of greater extinction for fainter (and therefore more distant) Wolf--Rayet stars.}
\end{minipage}
\end{figure*}

As an independent check of our distance calculations, we plotted $K_s$-band absorption vs \textit{2MASS} $K_s$ magnitude in figure~\ref{fig:ak}. Included is a least-absolute-deviation linear fit, which, as expected, shows the increase of absorption with $K_s$ magnitude (as a directly-observable analogue for distance) for both WN and WC stars, as well as a combined plot for all WR stars.

\begin{figure*}
\includegraphics[height=0.45\textheight]{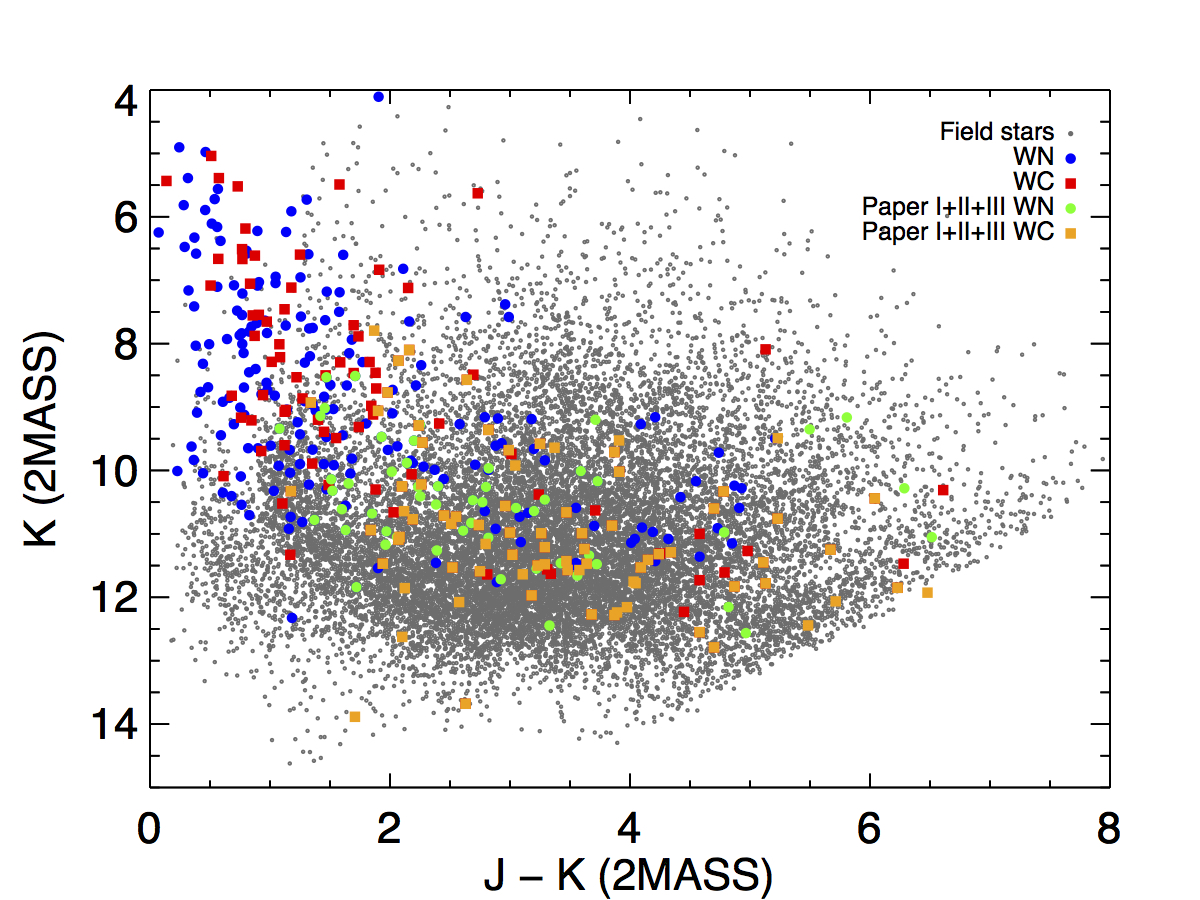}
\caption{\label{fig:kjk}  NIR colour-magnitude diagram, using \textit{2MASS} magnitudes of field and WR stars. This diagram demonstrates that the Shara et al survey is probing a more distant population of WRs than previous methods. The WRs we have identified are significantly fainter and redder than those in the literature, and would have been lost among the field stars by other selection methods.}
\end{figure*}

This imaging survey [Paper I, Paper II, and the present paper] has contributed 27 per cent of the currently identified population of Galactic WR stars. The majority of this contribution lies in regions that have not been explored previously in the literature, particularly in magnitude/colour space. Figure~\ref{fig:kjk} shows all currently-known WRs on an NIR colour-magnitude diagram. Before this survey, it might have been believed that such a diagnostic would be effective at isolating candidate WR stars, as the great majority of WRs from the literature are separated from the main field star region. However, when the results from this survey are included, it's clear that Wolf--Rayet stars are found among the crowded field regions in this diagram, and that this survey is effectively identifying faint, reddened, distant WR stars.

The set of new WR stars and other emission objects presented in this paper is not complete, nor was there any attempt at completeness; the criteria by which we selected WR candidates evolved continuously throughout the follow-up observations. Instead these new WR stars serve as a test for the image subtraction method as a means of identifying new WR stars in the survey images which had already been probed by Papers I and II. These new WR stars are a valuable addition to the catalogue of Galactic WR stars, and measures of completeness will be discussed in the next paper in this series, in preparation.

\section{Conclusions \& Looking Forward}\label{sec:end}
The imaging survey first introduced in Paper I has already produced 27 per cent of the known Galactic WR stars using photometric selection techniques \citep{2009AJ....138..402S,2012AJ....143..149S}.  Using new reductions and image-subtraction methods, we have shown in this paper that there are still many more Galactic WR star candidates to be discovered and confirmed, particularly in the Southern hemisphere.  The Galactic Centre region in particular has not been tapped, as the survey images are so crowded that special care must be taken with them. The pipeline described in this paper is capable of analysing all but the densest regions in our survey.

The WR population simulations presented in Paper I predicted several thousand WR stars in the Milky Way, which have not yet been identified. However, the results presented here are not inconsistent with those predictions, for a few reasons. First, due to the necessity of spectroscopic follow-up to confirm each WR star, even with a high success rate the number of observations necessary is very large. Second, the majority of WR stars still unidentified lie either on the far side of the Galaxy, and are thus very faint, or near the Galactic Centre in prohibitively crowded fields.

This publication makes use of data products from the Two Micron All Sky Survey, which is a joint project of the University of Massachusetts and the Infrared Processing and Analysis Centre/California Institute of Technology, funded by the National Aeronautics and Space Administration and the National Science Foundation.  This publication also makes use of data products from the Wide-field Infrared Survey Explorer, which is a joint project of the University of California, Los Angeles, and the Jet Propulsion Laboratory/California Institute of Technology, funded by the National Aeronautics and Space Administration. AFJM is grateful for financial assistance from NSERC (Canada) and FRQNT (Quebec). This research has made use of the NASA/ IPAC Infrared Science Archive, which is operated by the Jet Propulsion Laboratory, California Institute of Technology, under contract with the National Aeronautics and Space Administration.

GK and MS gratefully acknowledge support from Ethel and Hilary Lipsitz.

\nocite{*}
\bibliographystyle{mn2e}

\begin{thebibliography}{}
\makeatletter
\relax
\def\mn@urlcharsother{\let\do\@makeother \do\$\do\&\do\#\do\^\do\_\do\%\do\~}
\def\mn@doi{\begingroup\mn@urlcharsother \@ifnextchar [ {\mn@doi@}
  {\mn@doi@[]}}
\def\mn@doi@[#1]#2{\def\@tempa{#1}\ifx\@tempa\@empty \href
  {http://dx.doi.org/#2} {doi:#2}\else \href {http://dx.doi.org/#2} {#1}\fi
  \endgroup}
\def\mn@eprint#1#2{\mn@eprint@#1:#2::\@nil}
\def\mn@eprint@arXiv#1{\href {http://arxiv.org/abs/#1} {{\tt arXiv:#1}}}
\def\mn@eprint@dblp#1{\href {http://dblp.uni-trier.de/rec/bibtex/#1.xml}
  {dblp:#1}}
\def\mn@eprint@#1:#2:#3:#4\@nil{\def\@tempa {#1}\def\@tempb {#2}\def\@tempc
  {#3}\ifx \@tempc \@empty \let \@tempc \@tempb \let \@tempb \@tempa \fi \ifx
  \@tempb \@empty \def\@tempb {arXiv}\fi \@ifundefined
  {mn@eprint@\@tempb}{\@tempb:\@tempc}{\expandafter \expandafter \csname
  mn@eprint@\@tempb\endcsname \expandafter{\@tempc}}}

\bibitem[\protect\citeauthoryear{Acker et al.}{1992}]{1992secg.book.....A} 
Acker A., Marcout J., Ochsenbein F., Stenholm B., Tylenda R., Schohn C., European Southern Observatory, Garching (Germany), 1992, 1047 p.

\bibitem[\protect\citeauthoryear{Crowther}{2007}]{2007ARA&A..45..177C} Crowther P.~A., 2007, ARA\&A, 45, 177 


\bibitem[\protect\citeauthoryear{Crowther et al.}{2006}]{2006MNRAS.372.1407C} Crowther P.~A., Hadfield L.~J., Clark 
J.~S., Negueruela I., Vacca W.~D., 2006, MNRAS, 372, 1407 


\bibitem[\protect\citeauthoryear{Crowther, Morris, 
\& Smith}{2006}]{2006ApJ...636.1033C} Crowther P.~A., Morris P.~W., Smith J.~D., 2006, ApJ, 636, 1033 


\bibitem[\protect\citeauthoryear{Cushing, Vacca, 
\& Rayner}{2004}]{2004PASP..116..362C} Cushing M.~C., Vacca W.~D., Rayner J.~T., 2004, PASP, 116, 362 


\bibitem[\protect\citeauthoryear{Detmers et 
al.}{2008}]{2008A&A...484..831D} Detmers R.~G., Langer N., Podsiadlowski P., Izzard R.~G., 2008, A\&A, 484, 831 


\bibitem[\protect\citeauthoryear{Evans et al.}{2010}]{2010ApJS..189...37E} 
Evans I.~N., et al., 2010, ApJS, 189, 37 


\bibitem[\protect\citeauthoryear{Faherty et 
al.}{2014}]{2014AJ....147..115F} Faherty J.~K., Shara M.~M., Zurek D., 
Kanarek G., Moffat A.~F.~J., 2014, AJ, 147, 115 


\bibitem[\protect\citeauthoryear{Figer, McLean, 
\& Najarro}{1997}]{1997ApJ...486..420F} Figer D.~F., McLean I.~S., Najarro F., 1997, ApJ, 486, 420 


\bibitem[\protect\citeauthoryear{Greene 
\& Lada}{1996}]{1996AJ....112.2184G} Greene T.~P., Lada C.~J., 1996, AJ, 112, 2184 


\bibitem[\protect\citeauthoryear{Guerrero 
\& Chu}{2008}]{2008ApJS..177..238G} Guerrero M.~A., Chu Y.-H., 2008, ApJS, 177, 238 


\bibitem[\protect\citeauthoryear{Gvaramadze, Kniazev, 
\& Fabrika}{2010}]{2010MNRAS.405.1047G} Gvaramadze V.~V., Kniazev A.~Y., Fabrika S., 2010, MNRAS, 405, 1047 


\bibitem[\protect\citeauthoryear{Hadfield et 
al.}{2007}]{2007MNRAS.376..248H} Hadfield L.~J., van Dyk S.~D., Morris 
P.~W., Smith J.~D., Marston A.~P., Peterson D.~E., 2007, MNRAS, 376, 248 


\bibitem[\protect\citeauthoryear{Indebetouw et 
al.}{2005}]{2005ApJ...619..931I} Indebetouw R., et al., 2005, ApJ, 619, 931 


\bibitem[\protect\citeauthoryear{Kerber et 
al.}{2003}]{2003A&A...408.1029K} Kerber F., Mignani R.~P., Guglielmetti F., Wicenec A., 2003, A\&A, 408, 1029 


\bibitem[\protect\citeauthoryear{L{\'e}pine 
\& Moffat}{1999}]{1999ApJ...514..909L} L{\'e}pine S., Moffat A.~F.~J., 1999, ApJ, 514, 909 


\bibitem[\protect\citeauthoryear{Lan{\c c}on et 
al.}{2007}]{2007A&A...468..205L} Lan{\c c}on A., Hauschildt P.~H., Ladjal D., Mouhcine M., 2007, A\&A, 468, 205 


\bibitem[\protect\citeauthoryear{Lan{\c c}on 
\& Wood}{2000}]{2000A&AS..146..217L} Lan{\c c}on A., Wood P.~R., 2000, A\&AS, 146, 217 


\bibitem[\protect\citeauthoryear{Landsman}{1993}]{1993ASPC...52..246L} 
Landsman W.~B., 1993, ASPC, 52, 246 


\bibitem[\protect\citeauthoryear{Lang et al.}{2010}]{2010AJ....139.1782L} 
Lang D., Hogg D.~W., Mierle K., Blanton M., Roweis S., 2010, AJ, 139, 1782 


\bibitem[\protect\citeauthoryear{Mauerhan et 
al.}{2010}]{2010ApJ...710..706M} Mauerhan J.~C., Muno M.~P., Morris M.~R., 
Stolovy S.~R., Cotera A., 2010, ApJ, 710, 706 


\bibitem[\protect\citeauthoryear{Mauerhan, Van Dyk, 
\& Morris}{2009}]{2009AAS...21460509M} Mauerhan J., Van Dyk S., Morris P., 2009, AAS, 214, \#605.09 


\bibitem[\protect\citeauthoryear{Mauerhan, Van Dyk, 
\& Morris}{2011}]{2011AJ....142...40M} Mauerhan J.~C., Van Dyk S.~D., Morris P.~W., 2011, AJ, 142, 40 


\bibitem[\protect\citeauthoryear{Messineo et 
al.}{2012}]{2012A&A...537A..10M} Messineo M., Menten K.~M., Churchwell E., Habing H., 2012, A\&A, 537, AA10 


\bibitem[\protect\citeauthoryear{Miszalski et 
al.}{2008}]{2008MNRAS.384..525M} Miszalski B., Parker Q.~A., Acker A., 
Birkby J.~L., Frew D.~J., Kovacevic A., 2008, MNRAS, 384, 525 


\bibitem[\protect\citeauthoryear{Predehl 
\& Schmitt}{1995}]{1995A&A...293..889P} Predehl P., Schmitt J.~H.~M.~M., 1995, A\&A, 293, 889 


\bibitem[\protect\citeauthoryear{Rayner, Cushing, 
\& Vacca}{2009}]{2009ApJS..185..289R} Rayner J.~T., Cushing M.~C., Vacca W.~D., 2009, ApJS, 185, 289 


\bibitem[\protect\citeauthoryear{Rieke 
\& Lebofsky}{1985}]{1985ApJ...288..618R} Rieke G.~H., Lebofsky M.~J., 1985, ApJ, 288, 618 


\bibitem[\protect\citeauthoryear{Rosslowe 
\& Crowther}{2015}]{2015MNRAS.447.2322R} Rosslowe C.~K., Crowther P.~A., 2015, MNRAS, 447, 2322 


\bibitem[\protect\citeauthoryear{Shara et al.}{1999}]{1999AJ....118..390S} 
Shara M.~M., Moffat A.~F.~J., Smith L.~F., Niemela V.~S., Potter M., 
Lamontagne R., 1999, AJ, 118, 390 


\bibitem[\protect\citeauthoryear{Shara et al.}{2009}]{2009AJ....138..402S} 
Shara M.~M., et al., 2009, AJ, 138, 402 (Paper I)


\bibitem[\protect\citeauthoryear{Shara et al.}{2012}]{2012AJ....143..149S} 
Shara M.~M., Faherty J.~K., Zurek D., Moffat A.~F.~J., Gerke J., Doyon R., 
Artigau E., Drissen L., 2012, AJ, 143, 149 (Paper II)


\bibitem[\protect\citeauthoryear{Skinner et 
al.}{2006}]{2006Ap&SS.304...97S} Skinner S., G{\"u}del M., Schmutz W., Zhekov S., 2006, Ap\&SS, 304, 97 


\bibitem[\protect\citeauthoryear{Skrutskie et 
al.}{2006}]{2006AJ....131.1163S} Skrutskie M.~F., et al., 2006, AJ, 131, 
1163 


\bibitem[\protect\citeauthoryear{Smith et al.}{2012}]{2012AJ....144..166S} 
Smith J.~D.~T., Cushing M., Barletta A., McCarthy D., Kulesa C., Van Dyk 
S.~D., 2012, AJ, 144, 166 


\bibitem[\protect\citeauthoryear{Stetson}{1987}]{1987PASP...99..191S} 
Stetson P.~B., 1987, PASP, 99, 191


\bibitem[\protect\citeauthoryear{Stetson}{1994}]{1994PASP..106..250S} 
Stetson P.~B., 1994, PASP, 106, 250  


\bibitem[\protect\citeauthoryear{Stringfellow et 
al.}{2012}]{2012IAUS..282..267S} Stringfellow G.~S., Gvaramadze V.~V., 
Beletsky Y., Kniazev A.~Y., 2012, IAUS, 282, 267 


\bibitem[\protect\citeauthoryear{Urquhart et 
al.}{2009}]{2009A&A...501..539U} Urquhart J.~S., et al., 2009, A\&A, 501, 539 


\bibitem[\protect\citeauthoryear{Vacca, Cushing, 
\& Rayner}{2004}]{2004PASP..116..352V} Vacca W.~D., Cushing M.~C., Rayner J.~T., 2004, PASP, 116, 352 


\bibitem[\protect\citeauthoryear{van der 
Hucht}{2006}]{2006A&A...458..453V} van der Hucht K.~A., 2006, A\&A, 458, 453 


\bibitem[\protect\citeauthoryear{van der Hucht}{2001}]{2001yCat.3215....0V} 
van der Hucht K.~A., 2001, yCat, 3215, 0 


\bibitem[\protect\citeauthoryear{Wachter et 
al.}{2011}]{2011BSRSL..80..291W} Wachter S., Mauerhan J., van Dyk S., Hoard 
D.~W., Morris P., 2011, BSRSL, 80, 291 


\bibitem[\protect\citeauthoryear{Wolf \& Rayet}{Wolf \& Rayet}{1867}]{Wolf:jk}
Wolf C. J.~E.,  Rayet G.,  1867, Comptes Rendus, 65, 292


\bibitem[\protect\citeauthoryear{Woosley 
\& Bloom}{2006}]{2006ARA&A..44..507W} Woosley S.~E., Bloom J.~S., 2006, ARA\&A, 44, 507 


\bibitem[\protect\citeauthoryear{Wright et al.}{2010}]{2010AJ....140.1868W} 
Wright E.~L., et al., 2010, AJ, 140, 1868 

\makeatother
\end{thebibliography}

\begin{figure*}
\begin{minipage}[c]{\linewidth}\hfill
\includegraphics[width=0.9\linewidth]{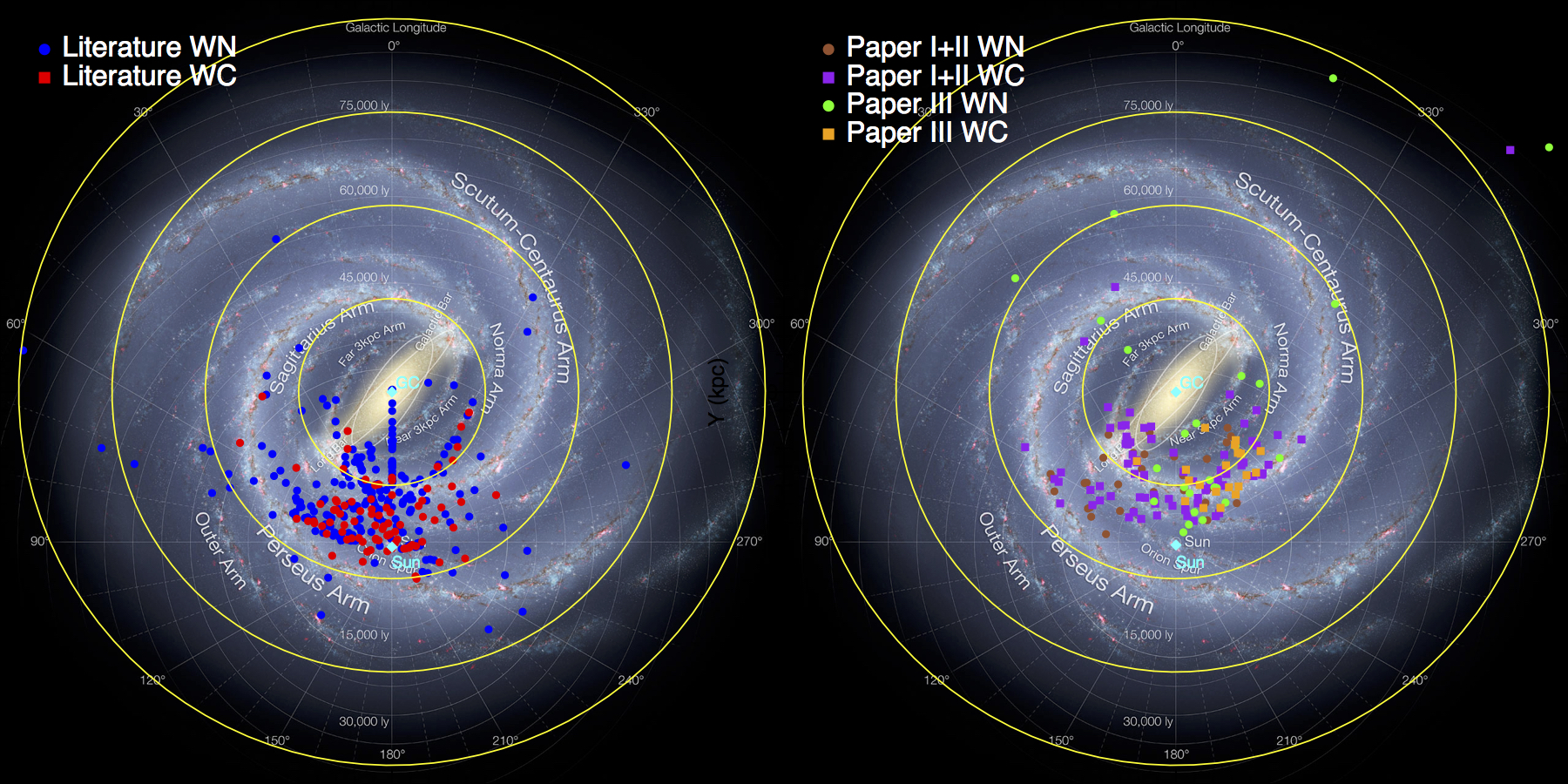}
\end{minipage}
\begin{minipage}[c]{0.70\linewidth}
\includegraphics[width=\linewidth]{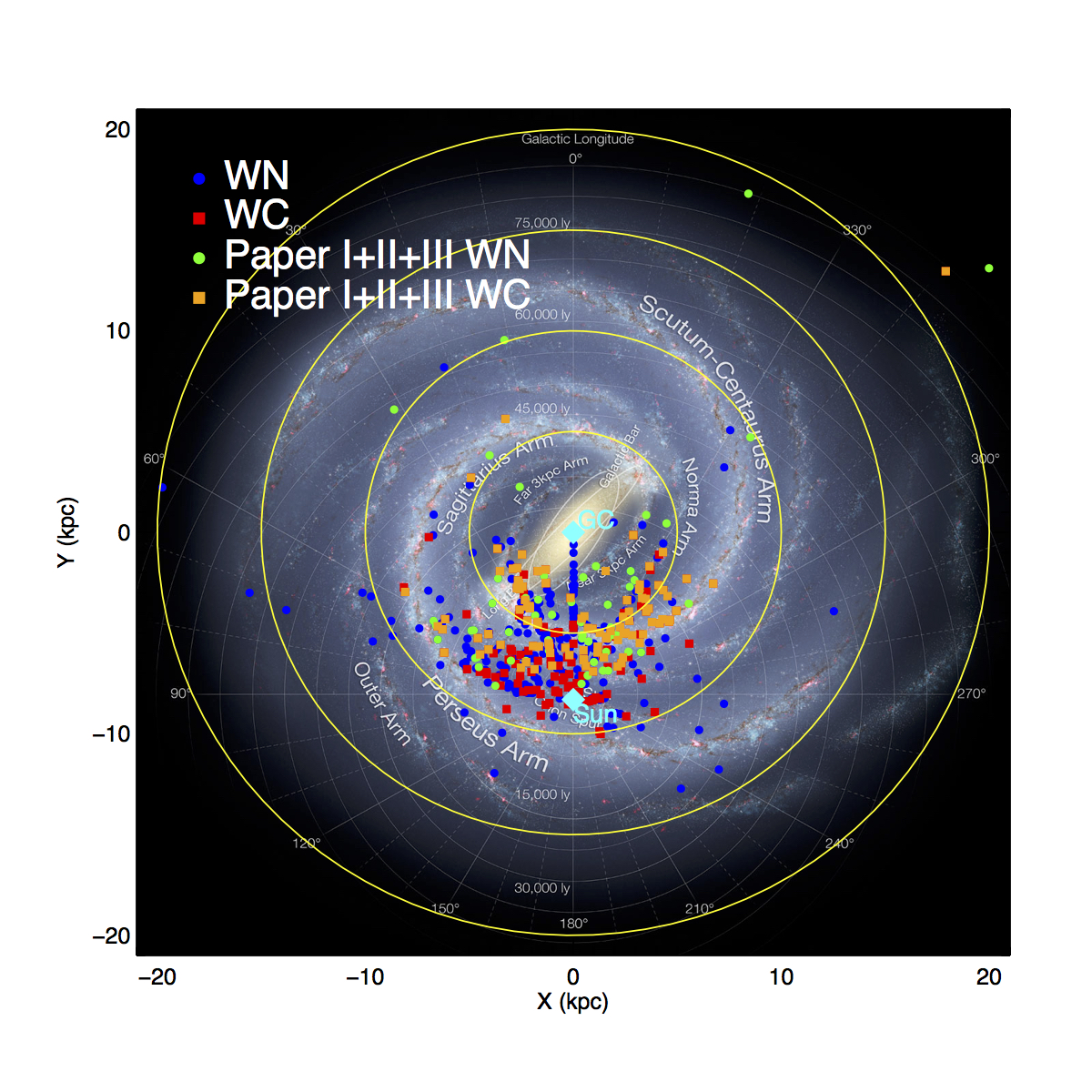}
\end{minipage}\hfill
\begin{minipage}[c]{0.28\linewidth}
\caption{\label{fig:galp}  The distribution of confirmed WR stars projected on to the plane of the Galaxy, superimposed over an artist's representation of the Milky Way (image credit: NASA/JPL--Caltech/ESO/R. Hurt). Top shows WRs from the literature only (left), and from Papers I, II, and this paper (right). This highlights the different regions of the Galaxy probed by the Shara et al. survey; in particular, note that the Shara survey  has more than doubled the number of confirmed WR stars on the far side of the Galaxy. All confirmed Galactic WR stars are shown in the bottom plot, where the distribution can clearly be seen to trace out the spiral arms on the near side of the Galaxy. Thus, as new WRs are identified on the far side, we can use that growing population as tracers of recent massive star formation, and infer a map of the spiral arms on the Galactic far side.}
\end{minipage}
\end{figure*}

\clearpage
\appendix
\begin{figure*}
\begin{minipage}[c]{0.48\linewidth}
\section[]{Finder charts for confirmed WR stars}
Included here are the first 9 finder charts for confirmed WR stars discussed in this paper, taken from the Shara et al. J and CONT2 (as an analogue for \textit{2MASS} $K_s$) survey images. The remainder of the finder charts are available as supplementary online-only data.
\vspace{16.7mm}

\includegraphics[width=\linewidth]{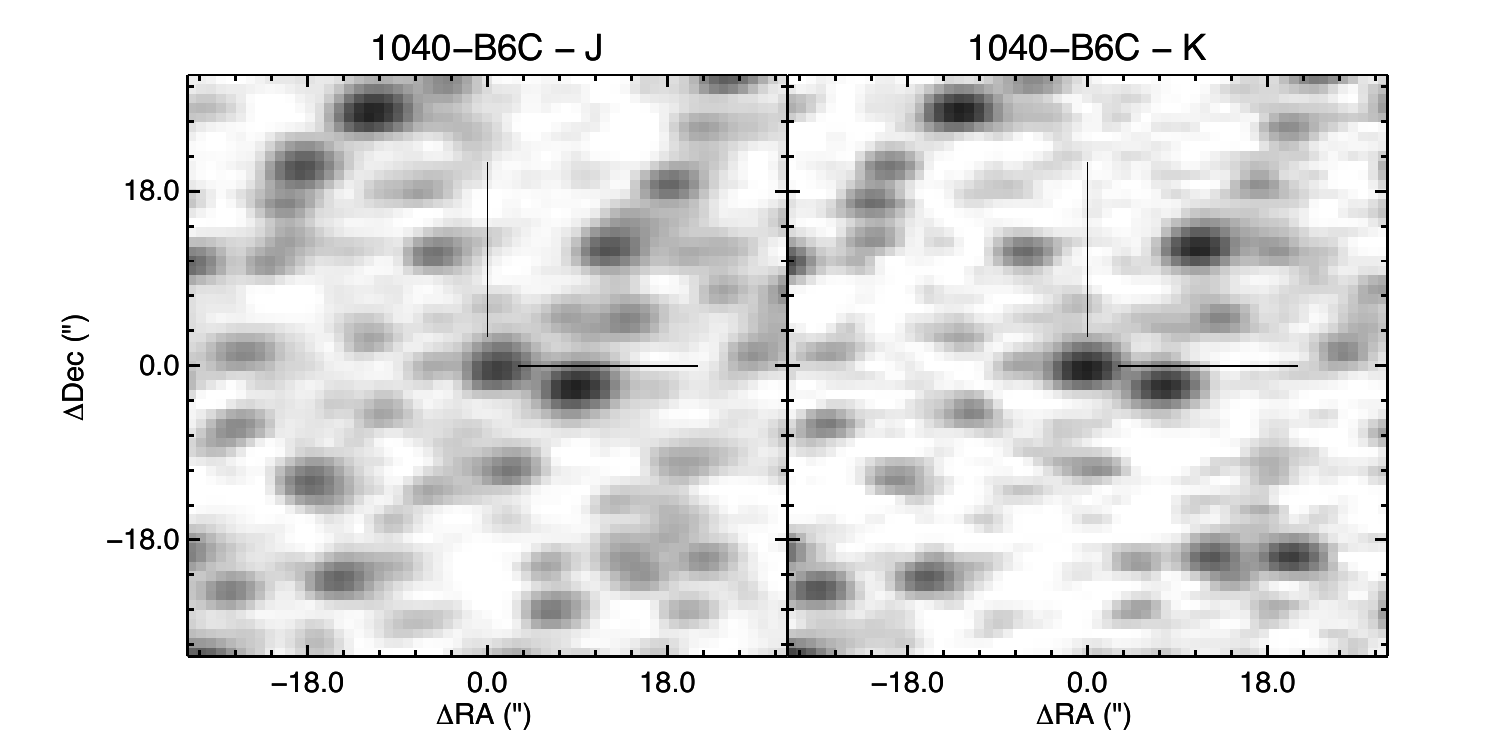}
\includegraphics[width=\linewidth]{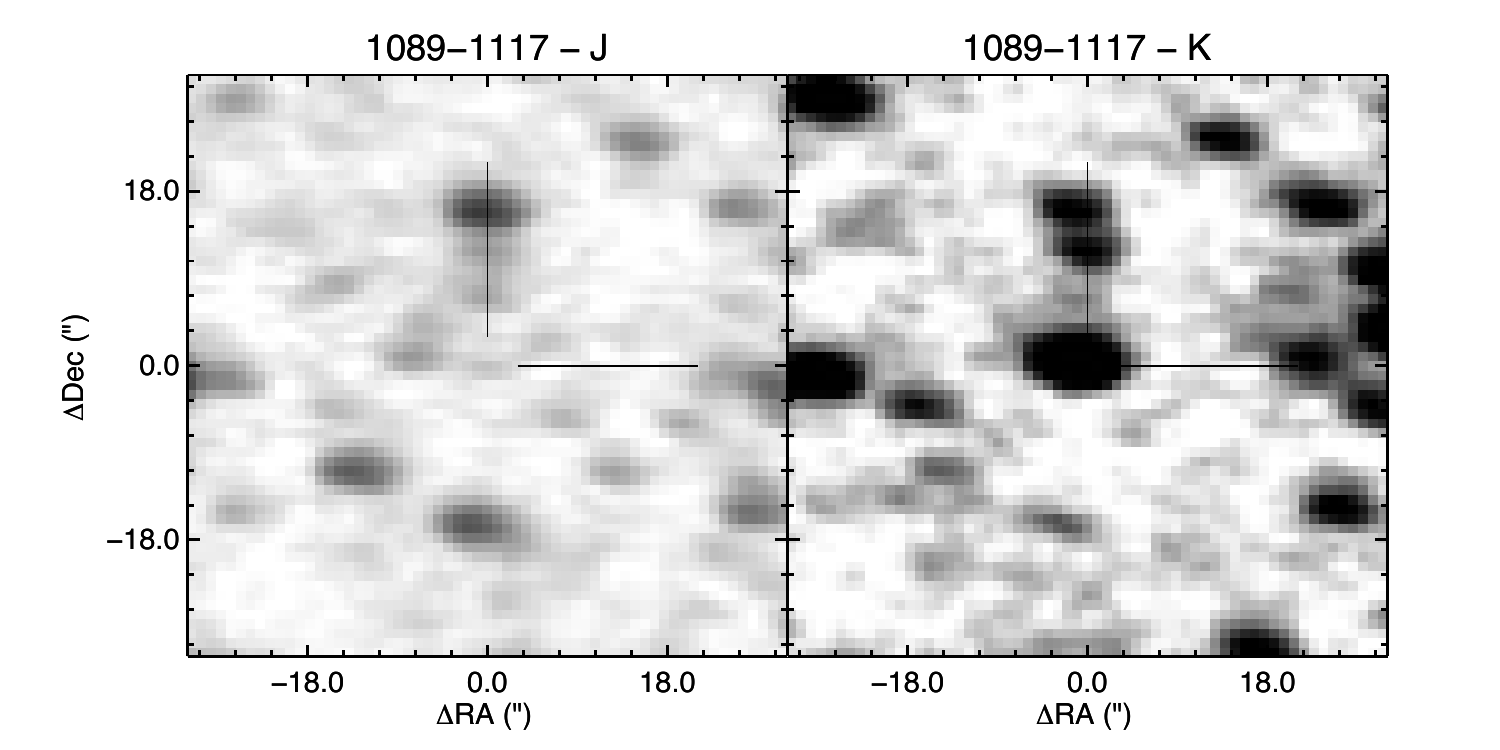}
\includegraphics[width=\linewidth]{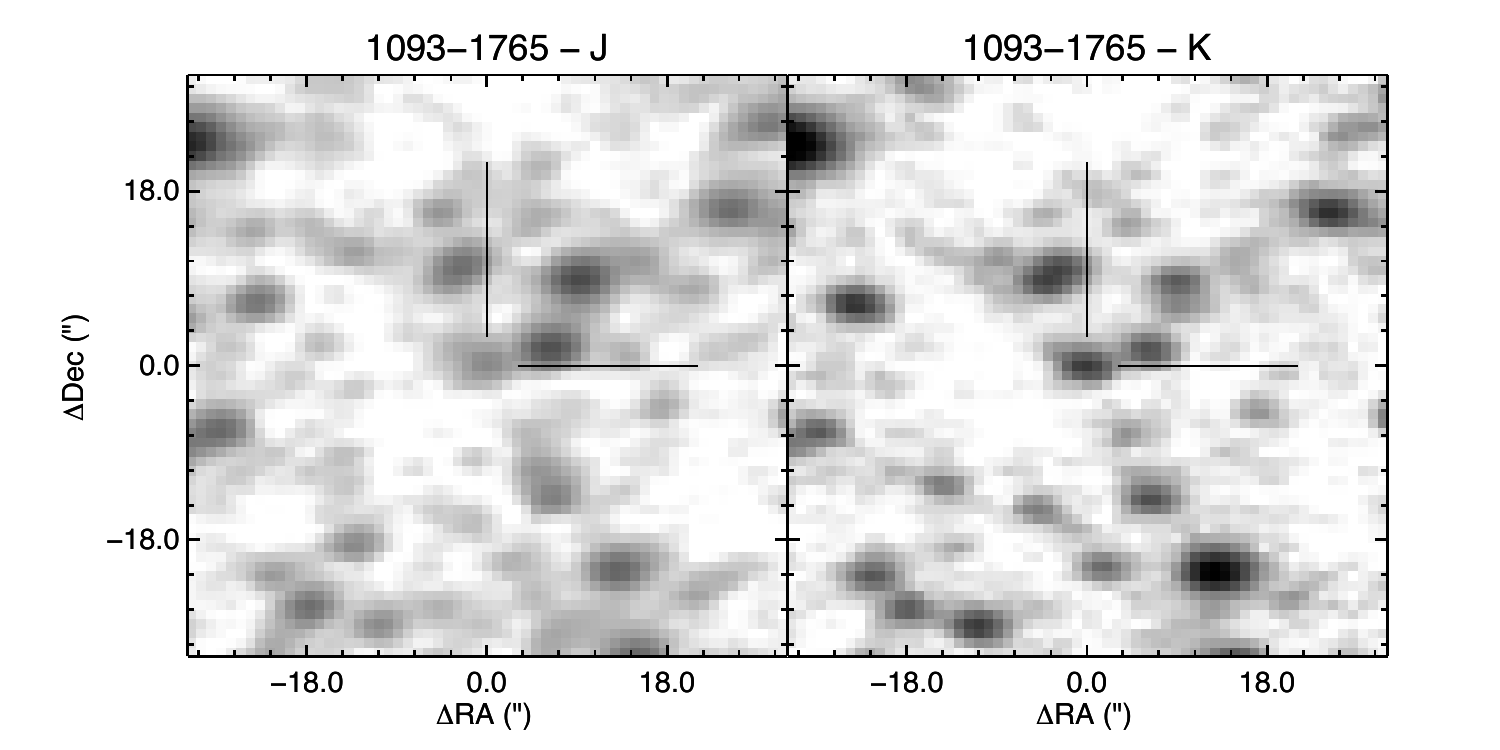}
\includegraphics[width=\linewidth]{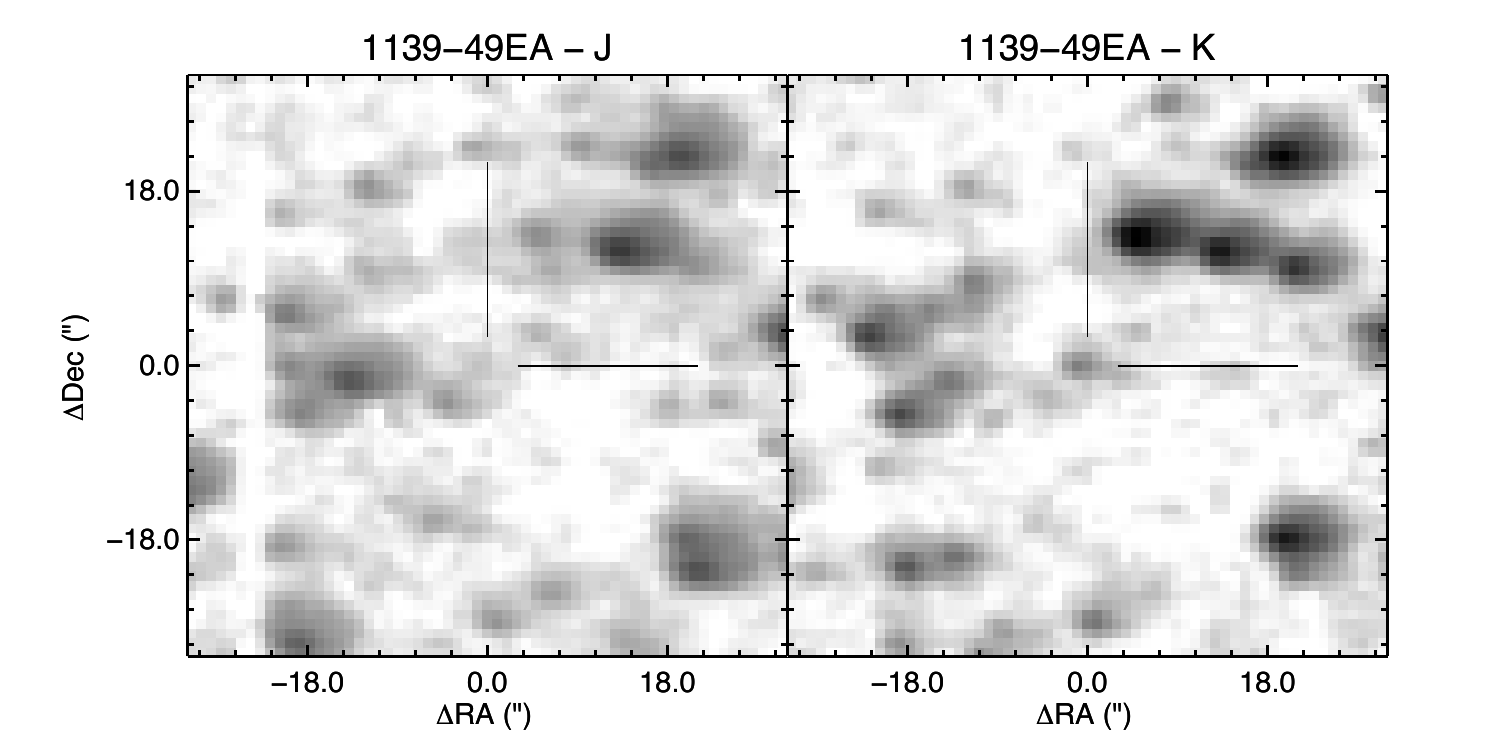}
\end{minipage}\hfill
\begin{minipage}[c]{0.48\linewidth}
\includegraphics[width=\linewidth]{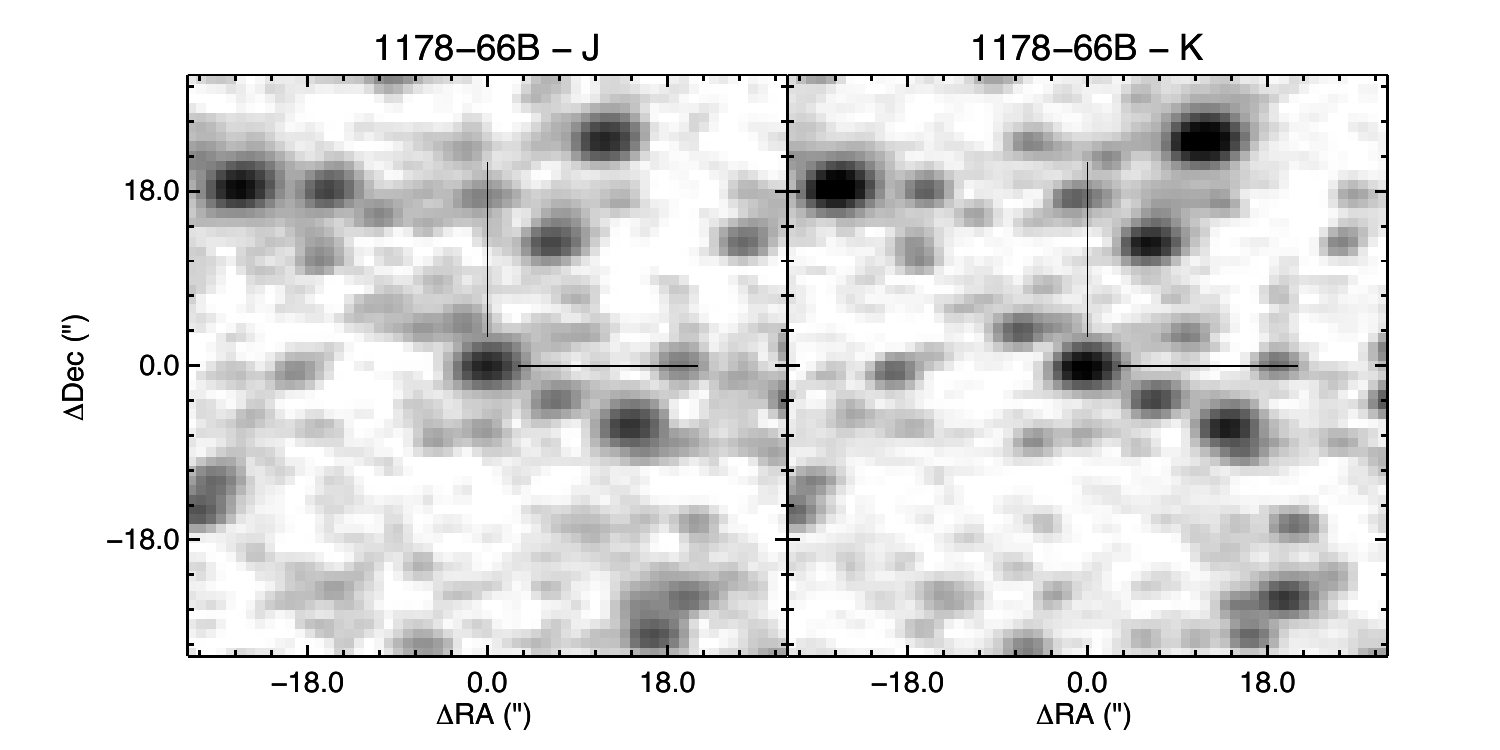}
\includegraphics[width=\linewidth]{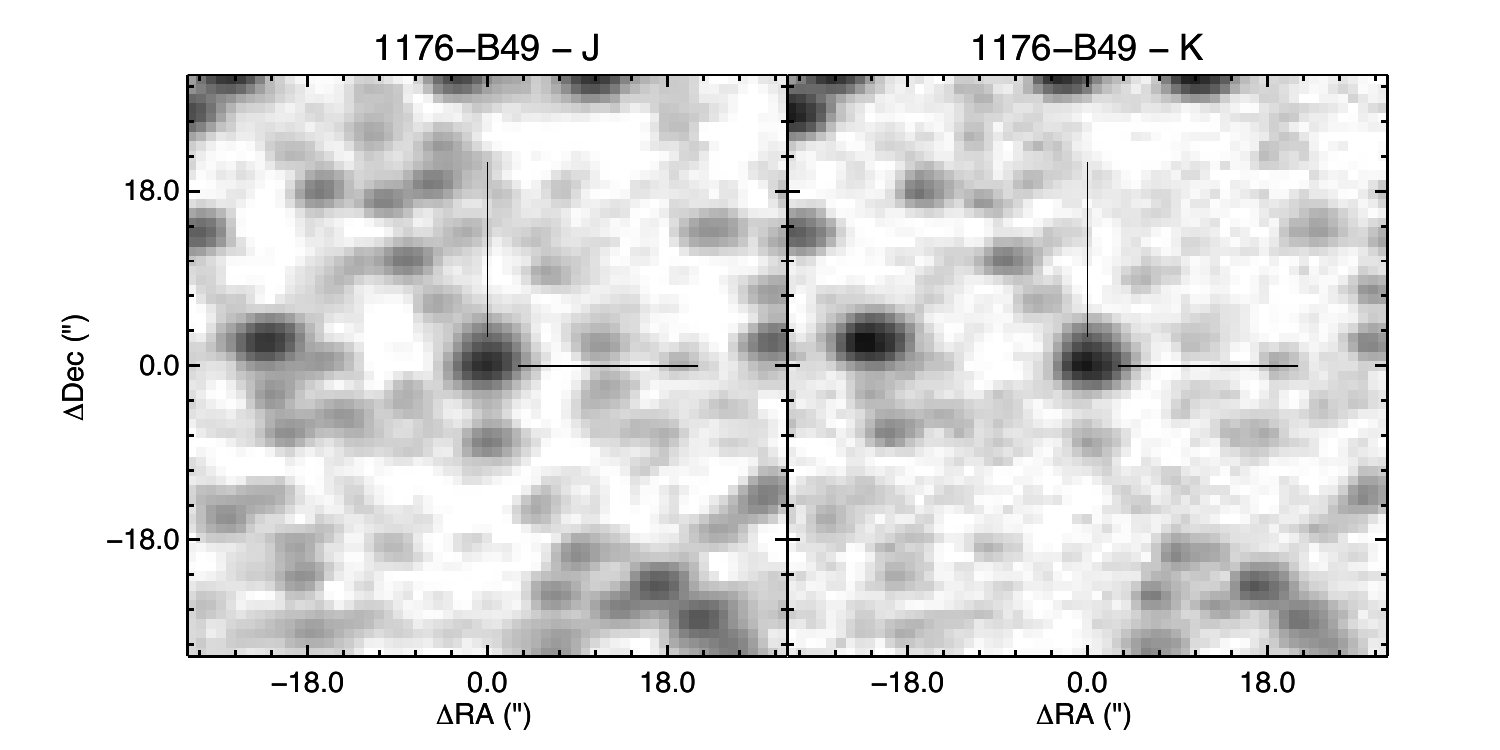}
\includegraphics[width=\linewidth]{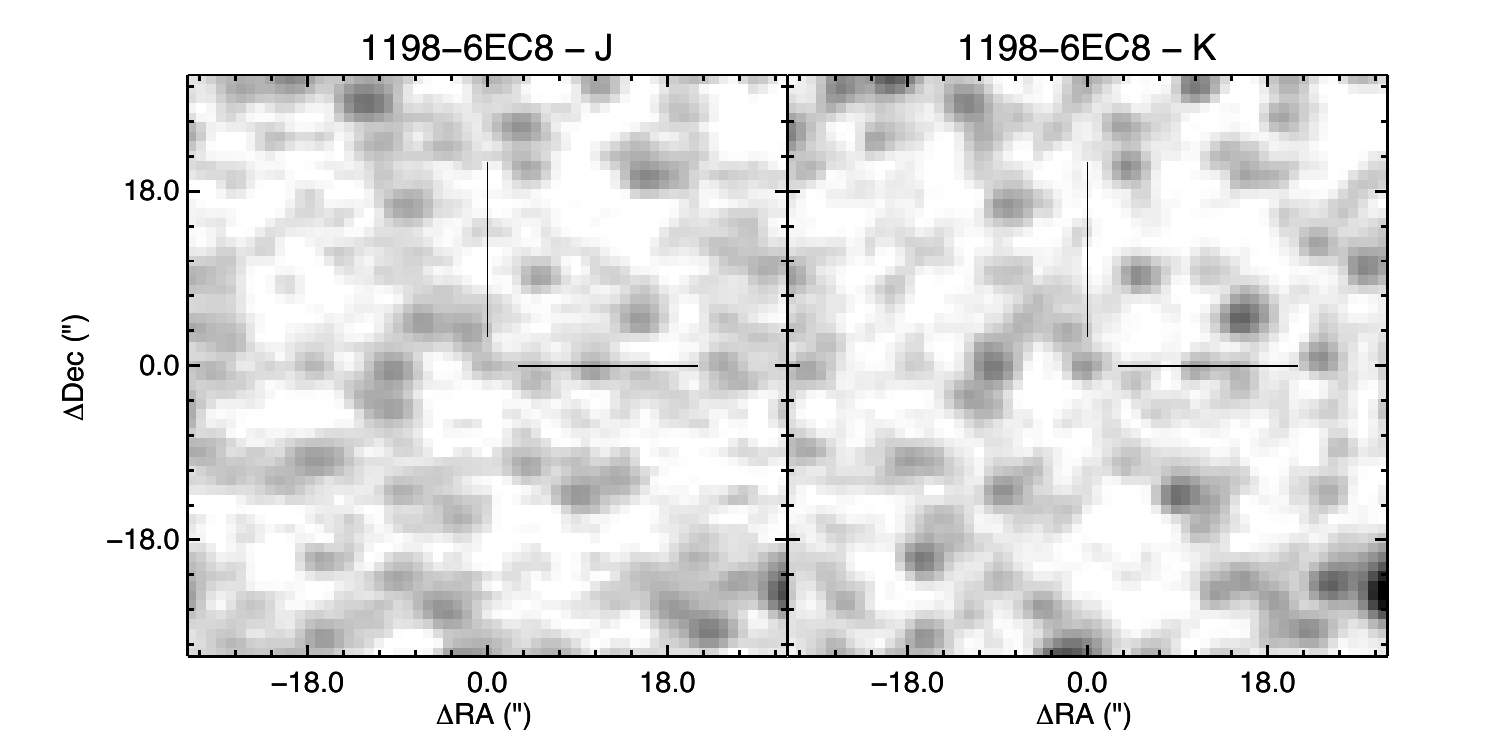}
\includegraphics[width=\linewidth]{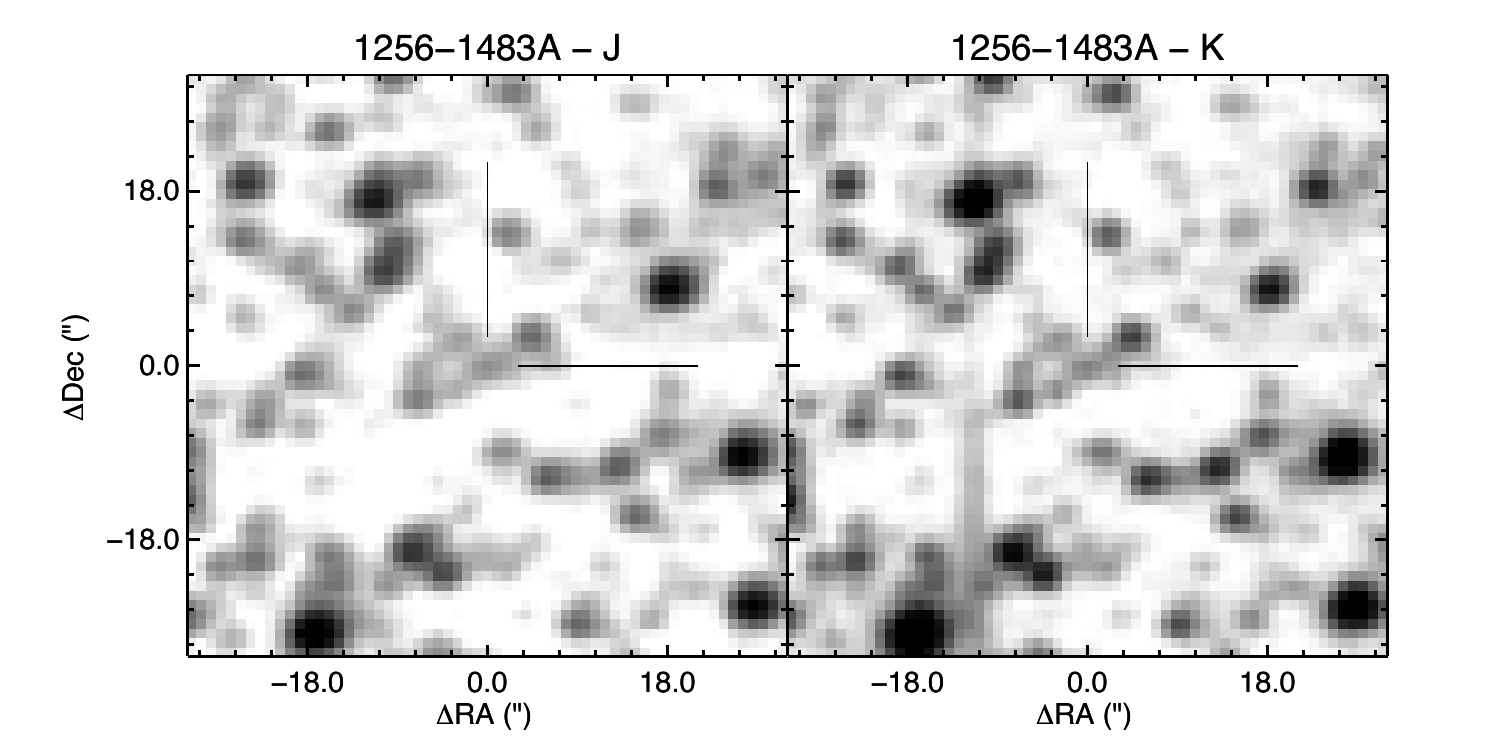}
\includegraphics[width=\linewidth]{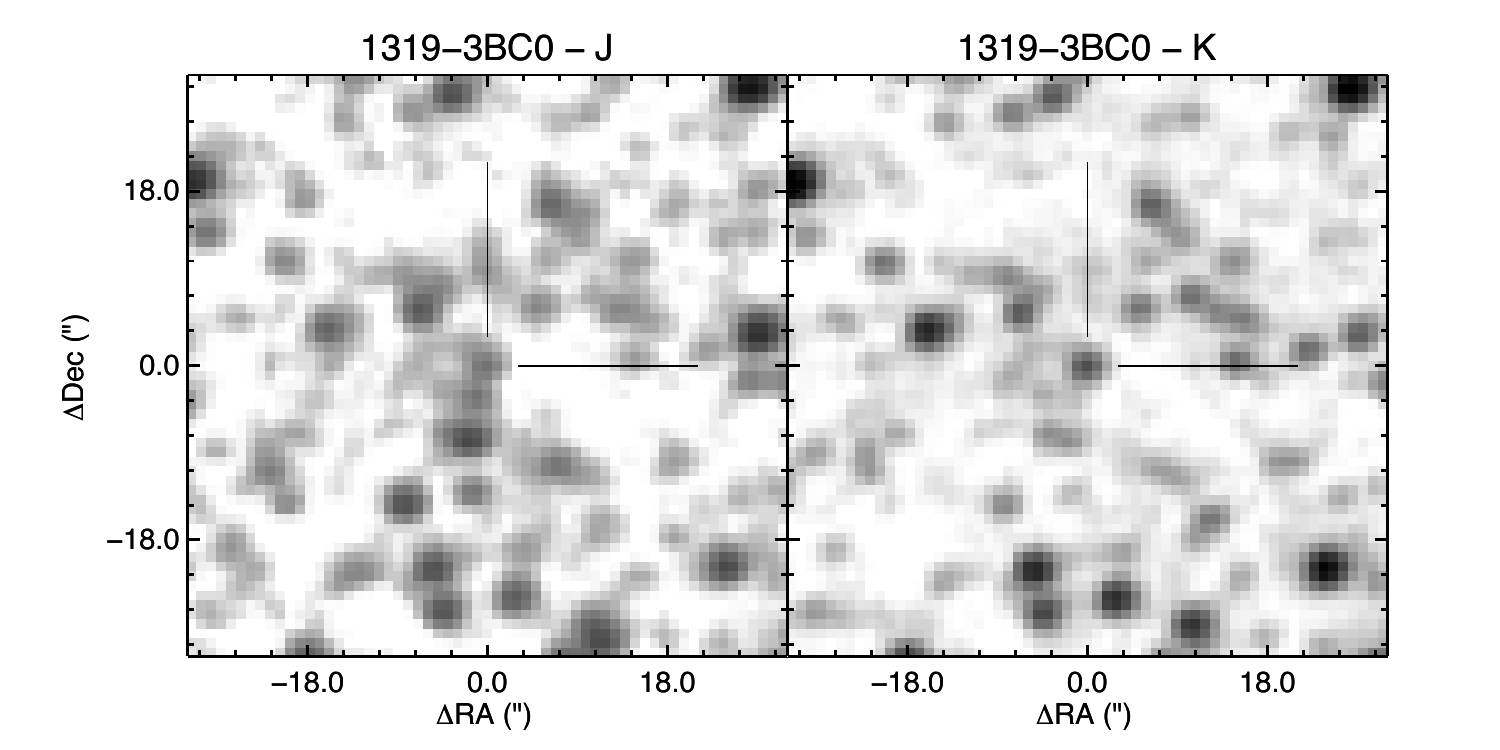}
\end{minipage}
\end{figure*}
\label{lastpage}
\end{document}